\documentclass[aps,prb,reprint,showkeys,superscriptaddress]{revtex4-1}
\usepackage{subcaption}
\usepackage{bm,graphicx,tabularx,array,booktabs,dcolumn,xcolor,microtype,multirow,amscd,amsmath,amssymb,amsfonts,physics,siunitx}
\usepackage[version=4]{mhchem}
\usepackage[utf8]{inputenc}
\usepackage[T1]{fontenc}
\usepackage{txfonts}

\usepackage[normalem]{ulem}
\definecolor{hughgreen}{RGB}{0, 128, 0}
\newcommand{\titou}[1]{\textcolor{black}{#1}}

\usepackage[
	colorlinks=true,
    citecolor=blue,
    linkcolor=blue,
    filecolor=blue,      
    urlcolor=blue,	
    breaklinks=true
	]{hyperref}
\newcommand{\latin}[1]{#1}
\newcommand{\ie}{\latin{i.e.}}
\newcommand{\eg}{\latin{e.g.}}
\newcommand{\etal}{\textit{et al.}}

\newcommand{\mc}{\multicolumn}

\newcommand{\bH}{\mathbf{H}}
\newcommand{\bV}{\mathbf{V}}

\newcommand{\cS}{\mathcal{S}}

\newcommand{\Ne}{N} 
\newcommand{\Nn}{M} 
\newcommand{\hI}{\Hat{I}}
\newcommand{\hH}{\Hat{H}}

\newcommand{\hV}{\Hat{V}}

\newcolumntype{Y}{>{\centering\arraybackslash}X}

\newcommand{\ta}{\theta^{\,\alpha}}
\newcommand{\tb}{\theta^{\,\beta}}
\newcommand{\ts}{\theta^{\sigma}}

\renewcommand{\i}{\mathrm{i}} 
\newcommand{\e}{\mathrm{e}} 
\newcommand{\rc}{r_{\text{c}}}
\newcommand{\lc}{\lambda_{\text{c}}}

\newcommand{\lep}{\lambda_{\text{EP}}}

\newcommand{\Emp}{E_{\text{MP}}}

\newcommand{\bbC}{\mathbb{C}}

\newcommand{\Lup}{\mathcal{L}^{\uparrow}}
\newcommand{\Ldown}{\mathcal{L}^{\downarrow}}
\newcommand{\Lsi}{\mathcal{L}^{\sigma}}
\newcommand{\Rup}{\mathcal{R}^{\uparrow}}
\newcommand{\Rdown}{\mathcal{R}^{\downarrow}}
\newcommand{\Rsi}{\mathcal{R}^{\sigma}}
\newcommand{\vhf}{\Hat{v}_{\text{HF}}}
\newcommand{\whf}{\Psi_{\text{HF}}}

\newcommand{\LCPQ}{Laboratoire de Chimie et Physique Quantiques (UMR 5626), Universit\'e de Toulouse, CNRS, UPS, France.}
\newcommand{\UOX}{Physical and Theoretical Chemical Laboratory, Department of Chemistry, University of Oxford, Oxford, OX1 3QZ, U.K.}
\begin{document}	

\title{Perturbation Theory in the Complex Plane: Exceptional Points and Where to Find Them}

\author{Antoine \surname{Marie}}
\affiliation{\LCPQ}
\author{Hugh G.~A.~\surname{Burton}}
\email[Corresponding author: ]{hugh.burton@chem.ox.ac.uk}
\affiliation{\UOX}
\author{Pierre-Fran\c{c}ois \surname{Loos}}
\email[Corresponding author: ]{loos@irsamc.ups-tlse.fr}
\affiliation{\LCPQ}

\begin{abstract}
We explore the non-Hermitian extension of quantum chemistry in the complex plane and its link with perturbation theory.
We observe that the physics of a quantum system is intimately connected to the position of complex-valued energy singularities, known as exceptional points.
After presenting the fundamental concepts of non-Hermitian quantum chemistry in the complex plane, including the mean-field Hartree--Fock approximation and Rayleigh--Schr\"odinger perturbation theory, we provide a historical overview of the various research activities that have been performed on the physics of singularities.
In particular, we highlight seminal work on the convergence behaviour of perturbative series obtained within M{\o}ller--Plesset perturbation theory, and its links with quantum phase transitions.
We also discuss several resummation techniques (such as Pad\'e and quadratic approximants) that can improve the overall accuracy of the M{\o}ller--Plesset perturbative series in both convergent and divergent cases.
Each of these points is illustrated using the Hubbard dimer at half filling, which proves to be a versatile model for understanding the subtlety of analytically-continued perturbation theory in the complex plane.
\end{abstract}


\maketitle

\section{Introduction}
\label{sec:intro}

Perturbation theory isn't usually considered in the complex plane.
Normally, it is applied using real numbers as one of very few available tools for 
describing realistic quantum systems.
In particular, time-independent Rayleigh--Schr\"odinger perturbation theory\cite{RayleighBook,Schrodinger_1926} 
has emerged as an instrument of choice among the vast array of methods developed for this purpose.%
\cite{SzaboBook,JensenBook,CramerBook,HelgakerBook,ParrBook,FetterBook,ReiningBook}
However, the properties of perturbation theory in the complex plane
are essential for understanding the quality of perturbative approximations on the real axis.

In electronic structure theory, the workhorse of time-independent perturbation theory is M\o{}ller--Plesset (MP) 
theory,\cite{Moller_1934} which remains one of the most popular methods for computing the electron 
correlation energy.\cite{Wigner_1934,Lowdin_1958}
This approach estimates the exact electronic energy by constructing a perturbative correction on top
of a mean-field Hartree--Fock (HF) approximation.\cite{SzaboBook}
The popularity of MP theory stems from its black-box nature, size-extensivity, and relatively low computational scaling, 
making it easily applied in a broad range of molecular research.\cite{HelgakerBook}
However, it is now widely recognised that the series of MP approximations (defined for a given perturbation
order $n$ as MP$n$) can show erratic, slow, or divergent behaviour that limit its systematic improvability.%
\cite{Laidig_1985,Knowles_1985,Handy_1985,Gill_1986,Laidig_1987,Nobes_1987,Gill_1988,Gill_1988a,Lepetit_1988,Malrieu_2003} 
As a result, practical applications typically employ only the lowest-order MP2 approach, while 
the successive MP3, MP4, and MP5 (and higher order) terms are generally not considered to offer enough improvement
to justify their increased cost.
Turning the MP approximations into a convergent and 
systematically improvable series largely remains an open challenge.

Our conventional view of electronic structure theory is centred around the Hermitian notion of quantised energy levels,
where the different electronic states of a molecule are discrete and energetically ordered.
The lowest energy state defines the ground electronic state, while higher energy states
represent electronic excited states.
However, an entirely different perspective on quantisation can be found by analytically continuing
quantum mechanics into the complex domain.
In this inherently non-Hermitian framework, the energy levels emerge as individual \textit{sheets} of a complex
multi-valued function and can be connected as one continuous \textit{Riemann surface}.\cite{BenderPTBook}
This connection is possible because the orderability of real numbers is lost when energies are extended to the
complex domain.
As a result, our quantised view of conventional quantum mechanics only arises from
restricting our domain to Hermitian approximations.

Non-Hermitian Hamiltonians already have a long history in quantum chemistry and have been extensively used to 
describe metastable resonance phenomena.\cite{MoiseyevBook}
Through the methods of complex-scaling\cite{Moiseyev_1998} and complex absorbing 
potentials,\cite{Riss_1993,Ernzerhof_2006,Benda_2018} outgoing resonances can be stabilised as square-integrable
wave functions.
\titou{In these situations, the energy becomes complex-valued, with the real and imaginary components allowing
the resonance energy and lifetime to be computed respectively.}
We refer the interested reader to the excellent book by Moiseyev for a general overview. \cite{MoiseyevBook}

The Riemann surface for the electronic energy $E(\lambda)$ with a coupling parameter $\lambda$ can be 
constructed by analytically continuing the function into the complex $\lambda$ domain.
In the process, the ground and excited states become smoothly connected and form a continuous complex-valued
energy surface. 
\textit{Exceptional points} (EPs) can exist on this energy surface, corresponding to branch point 
singularities where two (or more) states become exactly degenerate.%
\cite{MoiseyevBook,Heiss_1988,Heiss_1990,Heiss_1999,Berry_2011,Heiss_2012,Heiss_2016,Benda_2018}
While EPs can be considered as the non-Hermitian analogues of conical intersections,\cite{Yarkony_1996} 
the behaviour of their eigenvalues near a degeneracy could not be more different.
Incredibly, following the eigenvalues around an EP leads to the interconversion of the degenerate states, 
and multiple loops around the EP are required to recover the initial energy.\cite{MoiseyevBook,Heiss_2016,Benda_2018}
In contrast, encircling a conical intersection leaves the states unchanged.
Furthermore, while the eigenvectors remain orthogonal at a conical intersection, the eigenvectors at an EP
become identical and result in a \textit{self-orthogonal} state. \cite{MoiseyevBook}
An EP effectively creates a ``portal'' between ground and excited-states in the complex plane.%
\cite{Burton_2019,Burton_2019a}
This transition between states has been experimentally observed in electronics, 
microwaves, mechanics, acoustics, atomic systems and optics.\cite{Bittner_2012,Chong_2011,Chtchelkatchev_2012,Doppler_2016,Guo_2009,Hang_2013,Liertzer_2012,Longhi_2010,Peng_2014, Peng_2014a,Regensburger_2012,Ruter_2010,Schindler_2011,Szameit_2011,Zhao_2010,Zheng_2013,Choi_2018,El-Ganainy_2018}

The MP energy correction can be considered as a function of the perturbation parameter $\lambda$.
When the domain of $\lambda$ is extended to the complex plane, EPs can also occur in the MP energy.
Although these EPs are generally complex-valued, 
their positions are intimately related to the 
convergence of the perturbation expansion on the real axis.%
\cite{BenderBook,Olsen_1996,Olsen_2000,Olsen_2019,Mihalka_2017a,Mihalka_2017b,Mihalka_2019}
Furthermore, the existence of an avoided crossing on the real axis is indicative of a nearby EP
in the complex plane.
Our aim in this article is to provide a comprehensive review of the fundamental relationship between EPs
and the convergence properties of the MP series.
In doing so, we will demonstrate how understanding the MP energy in the complex plane can 
be harnessed to significantly improve estimates of the exact energy using only the lowest-order terms
in the MP series.

In Sec.~\ref{sec:EPs}, we introduce the key concepts such as Rayleigh--Schr\"odinger perturbation theory and the mean-field HF approximation, and discuss their non-Hermitian analytic continuation into the complex plane.
Section \ref{sec:MP} presents MP perturbation theory and we report a comprehensive historical overview of the research that
has been performed on the physics of MP singularities.
In Sec.~\ref{sec:Resummation}, we discuss several resummation techniques for improving the accuracy
of low-order MP approximations, including Pad\'e and quadratic approximants.
Finally, we draw our conclusions in Sec.~\ref{sec:ccl} and highlight our perspective on directions for 
future research.
Throughout this review, we present illustrative and pedagogical examples based on the ubiquitous 
Hubbard dimer, reinforcing the amazing versatility of this powerful simplistic model.

\section{Exceptional Points in Electronic Structure}
\label{sec:EPs}

\subsection{Time-Independent Schr\"odinger Equation}
\label{sec:TDSE}
Within the Born-Oppenheimer approximation, the exact molecular Hamiltonian with $\Ne$ electrons and 
$\Nn$ (clamped) nuclei is defined for a given nuclear framework as
\begin{equation}\label{eq:ExactHamiltonian}
    \hH(\vb{R}) = 
    - \frac{1}{2} \sum_{i}^{\Ne} \grad_i^2 
    - \sum_{i}^{\Ne} \sum_{A}^{\Nn} \frac{Z_A}{\abs{\vb{r}_i-\vb{R}_A}} 
    + \sum_{i<j}^{\Ne}\frac{1}{\abs{\vb{r}_i-\vb{r}_j}},
\end{equation}
where $\vb{r}_i$ defines the position of the $i$th electron, $\vb{R}_{A}$ and $Z_{A}$ are the position
and charge of the $A$th nucleus respectively, and $\vb{R} = (\vb{R}_{1}, \dots, \vb{R}_{\Nn})$ is a
collective vector for the nuclear positions.
The first term represents the kinetic energy of the electrons, while 
the two following terms account for the electron-nucleus attraction and the electron-electron repulsion.

The exact many-electron wave function at a given nuclear geometry $\Psi(\vb{R})$ corresponds 
to the solution of the (time-independent) Schr\"{o}dinger equation
\begin{equation} 
    \hH(\vb{R})\, \Psi(\vb{R}) = E(\vb{R})\, \Psi(\vb{R}),
    \label{eq:SchrEq}
\end{equation} 
with the eigenvalues $E(\vb{R})$ providing the exact energies.
The energy $E(\vb{R})$ can be considered as a ``one-to-many'' function since each input nuclear geometry
yields several eigenvalues corresponding to the ground and excited states of the exact spectrum.
However, exact solutions to Eq.~\eqref{eq:SchrEq} are only possible in the simplest of systems, such as 
the one-electron hydrogen atom and some specific two-electron systems with well-defined mathematical 
properties.\cite{Taut_1993,Loos_2009b,Loos_2010e,Loos_2012}
In practice, approximations to the exact Schr\"{o}dinger equation must be introduced, including
perturbation theories and the Hartree--Fock approximation considered in this review.
In what follows, we will drop the parametric dependence on the nuclear geometry and, 
unless otherwise stated, atomic units will be used throughout.

\subsection{Exceptional Points in the Hubbard Dimer}
\label{sec:example}

\begin{figure*}[t]
	\begin{subfigure}{0.49\textwidth}
	\includegraphics[height=0.65\textwidth]{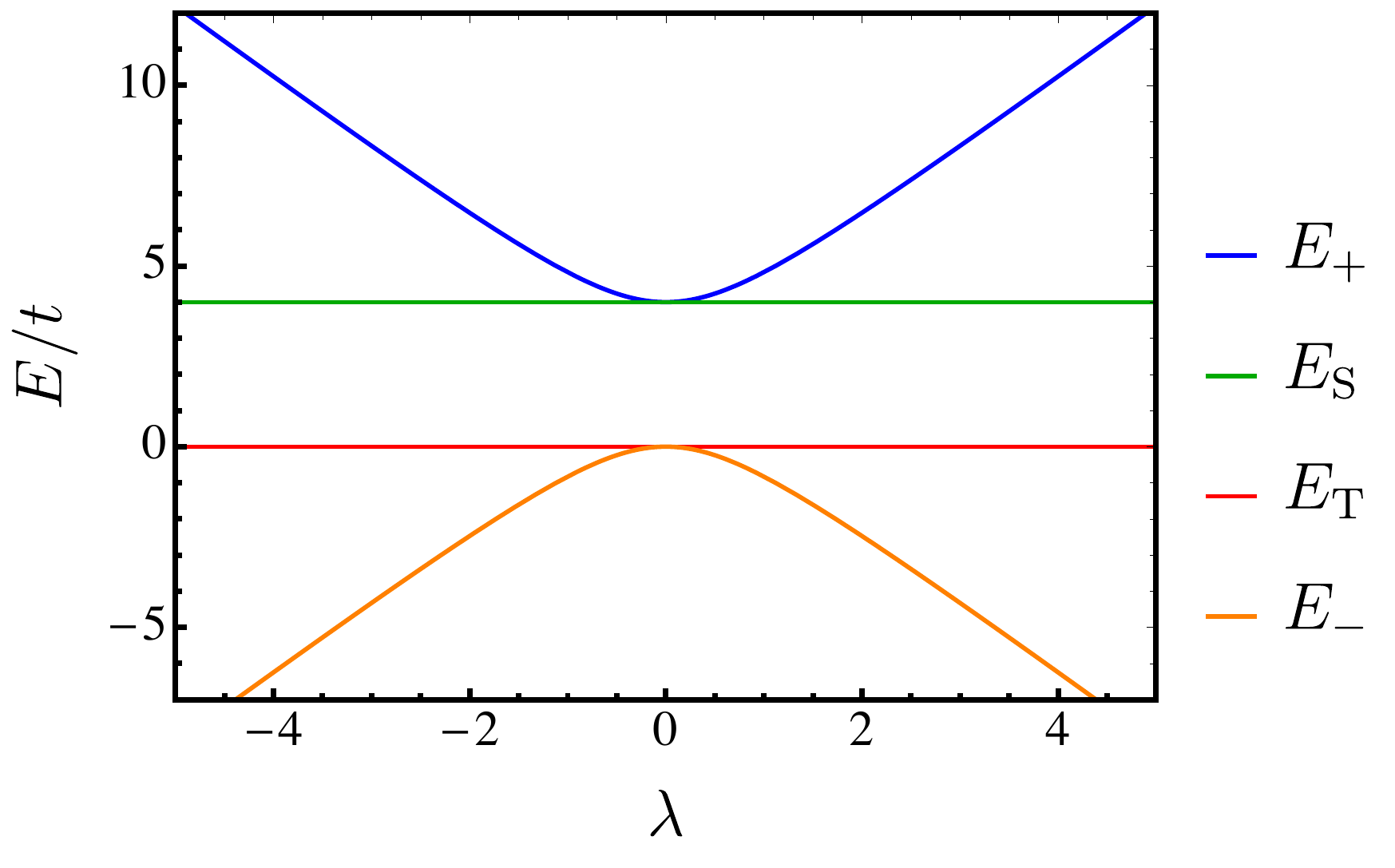}
	\subcaption{Real axis \label{subfig:FCI_real}}
    \end{subfigure}
	\begin{subfigure}{0.49\textwidth}
	\includegraphics[height=0.65\textwidth]{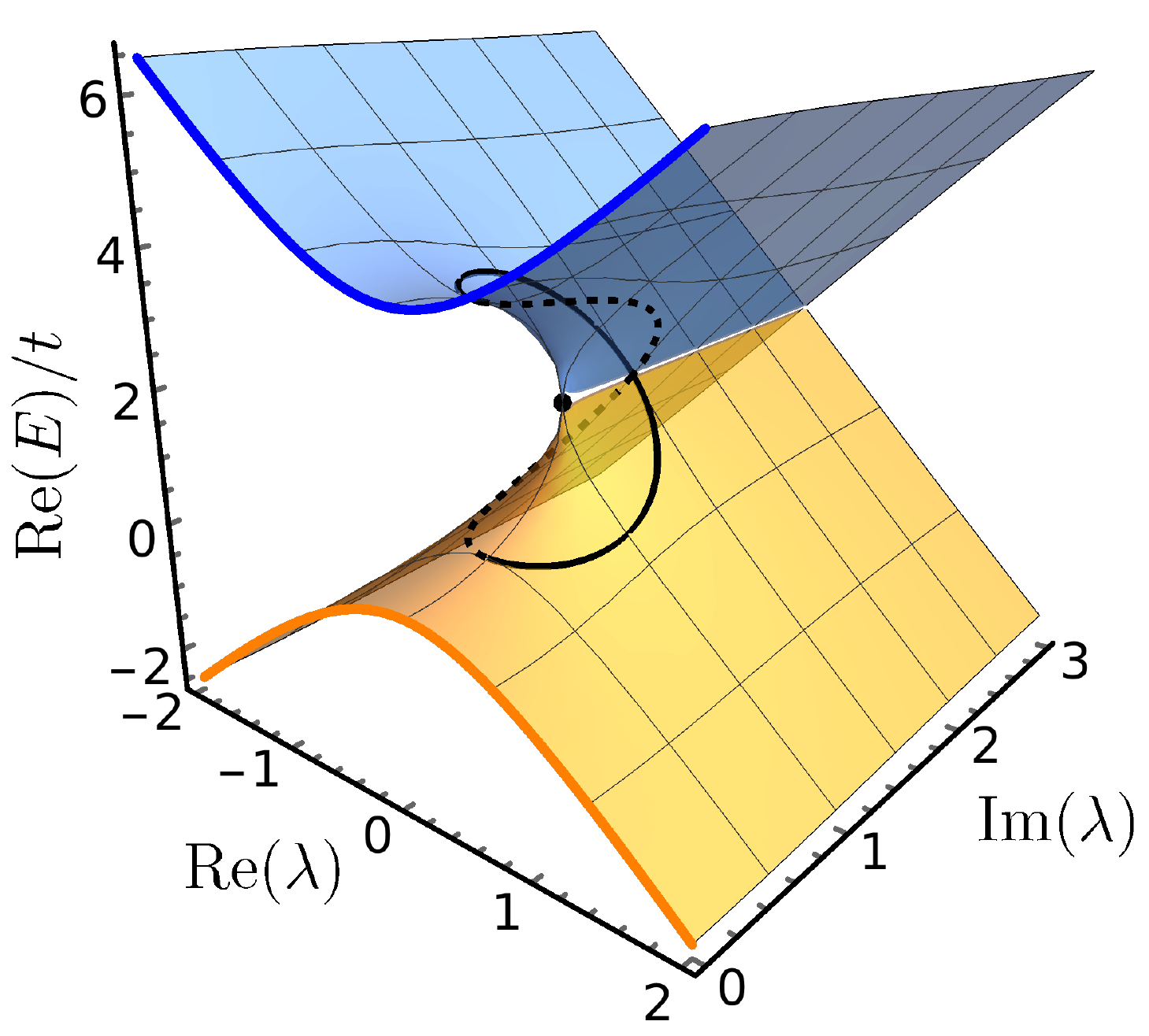}
	\subcaption{Complex plane \label{subfig:FCI_cplx}}
    \end{subfigure}
	\caption{%
	Exact energies for the Hubbard dimer ($U=4t$) as functions of $\lambda$ on the real axis (\subref{subfig:FCI_real}) and in the complex plane (\subref{subfig:FCI_cplx}).
    Only the \titou{real component of the}  interacting closed-shell singlet \titou{energies} are shown in the complex plane, 
    becoming degenerate at the EP (black dot).
    Following a contour around the EP (black solid) interchanges the states, while a second rotation (black dashed)
    returns the states to their original energies.
	\label{fig:FCI}}
\end{figure*}

To illustrate the concepts discussed throughout this article, we consider the symmetric Hubbard dimer at half filling, \ie, with two opposite-spin fermions.
Analytically solvable models are essential in theoretical chemistry and physics as their mathematical simplicity compared to realistic systems (e.g., atoms and molecules) allows new concepts and methods to be 
easily tested while retaining the key physical phenomena.

Using the (localised) site basis, the Hilbert space of the Hubbard dimer comprises the four configurations
\begin{align*}
& \ket{\Lup \Ldown}, &  & \ket{\Lup\Rdown}, &  & \ket{\Rup\Ldown}, &  & \ket{\Rup\Rdown},
\end{align*}
where $\Lsi$ ($\Rsi$) denotes an electron with spin $\sigma$ on the left (right) site.
The exact, or full configuration interaction (FCI), Hamiltonian is then 
\begin{equation}
\label{eq:H_FCI}
	\bH = 
	\begin{pmatrix}
		U &	- t & -  t & 0	\\
	   -  t &  0 &  0 & -  t \\
       -  t &  0 &  0 & -  t \\
        0 & -  t & -  t & U \\
	\end{pmatrix},
\end{equation}
where $t$ is the hopping parameter and $U$ is the on-site Coulomb repulsion.
We refer the interested reader to Refs.~\onlinecite{Carrascal_2015,Carrascal_2018} for more details about this system.
The parameter $U$ controls the strength of the electron correlation.
In the weak correlation regime (small $U$), the kinetic energy dominates and the electrons are delocalised over both sites.
In the large-$U$ (or strong correlation) regime, the electron repulsion term becomes dominant 
and the electrons localise on opposite sites to minimise their Coulomb repulsion. 
This phenomenon is often referred to as Wigner crystallisation. \cite{Wigner_1934}

To illustrate the formation of an EP, we scale the off-diagonal coupling strength by introducing the complex parameter $\lambda$ through the transformation $t \to \lambda t$ to give the parameterised Hamiltonian $\hH(\lambda)$.
When $\lambda$ is real, the Hamiltonian is Hermitian with the distinct (real-valued) (eigen)energies
\begin{subequations}
\begin{align}
E_{\mp} &= \frac{1}{2} \qty(U \mp \sqrt{ (4 \lambda t)^2 + U^2 } ),
\label{eq:singletE}
\\
E_{\text{T}} &= 0,
\\
E_{\text{S}} &= U.
\end{align}
\end{subequations}
While the open-shell triplet ($E_{\text{T}}$) and singlet ($E_{\text{S}}$) are independent of $\lambda$, the closed-shell singlet ground state ($E_{-}$) and doubly-excited state ($E_{+}$) couple strongly to form an avoided crossing at $\lambda=0$ (see Fig.~\ref{subfig:FCI_real}).
\titou{In contrast, when $\lambda$ is complex, the energies may become complex-valued, with the real components shown in 
Fig.~\ref{subfig:FCI_cplx}.
Although the imaginary component of the energy is linked to resonance lifetimes elsewhere in non-Hermitian 
quantum mechanics, \cite{MoiseyevBook} its physical interpretation in the current context is unclear.
Throughout this work, we will generally consider and plot only the real component of any complex-valued energies.
}

At non-zero values of $U$ and $t$, these closed-shell singlets can only become degenerate at a pair of complex conjugate points in the complex $\lambda$ plane 
\begin{equation}
\lambda_{\text{EP}} = \pm  \i \frac{U}{4t},
\end{equation}
with energy
\begin{equation}
\label{eq:E_EP}
	E_\text{EP} = \frac{U}{2}.
\end{equation}
These $\lambda$ values correspond to so-called EPs and connect the ground and excited states in the complex plane.
\titou{Crucially, the ground- and excited-state wave functions at an EP become \emph{identical} rather than just degenerate.} 
Furthermore, the energy surface becomes non-analytic at $\lambda_{\text{EP}}$ and a square-root singularity forms with two branch cuts running along the imaginary axis from $\lambda_{\text{EP}}$  to $\pm \i \infty$ (see Fig.~\ref{subfig:FCI_cplx}).
\titou{Along these branch cuts, the real components of the energies are equivalent and appear to give a seam 
of intersection, but a strict degeneracy is avoided because the imaginary components are different.}

On the real $\lambda$ axis, these EPs lead to the singlet avoided crossing at $\lambda = \Re(\lambda_{\text{EP}})$.
The ``shape'' of this avoided crossing is related to the magnitude of $\Im(\lambda_{\text{EP}})$, with smaller values giving a ``sharper'' interaction.
In the limit $U/t \to 0$, the two EPs converge at $\lep = 0$ to create a conical intersection with 
a gradient discontinuity on the real axis.
\titou{This gradient discontinuity defines a critical point in the ground-state energy,
where a sudden change occurs in the electronic wave function, and can be considered as a zero-temperature quantum phase transition.} 
\cite{Heiss_1988,Heiss_2002,Borisov_2015,Sindelka_2017,CarrBook,Vojta_2003,SachdevBook,GilmoreBook} 

Remarkably, the existence of these square-root singularities means that following a complex contour around an EP in the complex $\lambda$ plane will interconvert the closed-shell ground and excited states (see Fig.~\ref{subfig:FCI_cplx}).
This behaviour can be seen by expanding the radicand in Eq.~\eqref{eq:singletE} as a Taylor series around $\lambda_{\text{EP}}$ to give
\begin{equation}
E_{\pm} \approx E_{\text{EP}} \pm \sqrt{32t^2 \lambda_{\text{EP}}} \sqrt{\lambda - \lambda_{\text{EP}}}.
\end{equation}
Parametrising the complex contour as $\lambda(\theta) = \lambda_{\text{EP}} + R \exp(\i \theta)$ gives the continuous energy pathways 
\begin{equation}
E_{\pm} \qty(\theta) \approx E_{\text{EP}} \pm \sqrt{32t^2 \lambda_{\text{EP}} R}\, \exp(\i \theta/2)
\end{equation}
such that $E_{\pm}(2\pi)  = E_{\mp}(0)$ and $E_{\pm}(4\pi)  = E_{\pm}(0)$.
As a result, completely encircling an EP leads to the interconversion of the two interacting states, while a second complete rotation returns the two states to their original energies.
Additionally, the wave functions  can pick up a geometric phase in the process, and four complete loops are required to recover their starting forms.\cite{MoiseyevBook}

To locate EPs in practice, one must simultaneously solve
\begin{subequations}
\begin{align}
	\label{eq:PolChar}
	\det[\hH(\lambda) - E \hI] & = 0,
	\\ 
	\label{eq:DPolChar}
	\pdv{E}\det[\hH(\lambda) - E \hI] & = 0,
\end{align}
\end{subequations}
where $\hI$ is the identity operator.\cite{Cejnar_2007}
Equation \eqref{eq:PolChar} is the well-known secular equation providing the (eigen)energies of the system. 
If the energy is also a solution of Eq.~\eqref{eq:DPolChar}, then this energy value is at least two-fold degenerate. 
These degeneracies can be conical intersections between two states with different symmetries 
for real values of $\lambda$,\cite{Yarkony_1996} or EPs between two states with the 
same symmetry for complex values of $\lambda$.

\subsection{Rayleigh--Schr\"odinger Perturbation Theory}

One of the most common routes to approximately solving the Schr\"odinger equation
is to introduce a perturbative expansion of the exact energy.
Within Rayleigh--Schr\"odinger perturbation theory, the time-independent Schr\"odinger equation 
is recast as 
\begin{equation} 
	\hH(\lambda) \Psi(\lambda) 
    = \qty(\hH^{(0)} + \lambda \hV ) \Psi(\lambda) 
    = E(\lambda) \Psi(\lambda),
    \label{eq:SchrEq-PT}
\end{equation}
where $\hH^{(0)}$ is a zeroth-order Hamiltonian and $\hV = \hH - \hH^{(0)}$ represents the perturbation operator.
Expanding the wave function and energy as power series in $\lambda$ as 
\begin{subequations}
\begin{align}
    \Psi(\lambda) &= \sum_{k=0}^{\infty} \lambda^{k}\,\Psi^{(k)},
    \label{eq:psi_expansion}
    \\
    E(\lambda) &= \sum_{k=0}^{\infty} \lambda^{k}\,E^{(k)},
    \label{eq:E_expansion}
\end{align}
\end{subequations}
solving the corresponding perturbation equations up to a given order $n$, and
setting $\lambda = 1$ then yields approximate solutions to Eq.~\eqref{eq:SchrEq}.

Mathematically, Eq.~\eqref{eq:E_expansion} corresponds to a Taylor series expansion of the exact energy
around the reference system $\lambda~=~0$.
The energy of the target ``physical'' system is recovered at the point $\lambda = 1$.
However, like all series expansions, Eq.~\eqref{eq:E_expansion} has a radius of convergence $\rc$. 
When $\rc < 1$, the Rayleigh--Schr\"{o}dinger expansion will diverge
for the physical system.
The value of $\rc$ can vary significantly between different systems and strongly depends on the particular decomposition
of the reference and perturbation Hamiltonians in Eq.~\eqref{eq:SchrEq-PT}.\cite{Mihalka_2017b}
%
From complex analysis, \cite{BenderBook} the radius of convergence for the energy can be obtained by looking for the 
non-analytic singularities of $E(\lambda)$ in the complex $\lambda$ plane.
This property arises from the following theorem: \cite{Goodson_2011}
\begin{quote}
\it
``The Taylor series about a point $z_0$ of a function over the complex $z$ plane will converge at a value $z_1$ 
if the function is non-singular at all values of $z$ in the circular region centred at $z_0$ with radius $\abs{z_1-z_0}$. 
If the function has a singular point $z_s$ such that $\abs{z_s-z_0} < \abs{z_1-z_0}$, 
then the series will diverge when evaluated at $z_1$.''
\end{quote}
As a result, the radius of convergence for a function is equal to the distance from the origin of the closest singularity
in the complex plane, referred to as the ``dominant'' singularity.
This singularity may represent a pole of the function, or a branch point (\eg, square-root or logarithmic)
in a multi-valued function.

For example, the simple function
\begin{equation} \label{eq:DivExample}
	f(x)=\frac{1}{1+x^4}.
\end{equation}
is smooth and infinitely differentiable for $x \in \mathbb{R}$, and one might expect that its Taylor series expansion would 
converge in this domain.
However, this series diverges for $x \ge 1$.
This divergence occurs because $f(x)$ has four poles in the complex 
($\e^{\i\pi/4}$, $\e^{-\i\pi/4}$, $\e^{\i3\pi/4}$, and $\e^{-\i3\pi/4}$) with a modulus equal to $1$, demonstrating
that complex singularities are essential to fully understand the series convergence on the real axis.\cite{BenderBook}

The radius of convergence for the perturbation series Eq.~\eqref{eq:E_expansion} is therefore dictated by the magnitude $r_c = \abs{\lambda_c}$ of the
singularity in $E(\lambda)$ that is closest to the origin.
Note that when $\abs{\lambda} = r_c$, one cannot \textit{a priori} predict if the series is convergent or not.
For example, the series $\sum_{k=1}^\infty \lambda^k/k$ diverges at $\lambda = 1$ but converges at $\lambda = -1$.

Like the exact system in Sec.~\ref{sec:example}, the perturbation energy $E(\lambda)$ represents
a ``one-to-many'' function with the output elements representing an approximation to both the ground and excited states.
The most common singularities on $E(\lambda)$ therefore correspond to non-analytic EPs in the complex 
$\lambda$ plane where two states become degenerate.
\titou{Additional singularities can also arise at critical points of the energy.
A critical point corresponds to the intersection of two energy surfaces
where the eigenstates remain distinct but a gradient discontinuity occurs in 
the ground-state energy.
In contrast, at a square-root branch point, both the energies and the associated wave functions
of the intersecting surfaces become identical.}
Later we will demonstrate how the choice of reference Hamiltonian controls the position of these EPs, and 
ultimately determines the convergence properties of the perturbation series.

\subsection{Hartree--Fock Theory}
\label{sec:HF}

In the HF approximation, the many-electron wave function is approximated as a single Slater determinant $\whf(\vb{x}_1,\ldots,\vb{x}_\Ne)$, where $\vb{x} = (\sigma,\vb{r})$ is a composite vector gathering spin and spatial coordinates.
This Slater determinant is defined as an antisymmetric combination of $\Ne$ (real-valued) occupied one-electron spin-orbitals $\phi_p(\vb{x})$, which are, by definition, eigenfunctions of the one-electron Fock operator 
\begin{equation}\label{eq:FockOp}
    \Hat{f}(\vb{x}) \phi_p(\vb{x}) = \qty[ \Hat{h}(\vb{x}) + \vhf(\vb{x}) ] \phi_p(\vb{x}) = \epsilon_p \phi_p(\vb{x}).
\end{equation}
Here the (one-electron) core Hamiltonian is
\begin{equation}
\label{eq:Hcore}
	\Hat{h}(\vb{x}) = -\frac{\grad^2}{2} + \sum_{A}^{\Nn} \frac{Z_A}{\abs{\vb{r}-\vb{R}_A}}
\end{equation}
and
\begin{equation}
    \vhf(\vb{x}) = \sum_i^{\Ne} \qty[ \Hat{J}_i(\vb{x}) - \Hat{K}_i(\vb{x}) ]
\end{equation}
is the HF mean-field electron-electron potential with 
\begin{subequations}
\begin{gather}
	\label{eq:CoulOp}
    \Hat{J}_i(\vb{x})\phi_j(\vb{x})=\qty(\int \phi_i(\vb{x}')\frac{1}{\abs{\vb{r} - \vb{r}'}}\phi_i(\vb{x}') \dd\vb{x}' ) \phi_j(\vb{x}),
	\\
	\label{eq:ExcOp}
\Hat{K}_i(\vb{x})\phi_j(\vb{x})=\qty(\int \phi_i(\vb{x}')\frac{1}{\abs{\vb{r} - \vb{r}'}}\phi_j(\vb{x}') \dd\vb{x}')\phi_i(\vb{x}),
\end{gather}
\end{subequations}
defining the Coulomb and exchange operators (respectively) in the spin-orbital basis.\cite{SzaboBook}
The HF energy is then defined as 
\begin{equation}
    \label{eq:E_HF}
    E_\text{HF} = \frac{1}{2} \sum_i^{\Ne} \qty( h_i + f_i ),
\end{equation}
with the corresponding matrix elements
\begin{align}
	h_i & = \mel{\phi_i}{\Hat{h}}{\phi_i},
    & 
    f_i & = \mel{\phi_i}{\Hat{f}}{\phi_i}.
\end{align}
The optimal HF wave function is identified by using the variational principle to minimise the HF energy.
For any system with more than one electron, the resulting Slater determinant is not an eigenfunction of the exact Hamiltonian $\hH$. 
However, it is by definition an eigenfunction of the approximate many-electron HF Hamiltonian constructed 
from the one-electron Fock operators as
\begin{equation}\label{eq:HFHamiltonian}
	\hH_{\text{HF}} = \sum_{i}^{\Ne} f(\vb{x}_i).
\end{equation}
From hereon, $i$ and $j$ denote occupied orbitals, $a$ and $b$ denote unoccupied (or virtual) orbitals, while $p$, $q$, $r$, and $s$ denote arbitrary orbitals.

In the most flexible variant of real HF theory (generalised HF) the one-electron orbitals can be complex-valued
and contain a mixture of spin-up and spin-down components.\cite{Mayer_1993,Jimenez-Hoyos_2011}
However, the application of HF theory with some level of constraint on the orbital structure is far more common.
Forcing the spatial part of the orbitals to be the same for spin-up and spin-down electrons leads to restricted HF (RHF) method, 
while allowing different orbitals for different spins leads to the so-called unrestricted HF (UHF) approach.\cite{StuberPaldus}
The advantage of the UHF approximation is its ability to correctly describe strongly correlated systems, 
such as antiferromagnetic phases\cite{Slater_1951} or the dissociation of the hydrogen dimer.\cite{Coulson_1949}
However, by allowing different orbitals for different spins, the UHF wave function is no longer required to be an eigenfunction of 
the total spin operator $\hat{\mathcal{S}}^2$, leading to ``spin-contamination''.

\subsection{Hartree--Fock in the Hubbard Dimer}
\label{sec:HF_hubbard}

\begin{figure}
    \includegraphics[width=\linewidth]{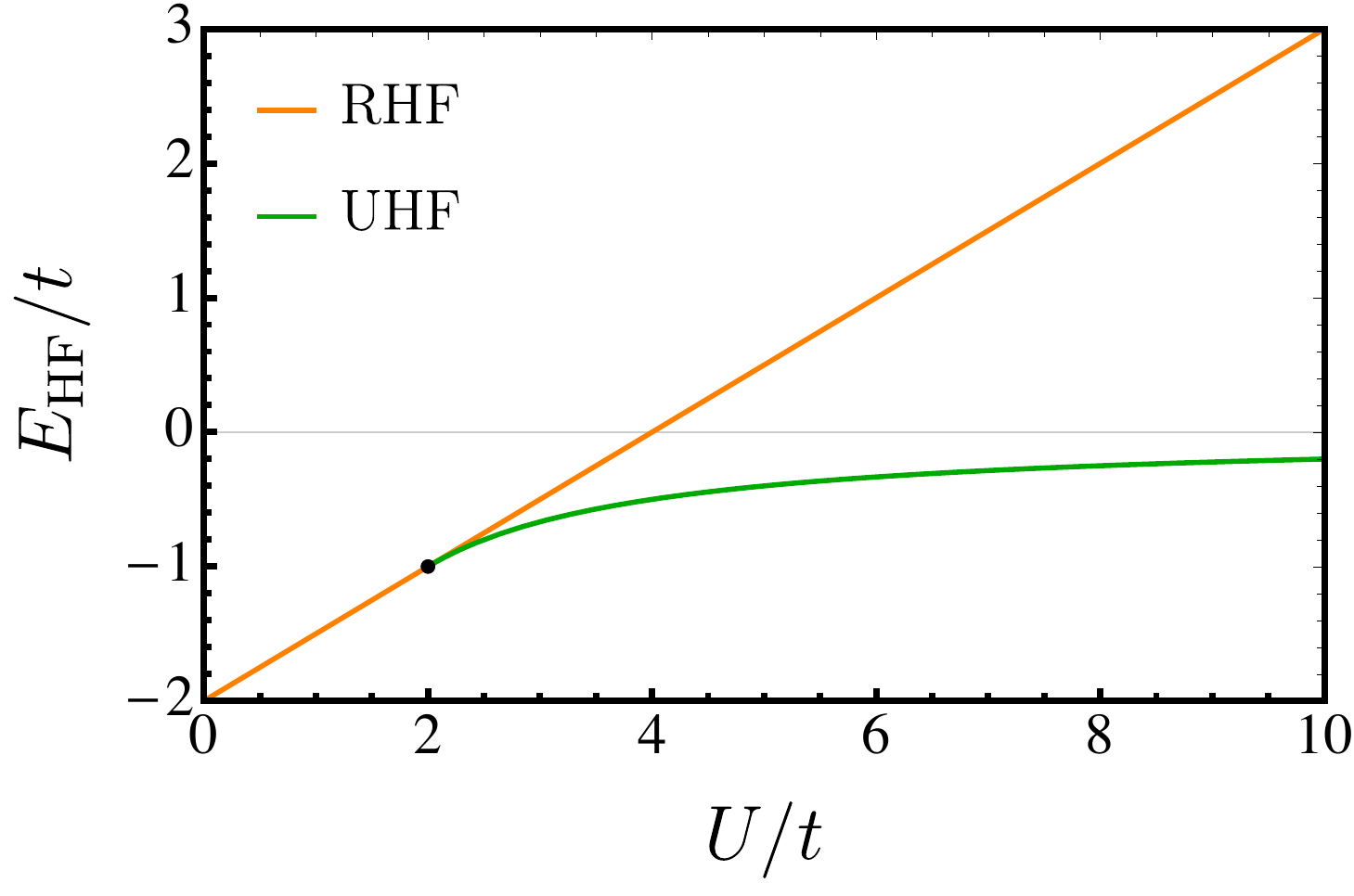}
    \caption{\label{fig:HF_real}
    RHF and UHF energies in the Hubbard dimer as a function of the correlation strength $U/t$. 
    The symmetry-broken UHF solution emerges at the coalescence point $U=2t$ (black dot), often known as the Coulson--Fischer point.}
\end{figure}

In the Hubbard dimer, the HF energy can be parametrised using two rotation angles $\ta$ and $\tb$ as
\begin{equation}
\label{eq:EHF}
E_\text{HF}(\ta, \tb) = -t\, \qty( \sin \ta + \sin \tb ) + \frac{U}{2} \qty( 1 + \cos \ta \cos \tb ),
\end{equation}
where we have introduced \titou{occupied $\phi_1^{\sigma}$} and \titou{unoccupied $\phi_2^{\sigma}$} molecular orbitals for 
the spin-$\sigma$ electrons as
\begin{subequations}
\begin{align}
	\label{eq:psi1}
    \titou{\phi_1^{\sigma}} & = \hphantom{-} \cos(\frac{\ts}{2}) \Lsi + \sin(\frac{\ts}{2}) \Rsi,
	\\
	\label{eq:psi2}
	\titou{\phi_2^{\sigma}} & = - \sin(\frac{\ts}{2}) \Lsi + \cos(\frac{\ts}{2}) \Rsi
\end{align}
\label{eq:RHF_orbs}
\end{subequations}
\titou{Equations \eqref{eq:EHF}, \eqref{eq:psi1}, and \eqref{eq:psi2} are valid for both RHF and UHF.}
In the weak correlation regime $0 \le U \le 2t$, the angles which minimise the HF energy, 
\ie, $\pdv*{E_\text{HF}}{\ts} = 0$, are 
\begin{equation}
	\ta_\text{RHF} = \tb_\text{RHF} = \pi/2,
\end{equation}
giving the molecular orbitals
\begin{align}
	\titou{\phi_{1,\text{RHF}}^{\sigma}} & = \frac{\Lsi + \Rsi}{\sqrt{2}},
	&
	\titou{\phi_{2,\text{RHF}}^{\sigma}} & = \frac{\Lsi - \Rsi}{\sqrt{2}},
\end{align}
and the ground-state RHF energy (Fig.~\ref{fig:HF_real})
\begin{equation}
	E_\text{RHF} \equiv E_\text{HF}(\ta_\text{RHF}, \tb_\text{RHF}) = -2t + \frac{U}{2}.
\end{equation}
\titou{Here, the molecular orbitals respectively transform 
according to the $\Sigma_\text{g}^{+}$ and $\Sigma_\text{u}^{+}$ irreducible representations of 
the $D_{\infty \text{h}}$ point group that represents the symmetric Hubbard dimer.
We can therefore consider these as symmetry-pure molecular orbitals.} 
However, in the strongly correlated regime $U>2t$, the closed-shell orbital restriction prevents RHF from 
modelling the correct physics with the two electrons on opposite sites.

\begin{figure*}[t]
	\begin{subfigure}{0.49\textwidth}
    \includegraphics[height=0.65\textwidth,trim={0pt 0pt 0pt -35pt},clip]{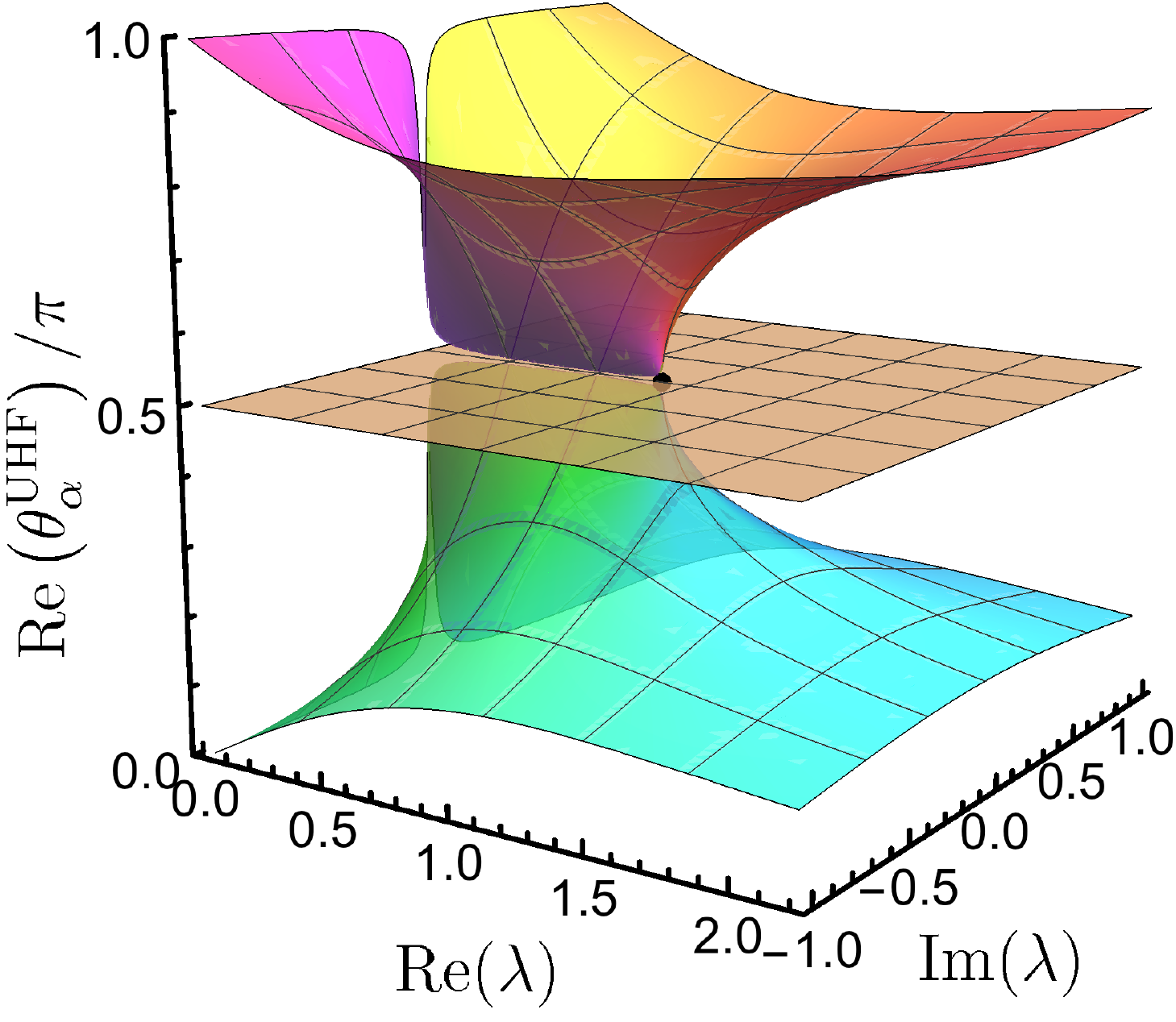}
	\subcaption{\label{subfig:UHF_cplx_angle}}
    \end{subfigure}
	\begin{subfigure}{0.49\textwidth}
	\includegraphics[height=0.65\textwidth]{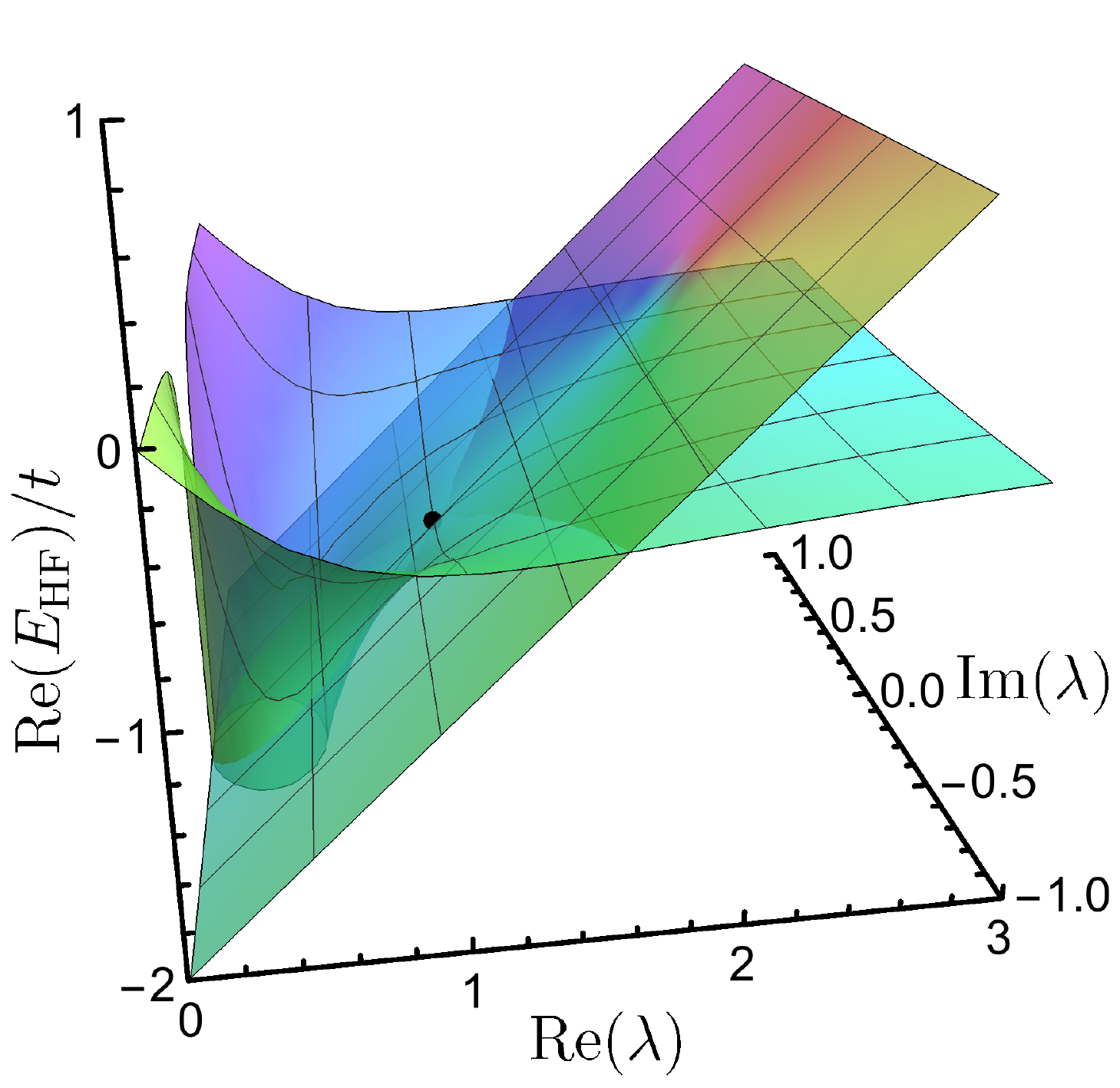}
	\subcaption{\label{subfig:UHF_cplx_energy}}
    \end{subfigure}
	\caption{%
    (\subref{subfig:UHF_cplx_angle}) Real component of the UHF angle $\ta_{\text{UHF}}$ for $\lambda \in \bbC$ in the Hubbard dimer for $U/t = 2$.
    Symmetry-broken solutions correspond to individual sheets and become equivalent at 
    the \textit{quasi}-EP $\lambda_{\text{c}}$ (black dot).
    The RHF solution is independent of $\lambda$, giving the constant plane at $\pi/2$.
    (\subref{subfig:UHF_cplx_energy}) The corresponding HF energy surfaces show a non-analytic 
    point at the \textit{quasi}-EP.
	\label{fig:HF_cplx}}
\end{figure*}

As the on-site repulsion is increased from 0, the HF approximation reaches a critical value at $U=2t$ where an alternative
UHF solution appears with a lower energy than the RHF one.
Note that the RHF wave function remains a genuine solution of the HF equations for $U \ge 2t$, but corresponds to a saddle point 
of the HF energy rather than a minimum.
This critical point is analogous to the infamous Coulson--Fischer point identified in the hydrogen dimer.\cite{Coulson_1949}
For $U \ge 2t$, the optimal orbital rotation angles for the UHF orbitals become
\begin{subequations}
\begin{align}
    \ta_\text{UHF} & = \arctan (-\frac{2t}{\sqrt{U^2 - 4t^2}}),
    \label{eq:ta_uhf}
	\\
    \tb_\text{UHF} & = \arctan (+\frac{2t}{\sqrt{U^2 - 4t^2}}),
    \label{eq:tb_uhf}
\end{align}
\end{subequations}
with the corresponding UHF ground-state energy (Fig.~\ref{fig:HF_real})
\begin{equation}
    E_\text{UHF} \equiv E_\text{HF}(\ta_\text{UHF}, \tb_\text{UHF}) = - \frac{2t^2}{U}.
\end{equation}

\titou{The molecular orbitals of the lower-energy UHF solution do not transform as an irreducible
representation of the $D_{\infty \text{h}}$ point group and therefore break spatial symmetry.
Allowing different orbitals for the different spins also means that the
overall wave function is no longer an eigenfunction of the  $\cS^2$ operator and can be considered to break spin symmetry.
This combined spatial and spin symmetry-breaking occurs for all $U \ge 2t$.}
Furthermore, time-reversal symmetry dictates that this UHF wave function must be degenerate with its spin-flipped counterpart, obtained 
by swapping $\ta_{\text{UHF}}$ and $\tb_{\text{UHF}}$ in Eqs.~\eqref{eq:ta_uhf} and \eqref{eq:tb_uhf}.
This type of symmetry breaking is also called a spin-density wave in the physics community as the system
``oscillates'' between the two symmetry-broken configurations. \cite{GiulianiBook}
Spatial symmetry breaking can also occur in RHF theory when a charge-density wave is formed from an oscillation 
between the two closed-shell configurations with both electrons localised on one site or the other.\cite{StuberPaldus,Fukutome_1981}

\subsection{Self-Consistency as a Perturbation} 

The inherent non-linearity in the Fock eigenvalue problem arises from self-consistency 
in the HF approximation, and is usually solved through an iterative approach.\cite{Roothaan_1951,Hall_1951}
Alternatively, the non-linear terms arising from the Coulomb and exchange operators can 
be considered as a perturbation from the core Hamiltonian \eqref{eq:Hcore} by introducing the
transformation $U \to \lambda\, U$, giving the parametrised Fock operator 
\begin{equation}
    \Hat{f}(\vb{x} ; \lambda) = \Hat{h}(\vb{x}) + \lambda\, \vhf(\vb{x}).
\end{equation}
The orbitals in the reference problem $\lambda=0$ correspond to the symmetry-pure eigenfunctions of the one-electron core
Hamiltonian, while self-consistent solutions at $\lambda = 1$ represent the orbitals of the true HF solution.

For real $\lambda$, the self-consistent HF energies at given (real) $U$ and $t$ values
in the Hubbard dimer directly mirror the energies shown in Fig.~\ref{fig:HF_real}, 
with coalesence points at 
\begin{equation}
    \lambda_{\text{c}} = \pm \frac{2t}{U}.
    \label{eq:scaled_fock}
\end{equation}
In contrast, when $\lambda$ becomes complex, the HF equations become non-Hermitian and 
each HF solution can be analytically continued for all $\lambda$ values using
the holomorphic HF approach.\cite{Hiscock_2014,Burton_2016,Burton_2018}
Remarkably, the coalescence point in this analytic continuation emerges as a 
\textit{quasi}-EP on the real $\lambda$ axis (Fig.~\ref{fig:HF_cplx}), where
the different HF solutions become equivalent but not self-orthogonal.\cite{Burton_2019}
By analogy with perturbation theory, the regime where this \textit{quasi}-EP occurs 
within $\lambda_{\text{c}} \le 1$ can be interpreted as an indication that 
the symmetry-pure reference orbitals no longer provide a qualitatively 
accurate representation for the true HF ground state at $\lambda = 1$.
For example, in the Hubbard dimer with $U > 2t$, one finds $\lambda_{\text{c}} < 1$ and the symmetry-pure orbitals
do not provide a good representation of the HF ground state.
In contrast, $U < 2t$ yields $\lambda_{\text{c}} > 1$ and corresponds to
the regime where the HF ground state is correctly represented by symmetry-pure orbitals.

We have recently shown that the complex scaled Fock operator \eqref{eq:scaled_fock}
also allows states of different symmetries to be interconverted by following a well-defined
contour in the complex $\lambda$-plane.\cite{Burton_2019}
In particular, by slowly varying $\lambda$ in a similar (yet different) manner
to an adiabatic connection in density-functional theory,\cite{Langreth_1975,Gunnarsson_1976,Zhang_2004} 
a ground-state wave function can be ``morphed'' into an excited-state wave function 
via a stationary path of HF solutions.
This novel approach to identifying excited-state wave functions demonstrates the fundamental 
role of \textit{quasi}-EPs in determining the behaviour of the HF approximation.
Furthermore, the complex-scaled Fock operator can be used routinely to construct analytic
continuations of HF solutions beyond the points where real HF solutions
coalesce and vanish.\cite{Burton_2019b}

\section{M{\o}ller--Plesset Perturbation Theory in the Complex Plane}
\label{sec:MP}

\subsection{Background Theory}

In electronic structure, the HF Hamiltonian \eqref{eq:HFHamiltonian} is often used as the zeroth-order Hamiltonian
to define M\o{}ller--Plesset (MP) perturbation theory.\cite{Moller_1934}
This approach can recover a large proportion of the electron correlation energy,\cite{Lowdin_1955a,Lowdin_1955b,Lowdin_1955c} 
and provides the foundation for numerous post-HF approximations.
With the MP partitioning, the parametrised perturbation Hamiltonian becomes
\begin{multline}\label{eq:MPHamiltonian}
    \hH(\lambda) =   
     \sum_{i}^{N} \qty[ - \frac{\grad_i^2}{2} - \sum_{A}^{M} \frac{Z_A}{\abs{\vb{r}_i-\vb{R}_A}} ]
    \\
    + (1-\lambda) \sum_{i}^{N} \vhf(\vb{x}_i)
    + \lambda\sum_{i<j}^{N}\frac{1}{\abs{\vb{r}_i-\vb{r}_j}}.
\end{multline}
Any set of orbitals can be used to define the HF Hamiltonian, although either the RHF or UHF orbitals are usually chosen to 
define the RMP or UMP series respectively.
The MP energy at a given order $n$ (\ie, MP$n$) is then defined as
\begin{equation}
	E_{\text{MP}n}= \sum_{k=0}^n E_{\text{MP}}^{(k)},
\end{equation}
where $E_{\text{MP}}^{(k)}$ is the $k$th-order MP correction and 
\begin{equation}
E_{\text{MP1}} =  E_{\text{MP}}^{(0)} + E_{\text{MP}}^{(1)} = E_\text{HF}.
\end{equation}
The second-order MP2 energy correction is given by
\begin{equation}\label{eq:EMP2}
    E_{\text{MP}}^{(2)} = \frac{1}{4} \sum_{ij} \sum_{ab} \frac{\abs{\mel{ij}{}{ab}}^2}{\epsilon_i + \epsilon_j - \epsilon_a - \epsilon_b},
\end{equation}
where $\mel{pq}{}{rs} = \braket{pq}{rs} - \braket{pq}{sr}$ are the anti-symmetrised two-electron integrals
in the molecular spin-orbital basis\cite{Gill_1994}
\begin{equation}
	\braket{pq}{rs} 
    = \iint \dd\vb{x}_1\dd\vb{x}_2
    \frac{\phi^{*}_p(\vb{x}_1)\phi^{*}_q(\vb{x}_2)\phi^{\vphantom{*}}_r(\vb{x}_1)\phi^{\vphantom{*}}_s(\vb{x}_2)}%
      {\abs{\vb{r}_1 - \vb{r}_2}}.
\end{equation}

While most practical calculations generally consider only the MP2 or MP3 approximations, higher order terms can 
be computed to understand the convergence of the MP$n$ series.\cite{Handy_1985}
\textit{A priori}, there is no guarantee that this series will provide the smooth convergence that is desirable for a
systematically improvable theory.
In fact, when the reference HF wave function is a poor approximation to the exact wave function, 
for example in multi-configurational systems, MP theory can yield highly oscillatory, 
slowly convergent, or catastrophically divergent results.\cite{Gill_1986,Gill_1988,Handy_1985,Lepetit_1988,Leininger_2000,Malrieu_2003}
Furthermore, the convergence properties of the MP series can depend strongly on the choice of restricted or
unrestricted reference orbitals.

Although practically convenient for electronic structure calculations, the MP partitioning is not 
the only possibility and alternative partitionings have been considered \cite{Surjan_2004} including: 
i) the Epstein-Nesbet (EN) partitioning which consists in taking the diagonal elements of $\hH$ as the zeroth-order Hamiltonian, \cite{Nesbet_1955,Epstein_1926} 
ii) the weak correlation partitioning in which the one-electron part is consider as the unperturbed Hamiltonian $\hH^{(0)}$ and the two-electron part is the perturbation operator $\hV$, and 
iii) the strong coupling partitioning where the two operators are inverted compared to the weak correlation partitioning. \cite{Seidl_2018,Daas_2020}
While an in-depth comparison of these different approaches can offer insight into 
their relative strengths and weaknesses for various situations, we will restrict our current discussion
to the convergence properties of the MP expansion.

\subsection{Early Studies of M{\o}ller--Plesset Convergence} 

Among the most desirable properties of any electronic structure technique is the existence of 
a systematic route to increasingly accurate energies. 
In the context of MP theory, one would like a monotonic convergence of the perturbation
series towards the exact energy such that the accuracy increases as each term in the series is added.
If such well-behaved convergence can be established, then our ability to compute individual 
terms in the series becomes the only barrier to computing the exact correlation in a finite basis set.
Unfortunately, the computational scaling of each term in the MP series increases with the perturbation
order, and practical calculations must rely on fast convergence
to obtain high-accuracy results using only the lowest order terms.

MP theory was first introduced to quantum chemistry through the pioneering
works of Bartlett \etal\ in the context of many-body perturbation theory,\cite{Bartlett_1975}
and Pople and co-workers in the context of determinantal expansions.\cite{Pople_1976,Pople_1978}
Early implementations were restricted to the fourth-order MP4 approach that was considered
to offer state-of-the-art quantitative accuracy.\cite{Pople_1978,Krishnan_1980}
However, it was quickly realised that the MP series often demonstrated very slow, oscillatory, 
or erratic convergence, with the UMP series showing particularly slow convergence.\cite{Laidig_1985,Knowles_1985,Handy_1985}
For example, RMP5 is worse than RMP4 for predicting the homolytic barrier fission of \ce{He2^2+} using a minimal basis set, 
while the UMP series monotonically converges but becomes increasingly slow beyond UMP5.\cite{Gill_1986}
The first examples of divergent MP series were observed in the \ce{N2} and \ce{F2} 
diatomics, where low-order RMP and UMP expansions give qualitatively wrong binding curves.\cite{Laidig_1987} 

The divergence of RMP expansions for stretched bonds can be easily understood from two perspectives.\cite{Gill_1988a}
Firstly, the exact wave function becomes increasingly multi-configurational as the bond is stretched, and the 
RHF wave function no longer provides a qualitatively correct reference for the perturbation expansion.
Secondly, the energy gap between the \titou{occupied and unoccupied} orbitals associated with the stretch becomes
increasingly small at larger bond lengths, leading to a divergence, for example, in the MP2 correction \eqref{eq:EMP2}.
In contrast, the origin of slow UMP convergence is less obvious as the reference UHF energy remains
qualitatively correct at large bond lengths and the orbital degeneracy is avoided.
Furthermore, this slow convergence can also be observed in molecules with a UHF ground state at the equilibrium
geometry (\eg, \ce{CN-}), suggesting a more fundamental link with spin-contamination 
in the reference wave function.\cite{Nobes_1987}

Using the UHF framework allows the singlet ground state wave function to mix with triplet wave functions, 
leading to spin contamination where the wave function is no longer an eigenfunction of the $\Hat{\cS}^2$ operator.
The link between slow UMP convergence and this spin-contamination was first systematically investigated
by Gill \etal\ using the minimal basis \ce{H2} model.\cite{Gill_1988}
In this work, the authors 
identified that the slow UMP convergence arises from its failure to correctly predict the amplitude of the
low-lying double excitation.
This erroneous description of the double excitation amplitude has the same origin as the spin-contamination in the reference
UHF wave function, creating the first direct link between spin-contamination and slow UMP convergence.\cite{Gill_1988}
%
Lepetit \etal\ later analysed the difference between perturbation convergence using the UMP 
and EN partitionings. \cite{Lepetit_1988}
They argued that the slow UMP convergence for stretched molecules arises from 
(i) the fact that the MP denominator (see Eq.~\ref{eq:EMP2})
tends to a constant value instead of vanishing, and (ii) the slow convergence of contributions from the 
singly-excited configurations that strongly couple to the doubly-excited configurations and first
appear at fourth-order.\cite{Lepetit_1988}
Drawing these ideas together, we believe that slow UMP convergence occurs because the single excitations must focus on removing
spin-contamination from the reference wave function, limiting their ability to fine-tune the amplitudes of the higher 
excitations that capture the correlation energy.

A number of spin-projected extensions have been derived to reduce spin-contamination in the wave function
and overcome the slow UMP convergence.
Early versions of these theories, introduced by Schlegel \cite{Schlegel_1986, Schlegel_1988} or 
Knowles and Handy,\cite{Knowles_1988a,Knowles_1988b} exploited the ``projection-after-variation'' philosophy,
where the spin-projection is applied directly to the UMP expansion.
These methods succeeded in accelerating the convergence of the projected MP series and were 
considered as highly effective methods for capturing the electron correlation at low computational cost.\cite{Knowles_1988b}
However, the use of projection-after-variation leads to gradient discontinuities in the vicinity of the UHF symmetry-breaking point,
and can result in spurious minima along a molecular binding curve.\cite{Schlegel_1986,Knowles_1988a}
More recent formulations of spin-projected perturbations theories have considered the  
``variation-after-projection'' framework using alternative definitions of the reference 
Hamiltonian.\cite{Tsuchimochi_2014,Tsuchimochi_2019}
These methods yield more accurate spin-pure energies without 
gradient discontinuities or spurious minima.

\subsection{Spin-Contamination in the Hubbard Dimer}
\label{sec:spin_cont}

\begin{figure*}
	\begin{subfigure}{0.32\textwidth}
	\includegraphics[height=0.75\textwidth]{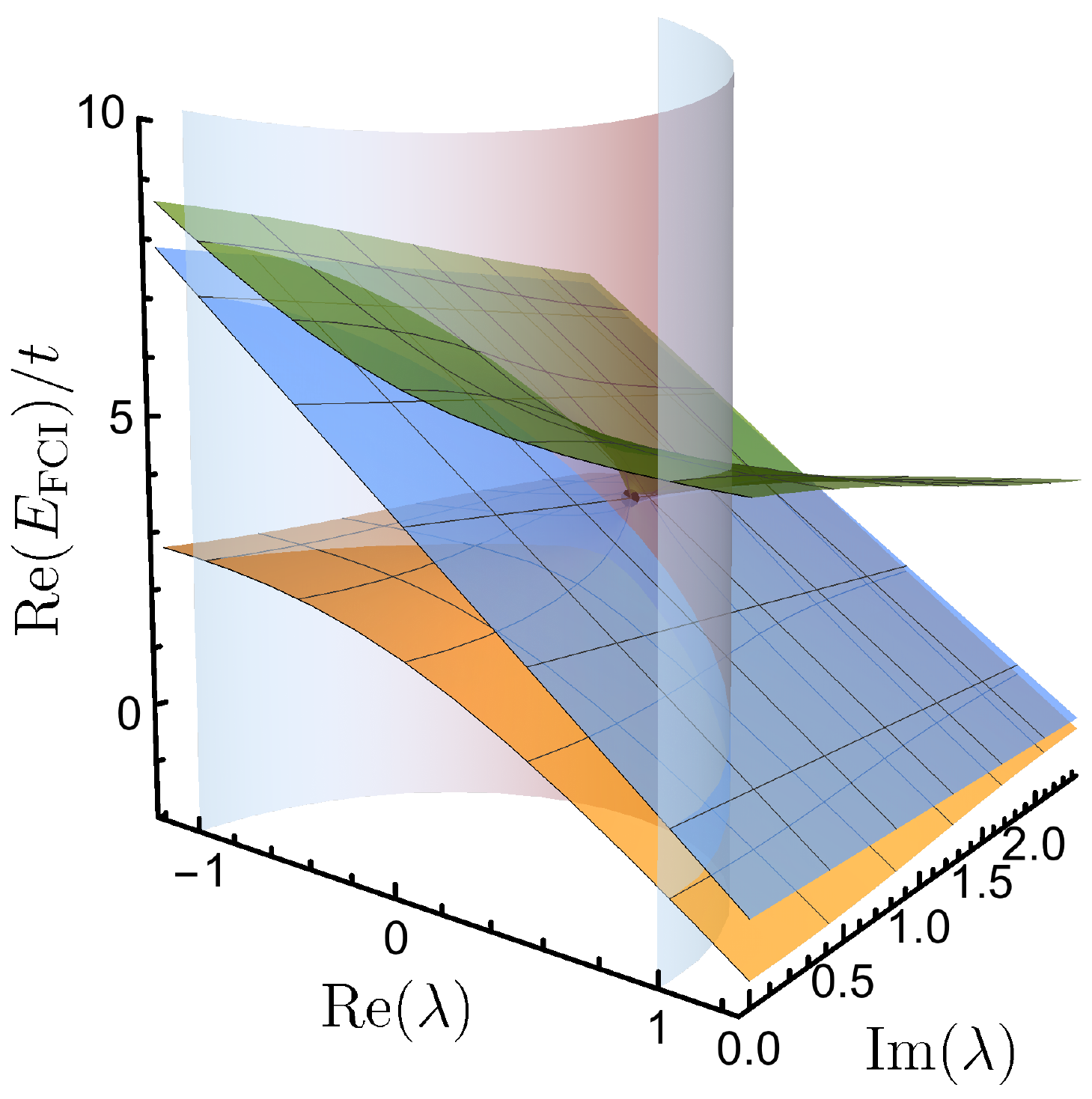}	
		\subcaption{\label{subfig:RMP_3.5} $U/t = 3.5$}
    \end{subfigure}
    \begin{subfigure}{0.32\textwidth}
	\includegraphics[height=0.75\textwidth]{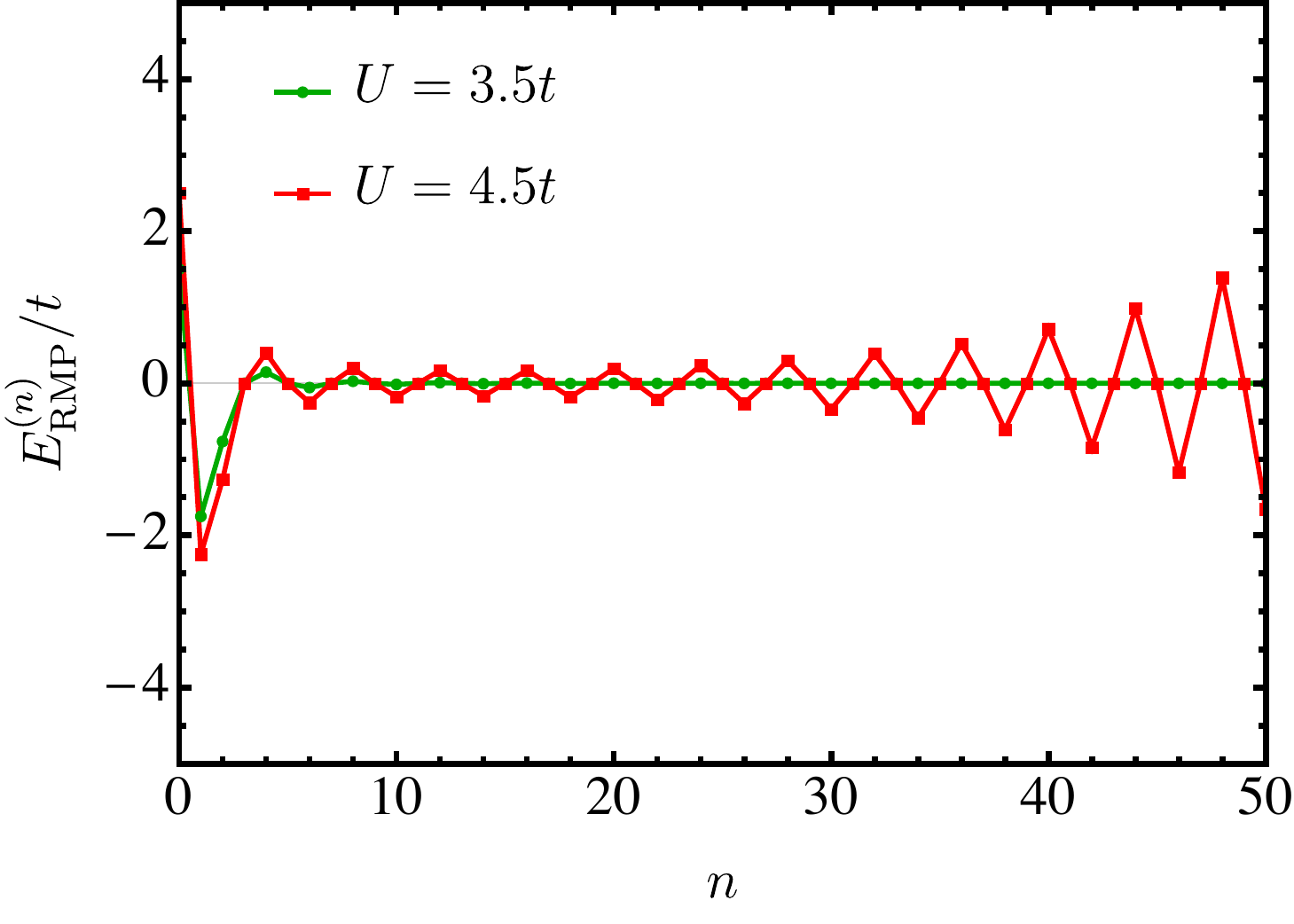}
		\subcaption{\label{subfig:RMP_cvg}}
    \end{subfigure}
    \begin{subfigure}{0.32\textwidth}
	\includegraphics[height=0.75\textwidth]{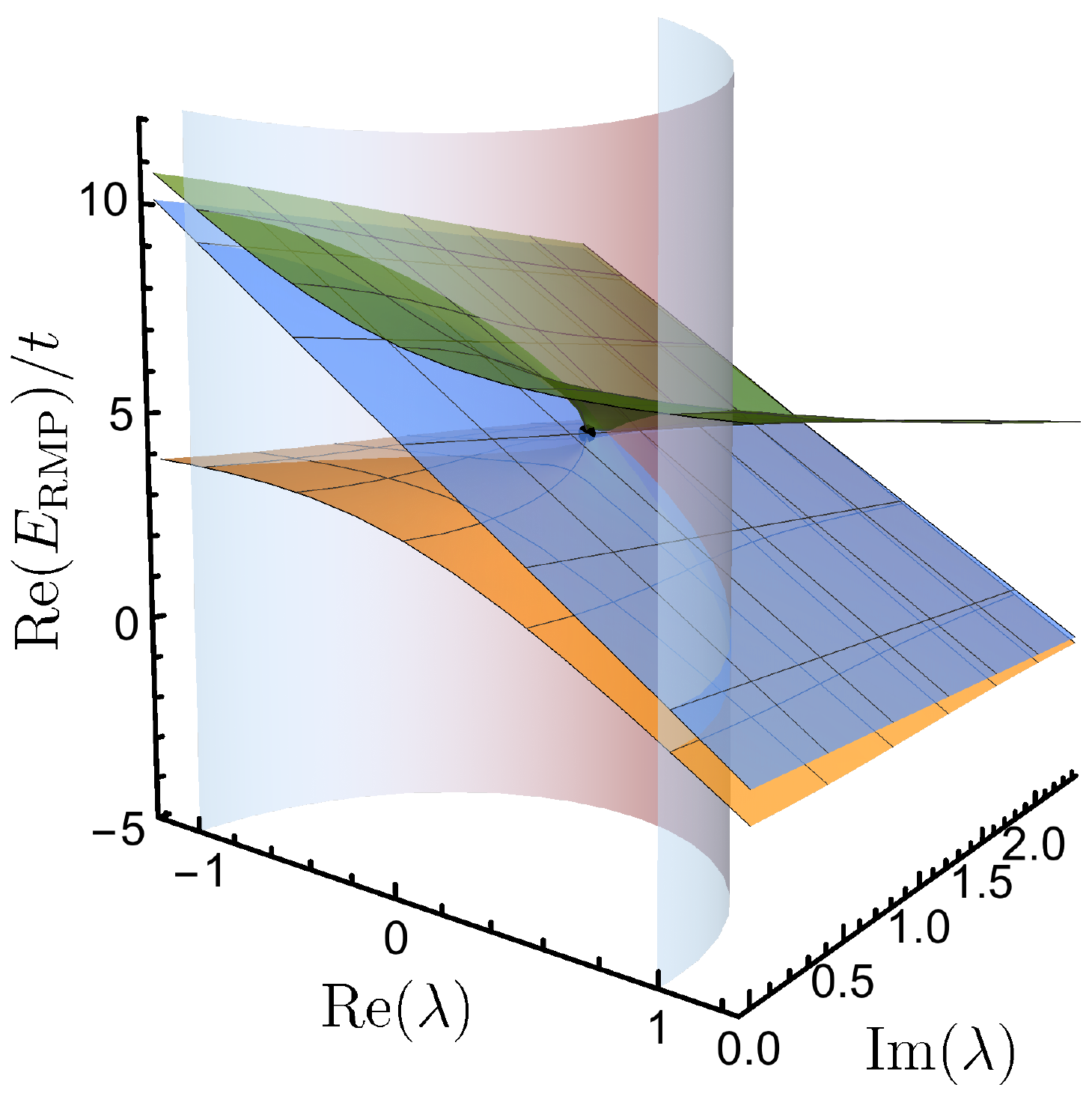}	
		\subcaption{\label{subfig:RMP_4.5} $U/t = 4.5$}
    \end{subfigure}
	\caption{
	Convergence of the RMP series as a function of the perturbation order $n$ for the Hubbard dimer at $U/t = 3.5$ (where $r_c > 1$) and $4.5$ (where $r_c < 1$).
	The Riemann surfaces associated with the exact energies of the RMP Hamiltonian \eqref{eq:H_RMP} are also represented for these two values of $U/t$ as functions of complex $\lambda$. 
	\label{fig:RMP}}
\end{figure*}

The behaviour of the RMP and UMP series observed in \ce{H2} can also be illustrated by considering
the analytic Hubbard dimer with a complex-valued perturbation strength.
In this system, the stretching of the \ce{H\bond{-}H} bond is directly mirrored by an increase in the ratio $U/t$.
Using the ground-state RHF reference orbitals leads to the parametrised RMP Hamiltonian
\begin{widetext}
\begin{equation}
\label{eq:H_RMP}
\bH_\text{RMP}\qty(\lambda) = 
	\begin{pmatrix}
		-2t + U - \lambda U/2	&	0					&	0					&	\lambda U/2	\\
		0						&	U - \lambda U/2 	&	\lambda U/2			&	0	\\
		0						&	\lambda U/2			&	U - \lambda U/2 	&	0	\\
		\lambda U/2 			&	0 					&	0					&	2t + U - \lambda U/2	\\
	\end{pmatrix},
\end{equation}
\end{widetext}
which yields the ground-state energy 
\begin{equation}
	\label{eq:E0MP}
	E_{-}(\lambda) = U - \frac{\lambda U}{2} - \frac{1}{2} \sqrt{(4t)^2 + \lambda ^2 U^2}.
\end{equation}
From this expression, the EPs can be identified as $\lep = \pm \i 4t / U$,
giving the radius of convergence
\begin{equation}
    \rc = \abs{\frac{4t}{U}}.
\end{equation}
Remarkably, these EPs are identical to the exact EPs discussed in Sec.~\ref{sec:example}.
The Taylor expansion of the RMP energy can then be evaluated to obtain the $k$th-order MP correction
\begin{equation}
	E_\text{RMP}^{(k)} = U \delta_{0,k} - \frac{1}{2} \frac{U^k}{(4t)^{k-1}} \mqty( 1/2 \\ k/2).
\end{equation}
 
The RMP series is convergent \titou{at $\lambda = 1$} for $U = 3.5\,t$ with $\rc > 1$, as illustrated for the individual terms at each 
perturbation order in Fig.~\ref{subfig:RMP_cvg}.
In contrast, for $U = 4.5t$ one finds $\rc < 1$, and the RMP series becomes divergent \titou{at $\lambda = 1$}.
The corresponding Riemann surfaces for $U = 3.5\,t$ and $4.5\,t$ are shown in Figs.~\ref{subfig:RMP_3.5} and 
\ref{subfig:RMP_4.5}, respectively, with the single EP at $\lep$ (black dot).
\titou{We illustrate the surface $\abs{\lambda} = 1$ using a vertical cylinder of unit radius to provide
a visual aid for determining if the series will converge at the physical case $\lambda =1$.} 
For the divergent case, $\lep$ lies inside this \titou{unit} cylinder, while in the convergent case $\lep$ lies
outside this cylinder.
In both cases, the EP connects the ground state with the doubly-excited state, and thus the convergence behaviour
for the two states using the ground-state RHF orbitals is identical.
\titou{Note that, when $\lep$ lies \emph{on} the unit cylinder, we cannot \textit{a priori} determine 
whether the perturbation series will converge or not.}

\begin{figure}[htb]
	\includegraphics[width=\linewidth]{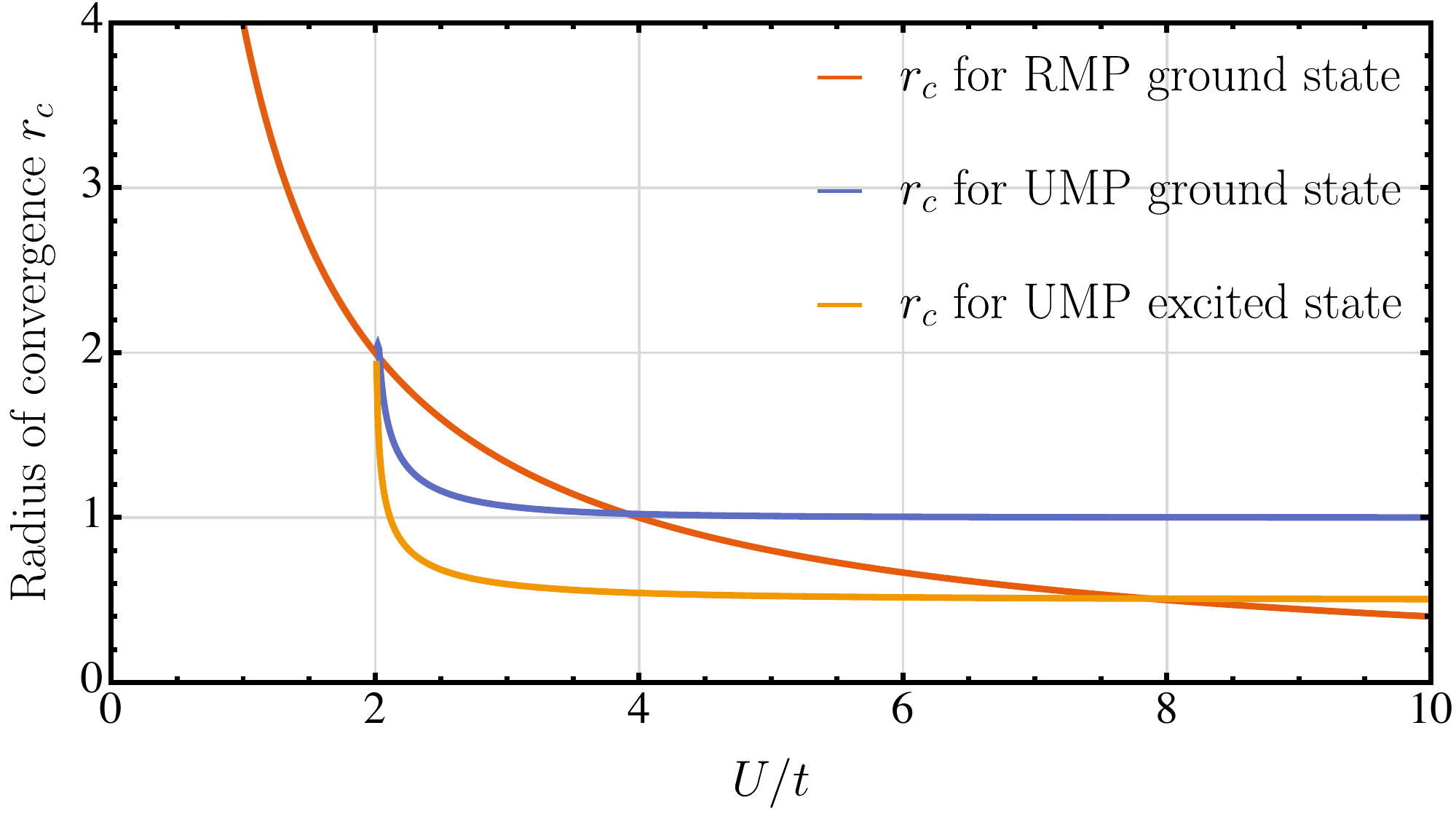}
	\caption{
	Radius of convergence $r_c$ for the RMP ground state (red), the UMP ground state (blue), and the UMP excited state (orange) 
    series of the Hubbard dimer as functions of the ratio $U/t$.
	\label{fig:RadConv}}
\end{figure}

\begin{figure*}
	\begin{subfigure}{0.32\textwidth}
	\includegraphics[height=0.75\textwidth]{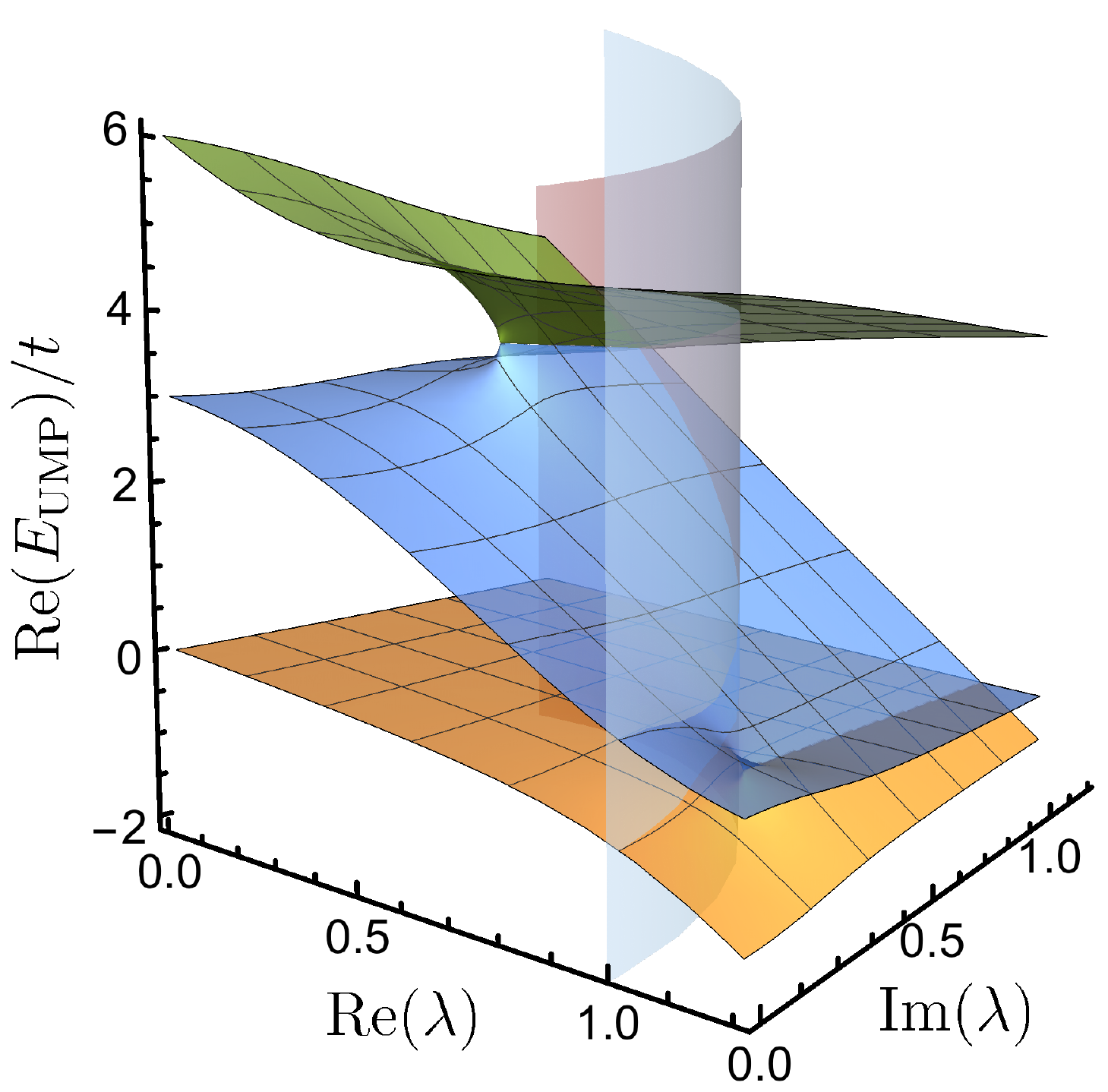}	
		\subcaption{\label{subfig:UMP_3} $U/t = 3$}
    \end{subfigure}
    \begin{subfigure}{0.32\textwidth}
	\includegraphics[height=0.75\textwidth]{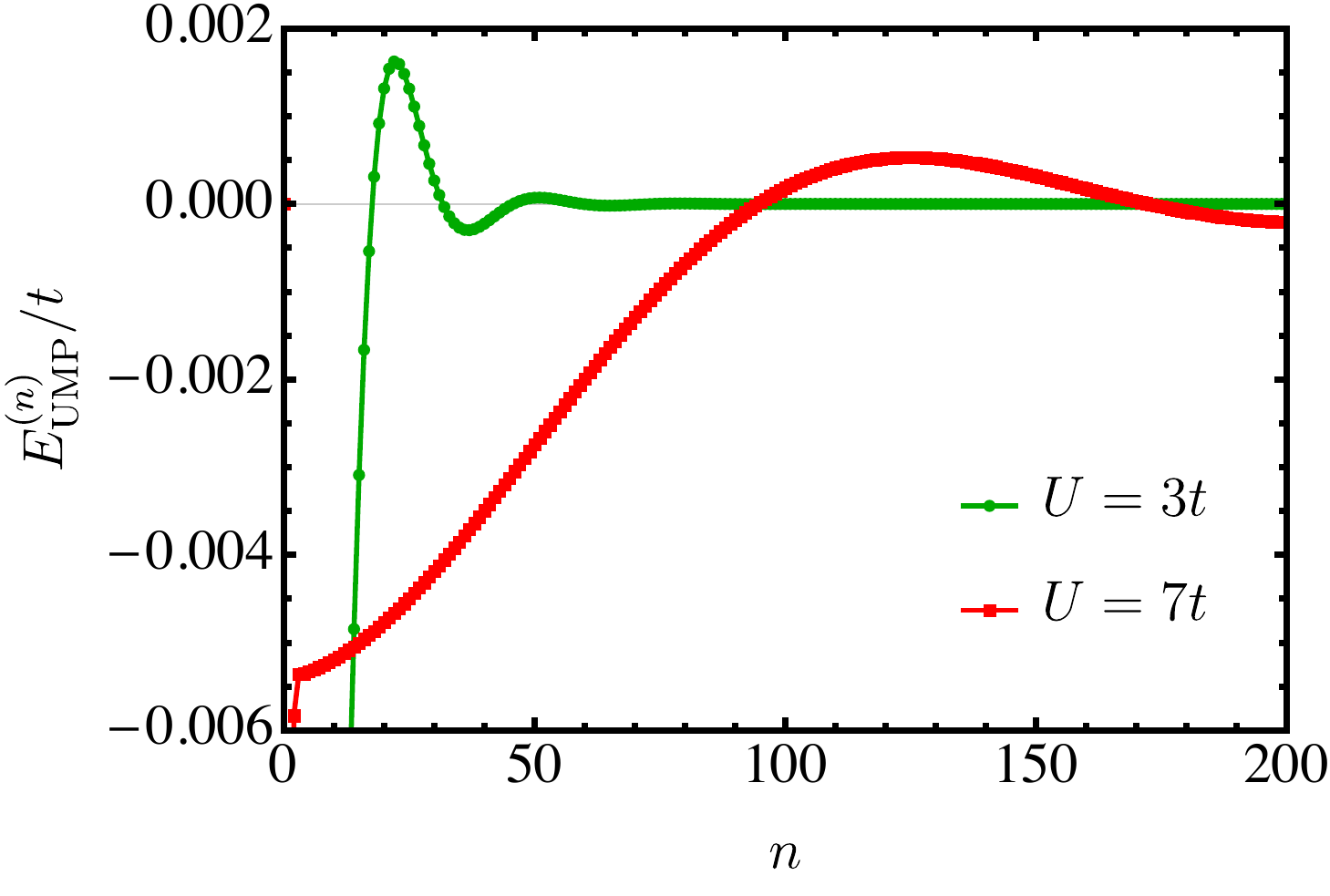}
		\subcaption{\label{subfig:UMP_cvg}}
    \end{subfigure}
    \begin{subfigure}{0.32\textwidth}
	\includegraphics[height=0.75\textwidth]{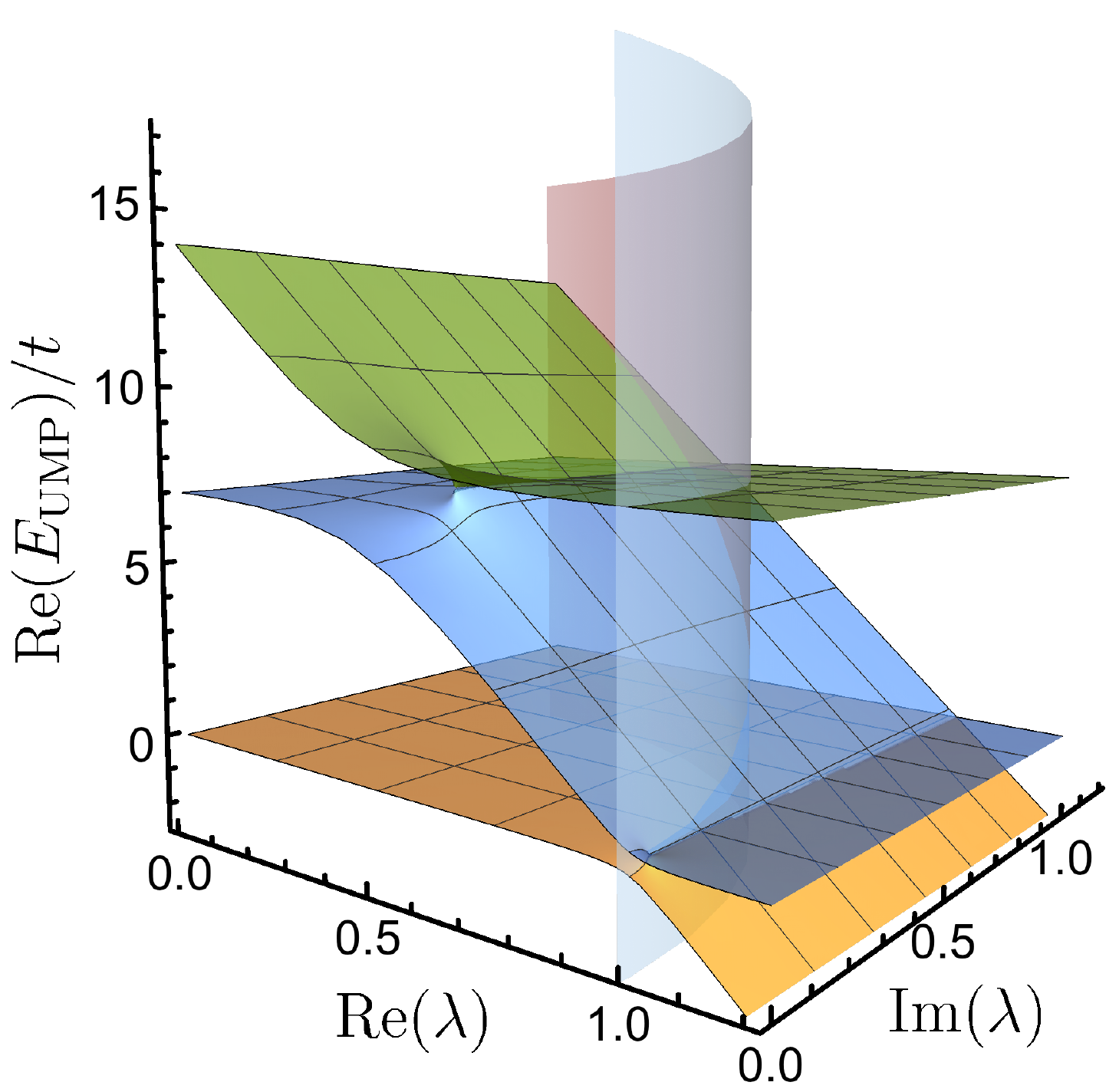}	
		\subcaption{\label{subfig:UMP_7} $U/t = 7$}
    \end{subfigure}	\caption{
	Convergence of the UMP series as a function of the perturbation order $n$ for the Hubbard dimer at $U/t = 3$ and $7$.
	The Riemann surfaces associated with the exact energies of the UMP Hamiltonian \eqref{eq:H_UMP} are also represented for these two values of $U/t$ as functions of $\lambda$.
	\label{fig:UMP}}
\end{figure*}

The behaviour of the UMP series is more subtle than the RMP series as the spin-contamination in the wave function
introduces additional coupling between the singly- and doubly-excited configurations.
Using the ground-state UHF reference orbitals in the Hubbard dimer yields the parametrised UMP Hamiltonian
\begin{widetext}
\begin{equation}
\label{eq:H_UMP}
\bH_\text{UMP}\qty(\lambda) = 
	\begin{pmatrix}
		-2t^2 \lambda/U	&	0									&	0									&	2t^2 \lambda/U		\\
		0				&	U - 2t^2 \lambda/U 					&	2t^2\lambda/U						&	2t \sqrt{U^2 - (2t)^2} \lambda/U	\\
		0				&	2t^2\lambda/U						&	U - 2t^2 \lambda/U 					&	-2t \sqrt{U^2 - (2t)^2} \lambda/U	\\
		2t^2 \lambda/U	&	2t \sqrt{U^2 - (2t)^2} \lambda/U 	&	-2t \sqrt{U^2 - (2t)^2} \lambda/U	&	2U(1-\lambda) + 6t^2\lambda/U		\\
	\end{pmatrix}.
\end{equation}
\end{widetext}
While a closed-form expression for the ground-state energy exists, it is cumbersome and we eschew reporting it.
Instead, the radius of convergence of the UMP series can be obtained numerically as a function of $U/t$, as shown
in Fig.~\ref{fig:RadConv}.
These numerical values reveal that the UMP ground-state series has $\rc > 1$ for all $U/t$ and always converges.
However, in the strong correlation limit (large $U/t$), this radius of convergence tends to unity, indicating that
the convergence of the corresponding UMP series becomes increasingly slow.
Furthermore, the doubly-excited state using the ground-state UHF orbitals has $\rc < 1$ for almost any value 
of $U/t$, reaching the limiting value of $1/2$ for $U/t \to \infty$. Hence, the 
excited-state UMP series will always diverge.
 
The convergence behaviour can be further elucidated by considering the full structure of the UMP energies 
in the complex $\lambda$-plane (see Figs.~\ref{subfig:UMP_3} and \ref{subfig:UMP_7}).
These Riemann surfaces are illustrated for $U = 3t$ and $7t$ alongside the perturbation terms at each order
in Fig.~\ref{subfig:UMP_cvg}.
At $U = 3t$, the RMP series is convergent \titou{at $\lambda = 1$}, while RMP becomes divergent \titou{at $\lambda = 1$} for $U=7t$.
The ground-state UMP expansion is convergent in both cases, although the rate of convergence is significantly slower 
for larger $U/t$ as the radius of convergence becomes increasingly close to one (Fig.~\ref{fig:RadConv}).

As the UHF orbitals break the \titou{spatial and} spin symmetry, new coupling terms emerge between the electronic states that
cause fundamental changes to the structure of EPs in the complex $\lambda$-plane.
For example, while the RMP energy shows only one EP between the ground and 
doubly-excited states (Fig.~\ref{fig:RMP}), the UMP energy has two pairs of complex-conjugate EPs: one connecting the ground state with the
singly-excited open-shell singlet, and the other connecting this single excitation to the 
doubly-excited second excitation (Fig.~\ref{fig:UMP}).
This new ground-state EP always appears outside the unit cylinder and guarantees convergence of the ground-state energy.
However, the excited-state EP is moved within the unit cylinder and causes the 
convergence of the excited-state UMP series to deteriorate.
Our interpretation of this effect is that the symmetry-broken orbital optimisation has redistributed the strong 
coupling between the ground- and doubly-excited states into weaker couplings between all states, and has thus
sacrificed convergence of the excited-state series so that the ground-state convergence can be maximised.

Since the UHF ground state already provides a good approximation to the exact energy, the ground-state sheet of
the UMP energy is relatively flat and the corresponding EP in the Hubbard dimer always lies outside the unit cylinder.
The slow convergence observed in stretched \ce{H2}\cite{Gill_1988} can then be seen as this EP 
moves increasingly close to the unit cylinder at large $U/t$ and $\rc$ approaches one (from above).
Furthermore, the majority of the UMP expansion in this regime is concerned with removing spin-contamination from the wave 
function rather than improving the energy.
It is well-known that the spin-projection needed to remove spin-contamination can require non-linear combinations
of highly-excited determinants,\cite{Lowdin_1955c} and thus it is not surprising that this process proceeds 
very slowly as the perturbation order is increased.

\subsection{Classifying Types of Convergence} 

As computational implementations of higher-order MP terms improved, the systematic investigation 
of convergence behaviour in a broader class of molecules became possible.
Cremer and He introduced an efficient MP6 approach and used it to analyse the RMP convergence of
29 atomic and molecular systems.\cite{Cremer_1996}
They established two general classes: ``class A'' systems that exhibit monotonic convergence; 
and ``class B'' systems for which convergence is erratic after initial oscillations. 
By analysing the different cluster contributions to the MP energy terms, they proposed that
class A systems generally include well-separated and weakly correlated electron pairs, while class B systems
are characterised by dense electron clustering in one or more spatial regions.\cite{Cremer_1996}
In class A systems, they showed that the majority of the correlation energy arises from pair correlation, 
with little contribution from triple excitations.
On the other hand, triple excitations have an important contribution in class B systems, including
orbital relaxation to doubly-excited configurations, and these contributions lead to oscillations of the total correlation energy.

Using these classifications, Cremer and He then introduced simple extrapolation formulas for estimating the 
exact correlation energy $\Delta E$ using terms up to MP6\cite{Cremer_1996}
\begin{subequations}
\begin{align}
	\label{eq:CrHeA}
\Delta E_{\text{A}}
    &= \Emp^{(2)} + \Emp^{(3)} + \Emp^{(4)}
     + \frac{\Emp^{(5)}}{1 - (\Emp^{(6)} / \Emp^{(5)})}, 
     \\
	\label{eq:CrHeB}
\Delta E_{\text{B}} 
    &= \Emp^{(2)} + \Emp^{(3)} + \qty(\Emp^{(4)} + \Emp^{(5)}) \exp(\Emp^{(6)} / \Emp^{(5)}).
\end{align}
\end{subequations}
These class-specific formulas reduced the mean absolute error from the FCI correlation energy by a
factor of four compared to previous class-independent extrapolations,
highlighting how one can leverage a deeper understanding of MP convergence to improve estimates of 
the correlation energy at lower computational costs. 
In Sec.~\ref{sec:Resummation}, we consider more advanced extrapolation routines that take account of EPs in the complex $\lambda$-plane.

In the late 90's, Olsen \etal\ discovered an even more concerning behaviour of the MP series. \cite{Olsen_1996} 
They showed that the series could be divergent even in systems that were considered to be well understood, 
such as \ce{Ne} or the \ce{HF} molecule. \cite{Olsen_1996, Christiansen_1996} 
Cremer and He had already studied these two systems and classified them as \textit{class B} systems.\cite{Cremer_1996} 
However, Olsen and co-workers performed their analysis in larger basis sets containing diffuse functions,
finding that the corresponding MP series becomes divergent at (very) high order.
The discovery of this divergent behaviour is particularly worrying as large basis sets 
are required to get meaningful and accurate energies.\cite{Loos_2019d,Giner_2019}
Furthermore, diffuse functions are particularly important for anions and/or Rydberg excited states, where the wave function 
is inherently more diffuse than the ground state.\cite{Loos_2018a,Loos_2020a}

Olsen \etal\ investigated the causes of these divergences and the different types of convergence by
analysing the relation between the dominant singularity (\ie, the closest singularity to the origin) 
and the convergence behaviour of the series.\cite{Olsen_2000} 
Their analysis is based on Darboux's theorem: \cite{Goodson_2011}
\begin{quote}
\textit{``In the limit of large order, the series coefficients become equivalent to 
    the Taylor series coefficients of the singularity closest to the origin. ''}
\end{quote}
\titou{Following this theory, a singularity in the unit circle is designated as an intruder state, 
with a front-door (or back-door) intruder state if the real part of the singularity is positive (or negative).}

Using their observations in Ref.~\onlinecite{Olsen_1996}, Olsen and collaborators proposed 
a simple method that performs a scan of the real axis to detect the avoided crossing responsible 
for the dominant singularities in the complex plane. \cite{Olsen_2000} 
By modelling this avoided crossing using a two-state Hamiltonian, one can obtain an approximation for
the dominant singularities as the EPs of the two-state matrix
\begin{equation}
	\label{eq:Olsen_2x2}
    \underbrace{\mqty(\alpha & \delta \\ \delta & \beta )}_{\bH} 
    = \underbrace{\mqty(\alpha + \alpha_{\text{s}} & 0 \\ 0 & \beta + \beta_{\text{s}} )}_{\bH^{(0)}} 
    + \underbrace{\mqty( -\alpha_{\text{s}} & \delta \\ \delta & - \beta_{\text{s}})}_{\bV},
\end{equation}
where the diagonal matrix is the unperturbed Hamiltonian matrix $\bH^{(0)}$ with level shifts
$\alpha_{\text{s}}$ and $\beta_{\text{s}}$, and $\bV$ represents the perturbation.

The authors first considered molecules with low-lying doubly-excited states with the same spatial
and spin symmetry as the ground state. \cite{Olsen_2000}
In these systems, the exact wave function has a non-negligible contribution from the doubly-excited states, 
and thus the low-lying excited states are likely to become intruder states. 
For \ce{CH_2} in a diffuse, yet rather small basis set, the series is convergent \titou{at $\lambda = 1$} at least up to the 50th order, and
the dominant singularity lies close (but outside) the unit circle, causing slow convergence of the series.
These intruder-state effects are analogous to the EP that dictates the convergence behaviour of 
the RMP series for the Hubbard dimer (Fig.~\ref{fig:RMP}).
Furthermore, the authors demonstrated that the divergence for \ce{Ne} is due to a back-door intruder state
that arise when the ground state undergoes sharp avoided crossings with highly diffuse excited states.
This divergence is related to a more fundamental critical point in the MP energy surface that we will
discuss in Sec.~\ref{sec:MP_critical_point}.

Finally, Ref.~\onlinecite{Olsen_1996} proved that the extrapolation formulas of Cremer and He \cite{Cremer_1996} 
[see Eqs.~\eqref{eq:CrHeA} and \eqref{eq:CrHeB}] are not mathematically motivated when considering the complex 
singularities causing the divergence, and therefore cannot be applied for all systems.
For example, the \ce{HF} molecule contains both back-door intruder states and low-lying doubly-excited states that
result in alternating terms up to 10th order. 
The series becomes monotonically convergent at higher orders since
the two pairs of singularities are approximately the same distance from the origin.

More recently, this two-state model has been extended to non-symmetric Hamiltonians as\cite{Olsen_2019}
\begin{equation}
	\underbrace{\mqty(\alpha & \delta_1 \\ \delta_2 & \beta)}_{\bH} = \underbrace{\mqty(\alpha & 0 \\ 0 & \beta + \gamma )}_{\bH^{(0)}} + \underbrace{\mqty( 0 & \delta_2 \\ \delta_1 & - \gamma)}_{\bV}.
\end{equation}
This extension allows various choices of perturbation to be analysed, including coupled cluster 
perturbation expansions \cite{Pawlowski_2019a,Pawlowski_2019b,Pawlowski_2019c,Pawlowski_2019d,Pawlowski_2019e} 
and other non-Hermitian perturbation methods.
Note that new forms of perturbation expansions only occur when the sign of $\delta_1$ and $\delta_2$ differ.
Using this non-Hermitian two-state model, the convergence of a perturbation series can be characterised 
according to a so-called ``archetype'' that defines the overall ``shape'' of the energy convergence.\cite{Olsen_2019} 
For Hermitian Hamiltonians, these archetypes can be subdivided into five classes 
(zigzag, interspersed zigzag, triadic, ripples, and geometric), 
while two additional archetypes (zigzag-geometric and convex-geometric) are observed in non-Hermitian Hamiltonians.
The geometric archetype appears to be the most common for MP expansions,\cite{Olsen_2019} but the 
ripples archetype corresponds to some of the early examples of MP convergence. \cite{Handy_1985,Lepetit_1988,Leininger_2000,Malrieu_2003}
The three remaining Hermitian archetypes seem to be rarely observed in MP perturbation theory.
In contrast, the non-Hermitian coupled cluster perturbation theory,%
\cite{Pawlowski_2019a,Pawlowski_2019b,Pawlowski_2019c,Pawlowski_2019d,Pawlowski_2019e} exhibits a range of archetypes
including the interspersed zigzag, triadic, ripple, geometric, and zigzag-geometric forms.
This analysis highlights the importance of the primary singularity in controlling the high-order convergence, 
regardless of whether this point is inside or outside the complex unit circle. \cite{Handy_1985,Olsen_2000}

\subsection{M{\o}ller--Plesset Critical Point}
\label{sec:MP_critical_point}

In the early 2000's, Stillinger reconsidered the mathematical origin behind the divergent series with odd-even
sign alternation.\cite{Stillinger_2000} 
This type of convergence behaviour corresponds to Cremer and He's class B systems with closely spaced
electron pairs and includes \ce{Ne}, \ce{HF}, \ce{F-}, and \ce{H2O}.\cite{Cremer_1996}
Stillinger proposed that these series diverge due to a dominant singularity
on the negative real $\lambda$ axis, corresponding to a multielectron autoionisation threshold.\cite{Stillinger_2000}
To understand Stillinger's argument, consider the parametrised MP Hamiltonian in the form
\begin{multline}
\label{eq:HamiltonianStillinger}
    \hH(\lambda) = 
    \sum_{i}^{\Ne} \Bigg[ 
    \overbrace{-\frac{1}{2}\grad_i^2 
    - \sum_{A}^{\Nn} \frac{Z_A}{\abs{\vb{r}_i-\vb{R}_A}}}^{\text{independent of $\lambda$}}
    \\
    + \underbrace{(1-\lambda)\vhf(\vb{x}_i)}_{\text{repulsive for $\lambda < 1$}}
    + \underbrace{\lambda\sum_{i<j}^{\Ne}\frac{1}{|\vb{r}_i-\vb{r}_j|}}_{\text{attractive for $\lambda < 0$}}
    \Bigg].
\end{multline}
The mean-field potential $\vhf$ essentially represents a negatively charged field with the spatial extent
controlled by the extent of the HF orbitals, usually located close to the nuclei.
When $\lambda$ is negative, the mean-field potential becomes increasingly repulsive, while the explicit two-electron 
Coulomb interaction becomes attractive.
There is therefore a negative critical point $\lc$ where it becomes energetically favourable for the electrons 
to dissociate and form a bound cluster at an infinite separation from the nuclei.\cite{Stillinger_2000}
This autoionisation effect is closely related to the critical point for electron binding in two-electron 
atoms (see Ref.~\onlinecite{Baker_1971}).
Furthermore, a similar set of critical points exists along the positive real axis, corresponding to single-electron ionisation
processes.\cite{Sergeev_2005}
While these critical points are singularities on the real axis, their exact mathematical form is difficult 
to identify and remains an open question.

To further develop the link between the critical point and types of MP convergence, Sergeev and Goodson investigated
the relationship with the location of the dominant singularity that controls the radius of convergence.\cite{Goodson_2004}
They demonstrated that the dominant singularity in class A systems corresponds to an EP with a positive real component, 
where the magnitude of the imaginary component controls the oscillations in the signs of successive MP 
terms.\cite{Goodson_2000a,Goodson_2000b}
In contrast, class B systems correspond to a dominant singularity on the negative real $\lambda$ axis representing
the MP critical point.
The divergence of class B systems, which contain closely spaced electrons (\eg, \ce{F-}), can then be understood as the 
HF potential $\vhf$ is relatively localised and the autoionization is favoured at negative 
$\lambda$ values closer to the origin.
With these insights, they regrouped the systems into new classes: i) $\alpha$ singularities which have ``large'' imaginary parts, 
and ii) $\beta$ singularities which have very small imaginary parts.\cite{Goodson_2004,Sergeev_2006} 

The existence of the MP critical point can also explain why the divergence observed by Olsen \etal\ in the \ce{Ne} atom 
and the \ce{HF} molecule occurred when diffuse basis functions were included.\cite{Olsen_1996}
Clearly diffuse basis functions are required for the electrons to dissociate from the nuclei, and indeed using
only compact basis functions causes the critical point to disappear.
While a finite basis can only predict complex-conjugate branch point singularities, the critical point is modelled
by a cluster of sharp avoided crossings between the ground state and high-lying excited states.\cite{Sergeev_2005}
Alternatively, Sergeev \etal\ demonstrated that the inclusion of a ``ghost'' atom also
allows the formation of the critical point as the electrons form a bound cluster occupying the ghost atom orbitals.\cite{Sergeev_2005}
This effect explains the origin of the divergence in the \ce{HF} molecule as the fluorine valence electrons jump to the hydrogen at
 a sufficiently negative $\lambda$ value.\cite{Sergeev_2005}
Furthermore, the two-state model of Olsen and collaborators \cite{Olsen_2000} was simply too minimal to understand the complexity of 
divergences caused by the MP critical point.

When a Hamiltonian is parametrised by a variable such as $\lambda$, the existence of abrupt changes in the 
eigenstates as a function of $\lambda$ indicate the presence of a zero-temperature quantum phase transition (QPT).%
\cite{Heiss_1988,Heiss_2002,Borisov_2015,Sindelka_2017,CarrBook,Vojta_2003,SachdevBook,GilmoreBook} 
Meanwhile, as an avoided crossing becomes increasingly sharp, the corresponding EPs move increasingly close to the real axis.
When these points converge on the real axis, they effectively ``annihilate'' each other and no longer behave as EPs.
Instead, they form a ``critical point'' singularity that resembles a conical intersection, and 
the convergence of a pair of complex-conjugate EPs on the real axis is therefore diagnostic of a QPT.\cite{Cejnar_2005, Cejnar_2007a}

Since the MP critical point corresponds to a singularity on the real $\lambda$ axis, it can immediately be
recognised as a QPT with respect to varying the perturbation parameter $\lambda$.
However, a conventional QPT can only occur in the thermodynamic limit, which here is analogous to the complete 
basis set limit.\cite{Kais_2006}
The MP critical point and corresponding $\beta$ singularities in a finite basis must therefore be modelled by pairs of
complex-conjugate EPs that tend towards the real axis, exactly as described by Sergeev \etal\cite{Sergeev_2005}
In contrast, $\alpha$ singularities correspond to large avoided crossings that are indicative of low-lying excited
states which share the symmetry of the ground state,\cite{Goodson_2004} and are thus not manifestations of a QPT.
Notably, since the exact MP critical point corresponds to the interaction between a bound state
and the continuum, its functional form is more complicated than a conical intersection and remains an open question.

\subsection{Critical Points in the Hubbard Dimer}
\label{sec:critical_point_hubbard}

\begin{figure*}[t]
	\begin{subfigure}{0.32\textwidth}
	\includegraphics[height=0.75\textwidth]{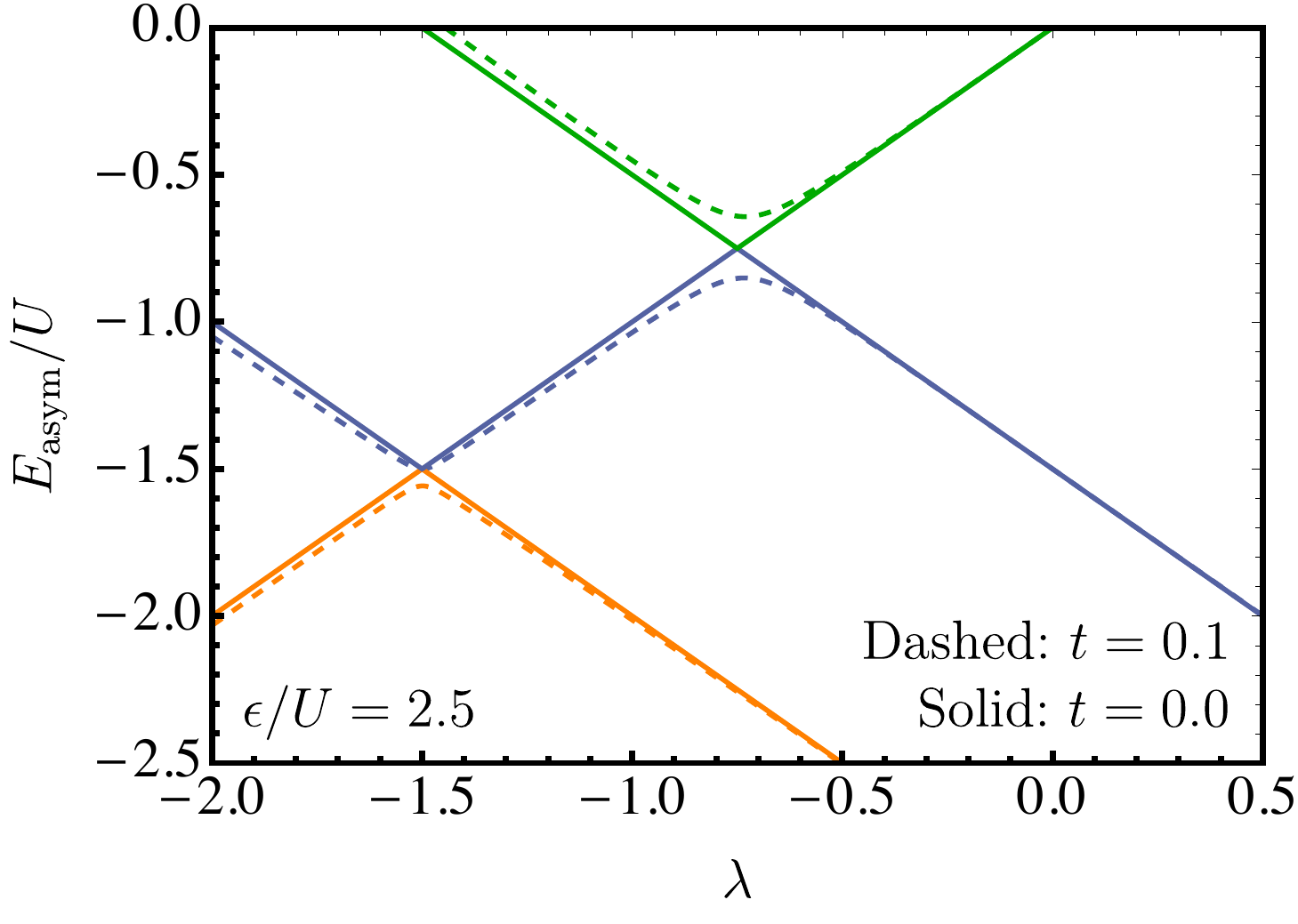}	
		\subcaption{\label{subfig:rmp_cp}}
    \end{subfigure}
    \begin{subfigure}{0.32\textwidth}
	\includegraphics[height=0.75\textwidth]{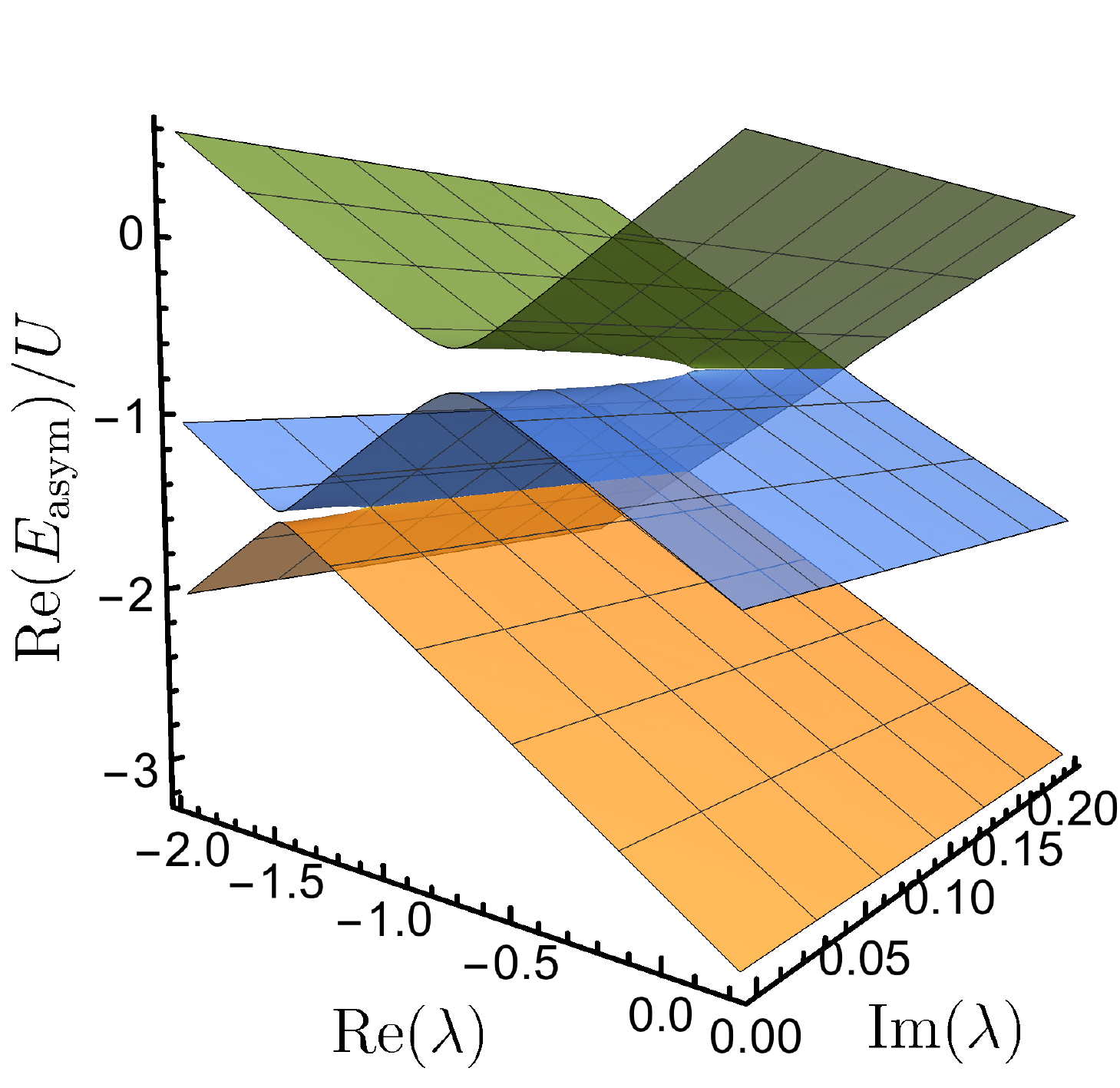}
		\subcaption{\label{subfig:rmp_cp_surf}}
    \end{subfigure}
    \begin{subfigure}{0.32\textwidth}
    \includegraphics[height=0.75\textwidth]{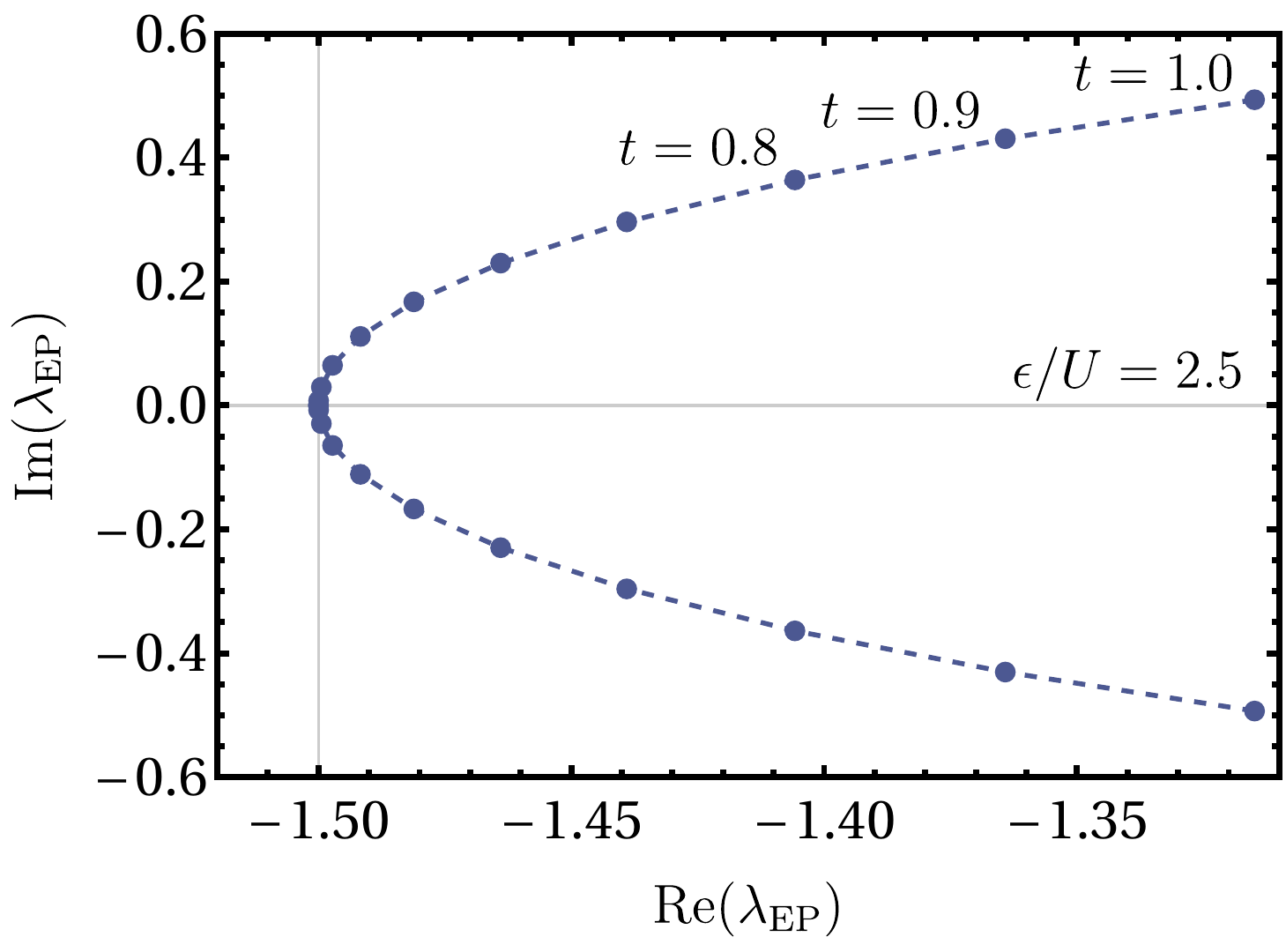}
		\subcaption{\label{subfig:rmp_ep_to_cp}}
    \end{subfigure}
	\caption{%
		RMP critical point using the asymmetric Hubbard dimer with $\epsilon = 2.5 U$.
		(\subref{subfig:rmp_cp}) Exact critical points with $t=0$ occur on the negative real $\lambda$ axis (dashed).
		(\subref{subfig:rmp_cp_surf}) Modelling a finite basis using $t=0.1$ yields complex-conjugate EPs close to the
		real axis, giving a sharp avoided crossing on the real axis (solid).
		(\subref{subfig:rmp_ep_to_cp}) Convergence of the ground-state EP onto the real axis in the limit $t \to 0$.
	\label{fig:RMP_cp}}
\end{figure*}

The simplified site basis of the Hubbard dimer makes explicitly modelling the ionisation continuum impossible.
Instead, we can use an asymmetric version of the Hubbard dimer \cite{Carrascal_2015,Carrascal_2018} 
where we consider one of the sites as a ``ghost atom'' that acts as a 
destination for ionised electrons being originally localised on the other site.
To mathematically model this scenario in this asymmetric Hubbard dimer, we introduce a one-electron potential $-\epsilon$ on the left site to 
represent the attraction between the electrons and the model ``atomic'' nucleus, where we define $\epsilon > 0$.
The reference Slater determinant for a doubly-occupied atom can be represented using RHF
orbitals [see Eq.~\eqref{eq:RHF_orbs}] with $\ta_{\text{RHF}} = \tb_{\text{RHF}} = 0$, 
which corresponds to strictly localising the two electrons on the left site.
With this representation, the parametrised asymmetric RMP Hamiltonian becomes
\begin{widetext}
\begin{equation}
\label{eq:H_asym}
\bH_\text{asym}\qty(\lambda) = 
\begin{pmatrix}
    2(U-\epsilon) - \lambda U &	-\lambda t				 &	-\lambda t	            &	0	        \\
    -\lambda t				  &	(U-\epsilon) - \lambda U &	0		                &	-\lambda t	\\
    -\lambda t				  &	0			             &	(U-\epsilon) -\lambda U	&	-\lambda t	\\
    0 				          &	-\lambda t 	 			 &	-\lambda t              &	\lambda U	\\
\end{pmatrix}.
\end{equation}
\end{widetext}

For the ghost site to perfectly represent ionised electrons, the hopping term between the two sites must vanish (\ie, $t=0$).
This limit corresponds to the dissociative regime in the asymmetric Hubbard dimer as discussed in Ref.~\onlinecite{Carrascal_2018}, 
and the RMP energies become
\begin{subequations}
\begin{align}
    E_{-} &= 2(U - \epsilon) - \lambda U,
    \\
    E_{\text{S}} &= (U - \epsilon) - \lambda U,
    \\ 
    E_{+} &= U \lambda,
\end{align}
\end{subequations}
as shown in Fig.~\ref{subfig:rmp_cp} (dashed lines).
The RMP critical point then corresponds to the intersection $E_{-} = E_{+}$, giving the critical $\lambda$ value
\begin{equation}
    \lc = 1 - \frac{\epsilon}{U}. 
\end{equation}
Clearly the radius of convergence $\rc = \abs{\lc}$ is controlled directly by the ratio $\epsilon / U$, 
with a convergent RMP series \titou{at $\lambda = 1$} occurring for $\epsilon > 2 U$.
The on-site repulsion $U$ controls the strength of the HF potential localised around the ``atomic site'', with a
stronger repulsion encouraging the electrons to be ionised at a less negative value of $\lambda$. 
Large $U$ can be physically interpreted as strong electron repulsion effects in electron dense molecules. 
In contrast, smaller $\epsilon$ gives a weaker attraction to the atomic site, 
representing strong screening of the nuclear attraction by core and valence electrons, 
and again a less negative $\lambda$ is required for ionisation to occur.
Both of these factors are common in atoms on the right-hand side of the periodic table, \eg, \ce{F},
\ce{O}, \ce{Ne}.
Molecules containing these atoms are therefore often class $\beta$ systems with
a divergent RMP series due to the MP critical point. \cite{Goodson_2004,Sergeev_2006} 

\begin{figure}[b]
	\includegraphics[width=\linewidth]{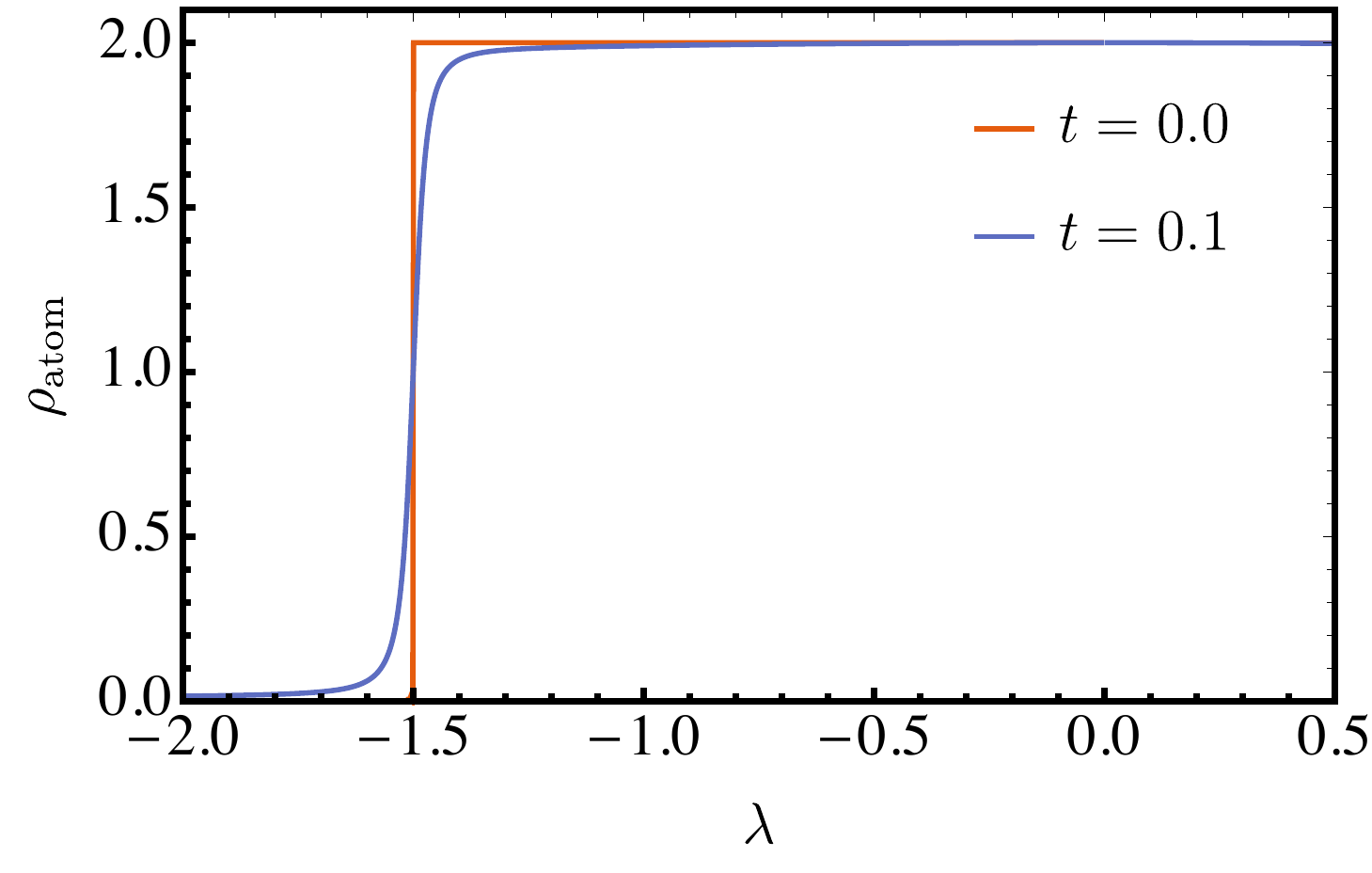}
	\caption{
\titou{Electron density $\rho_\text{atom}$ on the ``atomic'' site of the asymmetric Hubbard dimer with 
$\epsilon = 2.5 U$. 
The autoionisation process associated with the critical point is represented by the sudden drop on the negative $\lambda$ axis.
In the idealised limit $t=0$, this process becomes increasingly sharp and represents a zero-temperature QPT.} 
	\label{fig:rmp_dens}}
\end{figure}

The critical point in the exact case $t=0$ \titou{is represented by the gradient discontinuity in the 
ground-state energy} on the negative real $\lambda$ axis (Fig.~\ref{subfig:rmp_cp}: solid lines), 
mirroring the behaviour of a quantum phase transition.\cite{Kais_2006}
\titou{The autoionisation process is manifested by a sudden drop in the ``atomic site'' 
electron density $\rho_\text{atom}$ (Fig.~\ref{fig:rmp_dens}).} 
However, in practical calculations performed with a finite basis set, the critical point is modelled as a cluster
of branch points close to the real axis.
The use of a finite basis can be modelled in the asymmetric dimer by making the second site a less
idealised destination for the ionised electrons with a non-zero (yet small) hopping term $t$.
Taking the value $t=0.1$ (Fig.~\ref{subfig:rmp_cp}: dashed lines), the critical point becomes  
an avoided crossing with a complex-conjugate pair of EPs close to the real axis (Fig.~\ref{subfig:rmp_cp_surf}).
\titou{In contrast to the exact critical point with $t=0$, the ground-state energy remains
smooth through this avoided crossing, with a more gradual drop in the atomic site density.}
In the limit $t \to 0$, these EPs approach the real axis (Fig.~\ref{subfig:rmp_ep_to_cp}) \titou{and the 
avoided crossing becomes a gradient discontinuity},
mirroring Sergeev's discussion on finite basis
set representations of the MP critical point.\cite{Sergeev_2006}

\begin{figure*}[t]
	\begin{subfigure}{0.32\textwidth}
    \includegraphics[height=0.75\textwidth,trim={0pt 5pt -10pt 15pt},clip]{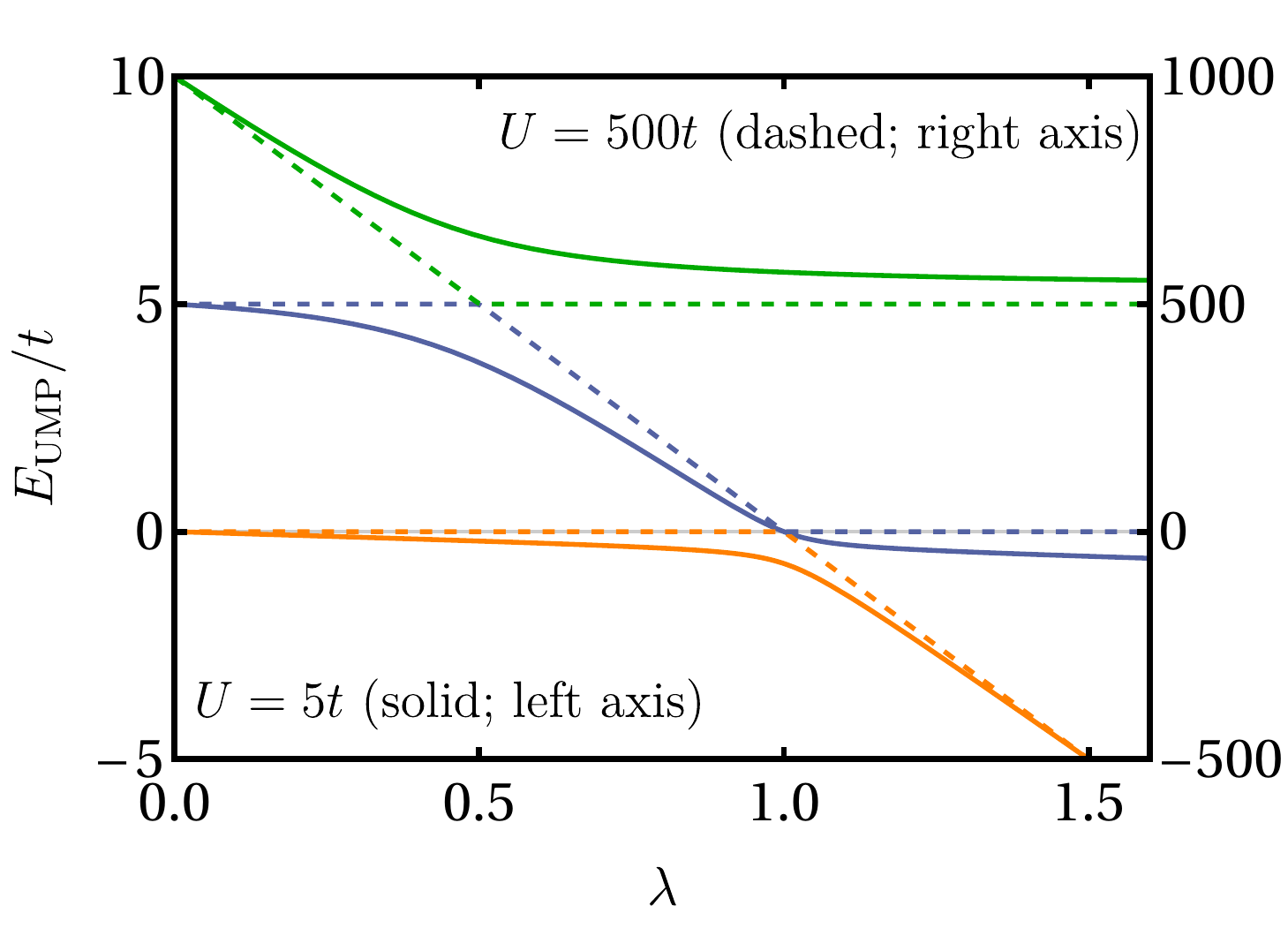}
		\subcaption{\label{subfig:ump_cp}}
    \end{subfigure}
    \begin{subfigure}{0.32\textwidth}
	\includegraphics[height=0.75\textwidth]{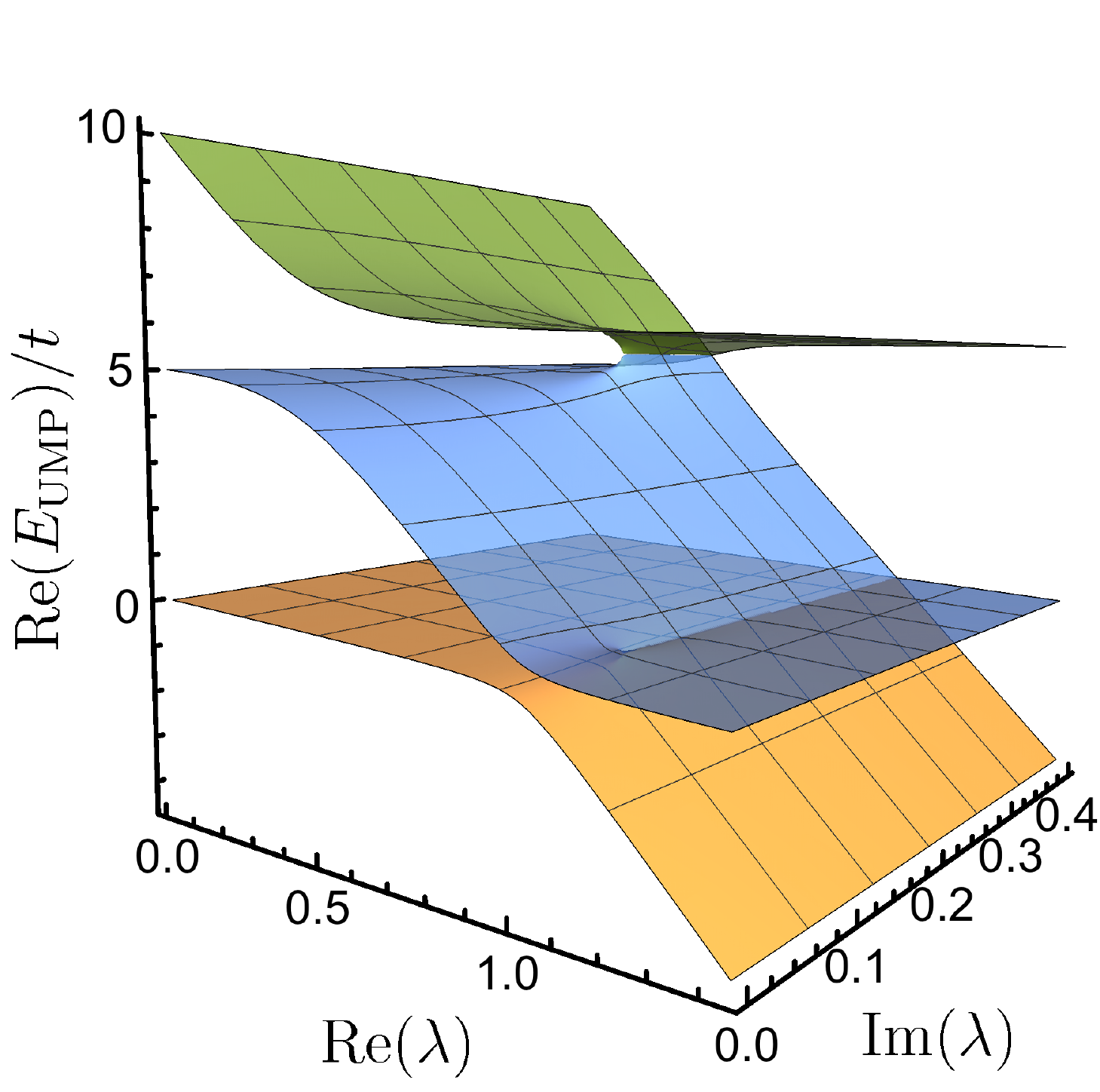}
		\subcaption{\label{subfig:ump_cp_surf}}
    \end{subfigure}
    \begin{subfigure}{0.32\textwidth}
	\includegraphics[height=0.75\textwidth]{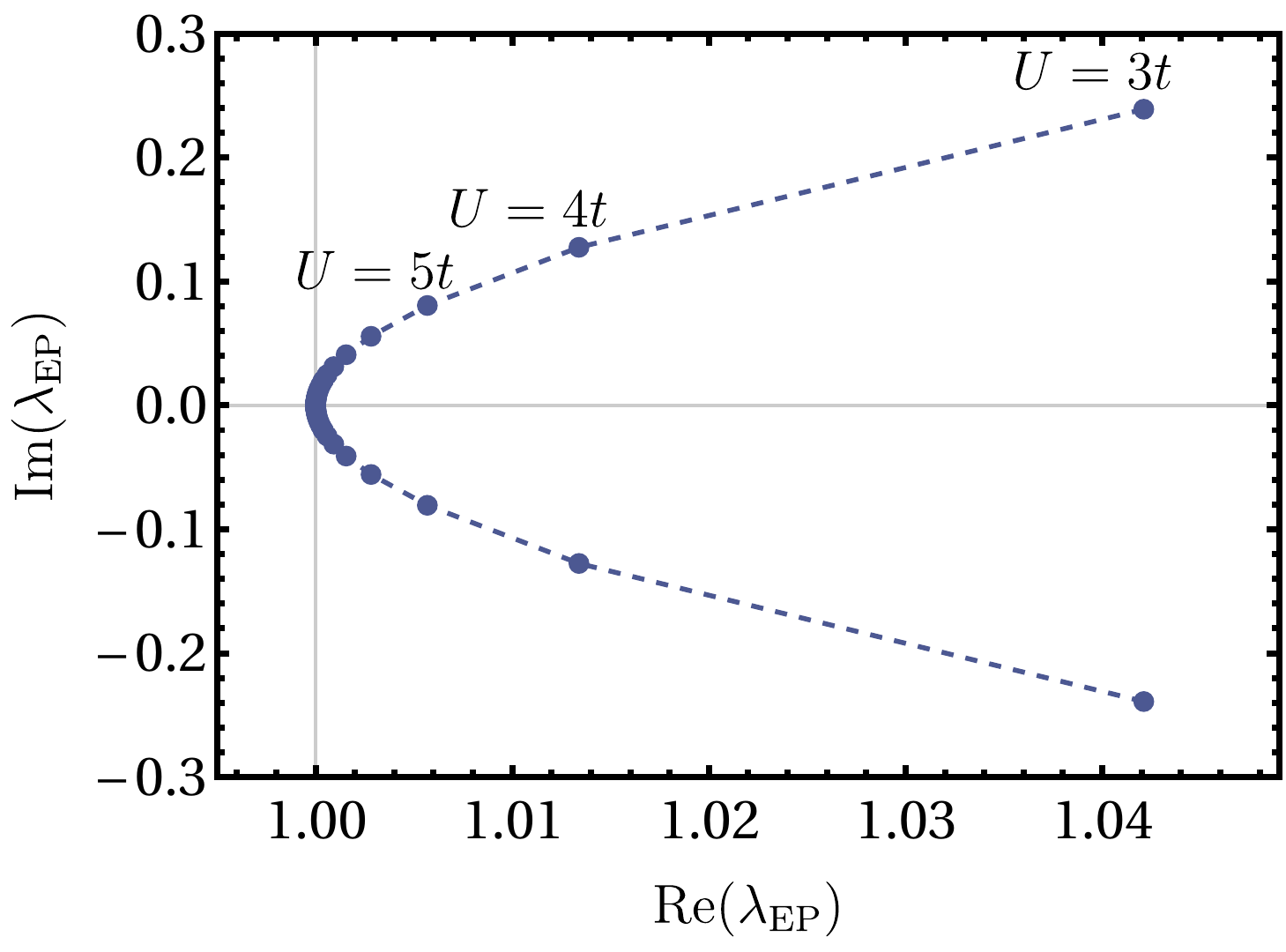}	
		\subcaption{\label{subfig:ump_ep_to_cp}}
    \end{subfigure}
\caption{%
    The UMP ground-state EP in the symmetric Hubbard dimer becomes a critical point in the strong correlation limit (\ie, large $U/t$).
    (\subref{subfig:ump_cp}) As $U/t$ increases, the avoided crossing on the real $\lambda$ axis
    becomes increasingly sharp.
    (\subref{subfig:ump_cp_surf}) \titou{The avoided crossing at $U=5t$ corresponds to EPs with non-zero imaginary components.}
    (\subref{subfig:ump_ep_to_cp}) Convergence of the EPs at $\lep$ onto the real axis for $U/t \to \infty$.
\label{fig:UMP_cp}}

\end{figure*}

Returning to the symmetric Hubbard dimer, we showed in Sec.~\ref{sec:spin_cont} that the slow
convergence of the strongly correlated UMP series
was due to a complex-conjugate pair of EPs just outside the radius of convergence.
These EPs have positive real components and small imaginary components (see Fig.~\ref{fig:UMP}), suggesting a potential 
connection to MP critical points and QPTs (see Sec.~\ref{sec:MP_critical_point}).
For $\lambda>1$, the HF potential becomes an attractive component in Stillinger's 
Hamiltonian displayed in Eq.~\eqref{eq:HamiltonianStillinger}, while the explicit electron-electron interaction
becomes increasingly repulsive.
Closed-shell critical points along the positive real $\lambda$ axis may then represent 
points where the two-electron repulsion overcomes the attractive HF potential 
and a single electron dissociates from the molecule (see Ref.~\onlinecite{Sergeev_2006}).

\begin{figure}[b]
	\includegraphics[width=\linewidth]{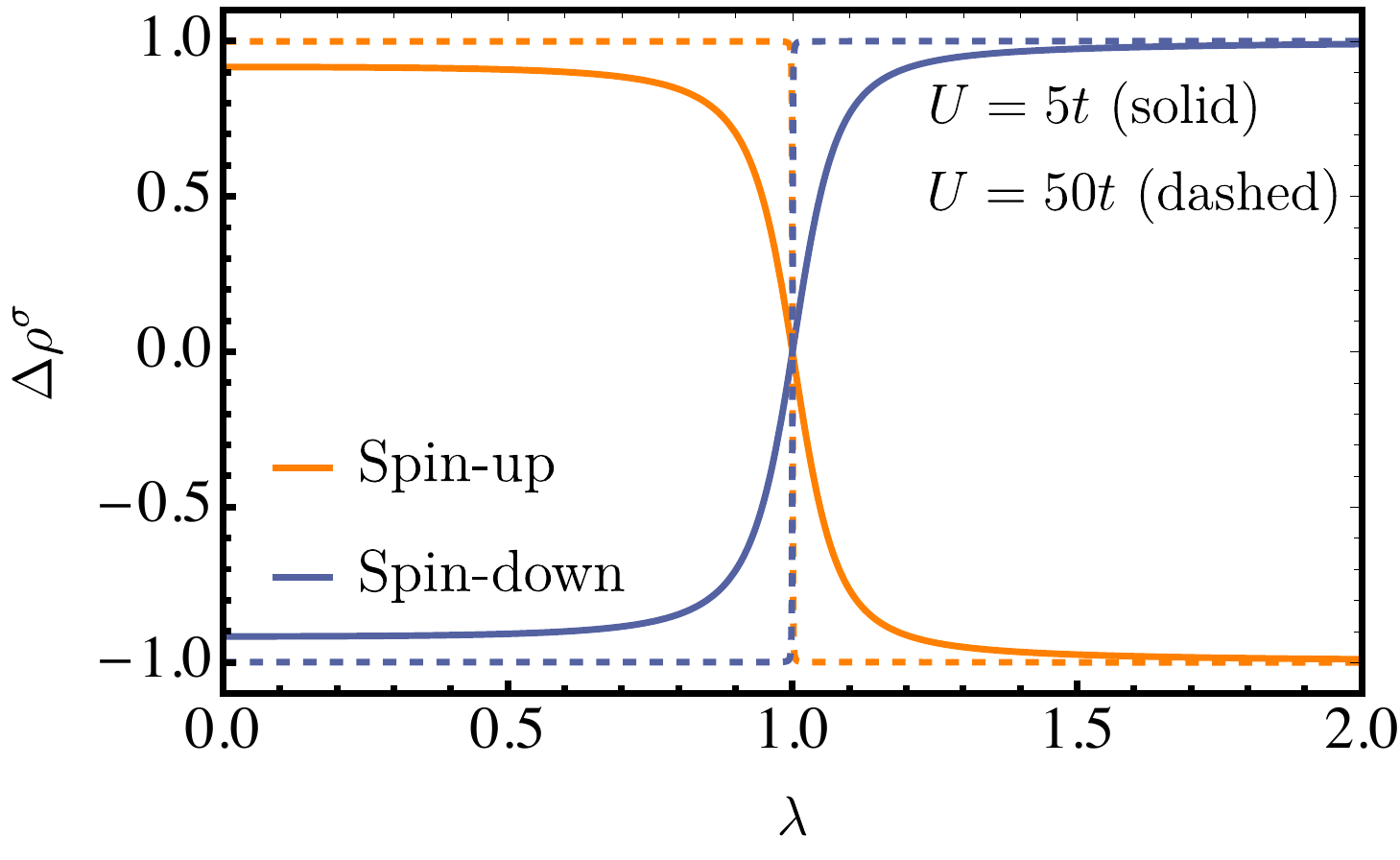}
	\caption{
\titou{Difference in the electron densities on the left and right sites for the UMP ground state in the symmetric Hubbard dimer
[see Eq.~\eqref{eq:ump_dens}]. 
    At $\lambda = 1$, the spin-up electron transfers from the right site to the left site, while the spin-down
electron transfers in the opposite direction. 
In the strong correlation limit (large $U/t$), this process becomes increasingly sharp and represents a
zero-temperature QPT.}
	\label{fig:ump_dens}}
\end{figure}

In contrast, spin symmetry-breaking in the UMP reference creates different HF potentials for the spin-up and spin-down electrons.
Consider one of the two reference UHF solutions where the spin-up and spin-down electrons are localised on the left and right sites respectively. 
The spin-up HF potential will then be a repulsive interaction from the spin-down electron 
density that is centred around the right site (and vice-versa).
As $\lambda$ becomes greater than 1 and the HF potentials become attractive, there will be a sudden
driving force for the electrons to swap sites.
This swapping process can also be represented as a double excitation, and thus an avoided crossing will occur
for $\lambda \geq 1$ (Fig.~\ref{subfig:ump_cp}).
While this appears to be an avoided crossing between the ground and first-excited state, 
the presence of an earlier excited-state avoided crossing means that the first-excited state qualitatively 
represents the reference double excitation for $\lambda > 1/2$.
\titou{We can visualise this swapping process by considering the difference in the
electron density on the left and right sites, defined for each spin as
\begin{equation}
\Delta \rho^{\sigma} = \rho_\mathcal{R}^{\sigma} - \rho_\mathcal{L}^{\sigma},
\label{eq:ump_dens}
\end{equation}
where $\rho_{\mathcal{L}}^{\sigma}$ ($\rho_{\mathcal{R}}^{\sigma}$) is the spin-$\sigma$ electron density
on the left (right) site. 
This density difference is shown for the UMP ground-state at $U = 5 t$ in Fig.~\ref{fig:ump_dens} (solid lines).
Here, the transfer of the spin-up electron from the right site to the left site can be seen as $\lambda$ passes through 1
(and similarly for the spin-down electron).}

The ``sharpness'' of the avoided crossing is controlled by the correlation strength $U/t$.
For small $U/t$, the HF potentials will be weak and the electrons will delocalise over the two sites,
both in the UHF reference and the exact wave function.
This delocalisation dampens the electron swapping process and leads to a ``shallow'' avoided crossing (solid lines in Fig.~\ref{subfig:ump_cp})
that corresponds to EPs with non-zero imaginary components (Fig.~\ref{subfig:ump_cp_surf}).
As $U/t$ becomes larger, the HF potentials become stronger and the on-site repulsion dominates the hopping
term to make electron delocalisation less favourable.
In other words, the electrons localise on individual sites to form a Wigner crystal.
These effects create a stronger driving force for the electrons to swap sites until, eventually, this swapping
occurs suddenly at $\lambda = 1$, \titou{as shown for $U= 50 t$ in Fig.~\ref{fig:ump_dens} (dashed lines).}
In this limit, the ground-state EPs approach the real axis (Fig.~\ref{subfig:ump_ep_to_cp}) and the avoided 
crossing creates a gradient discontinuity in the ground-state energy (dashed lines in Fig.~\ref{subfig:ump_cp}).
We therefore find that, in the strong correlation limit, the symmetry-broken ground-state EP becomes
a new type of MP critical point and represents a QPT as the perturbation parameter $\lambda$ is varied.
Remarkably, this argument explains why the dominant UMP singularity lies so close, but always outside, the 
radius of convergence (see Fig.~\ref{fig:RadConv}).

\section{Resummation Methods}
\label{sec:Resummation}

\begin{figure*}
    \includegraphics[height=0.23\textheight]{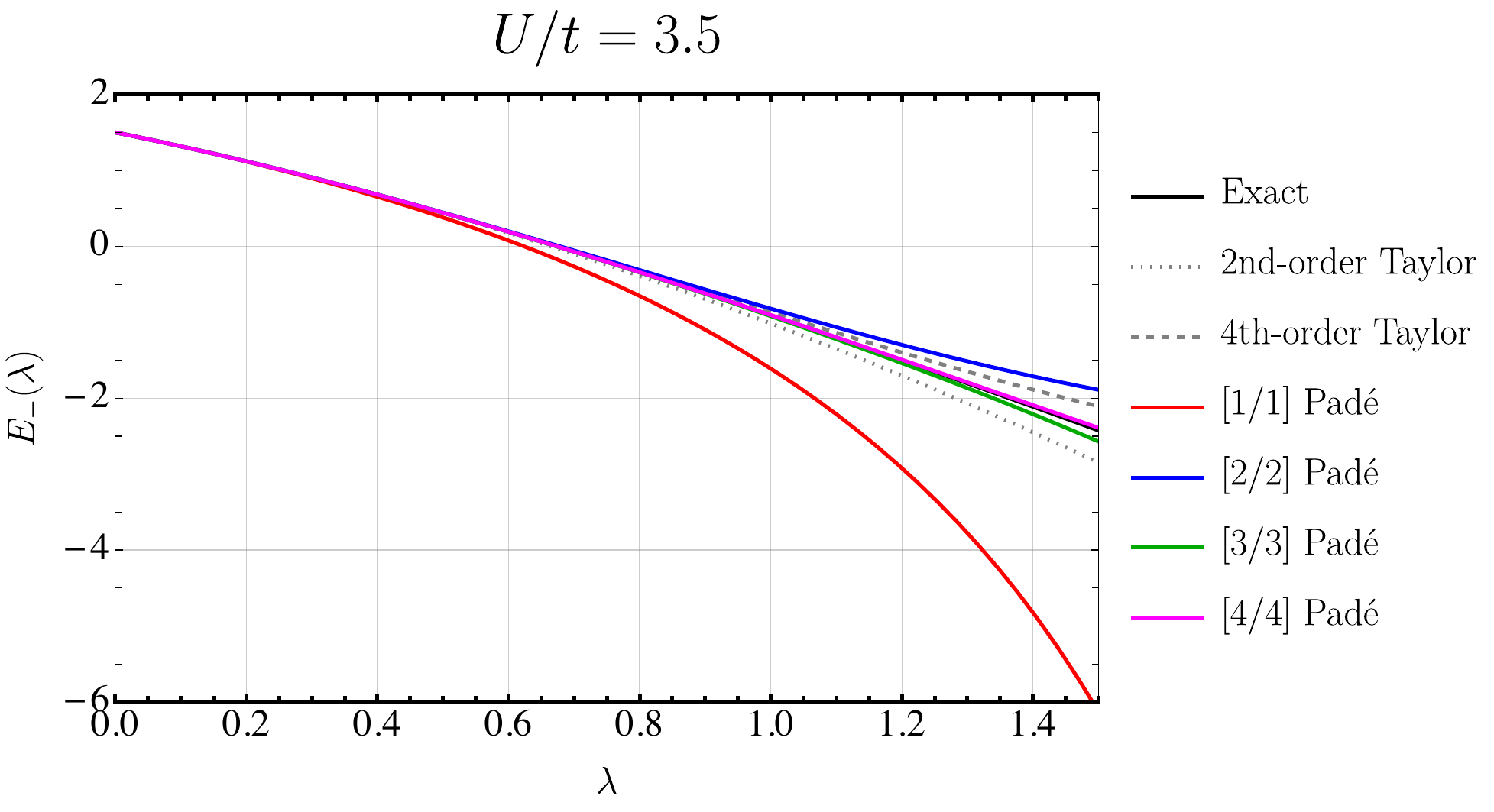}
    \includegraphics[height=0.23\textheight]{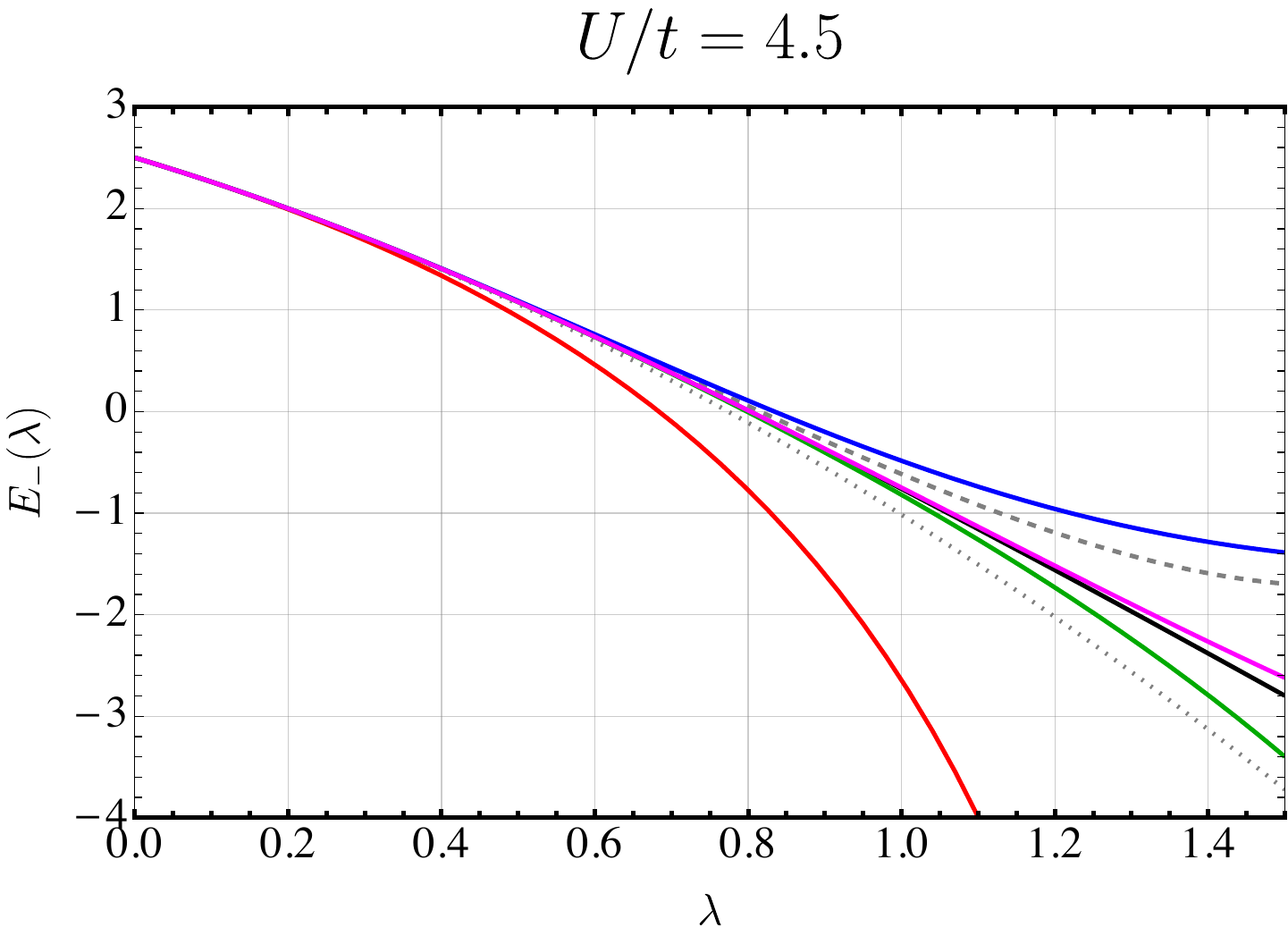}
    \caption{\label{fig:PadeRMP}
    RMP ground-state energy as a function of $\lambda$ in the Hubbard dimer obtained using various truncated Taylor series and approximants
    at $U/t = 3.5$ (left) and $U/t = 4.5$ (right).}
\end{figure*}

It is frequently stated that
\textit{``the most stupid thing to do with a series is to sum it.''}
Nonetheless, quantum chemists are basically doing this on a daily basis.
As we have seen throughout this review, the MP series can often show erratic, 
slow, or divergent behaviour.
In these cases, estimating the correlation energy by simply summing successive
low-order terms is almost guaranteed to fail.
Here, we discuss alternative tools that can be used to sum slowly convergent or divergent series.
These so-called ``resummation'' techniques form a vast field of research and thus we will
provide details for only the most relevant methods.
We refer the interested reader to more specialised reviews for additional information.%
\cite{Goodson_2011,Goodson_2019}

\subsection{Pad\'e Approximant}

The failure of a Taylor series for correctly modelling the MP energy function $E(\lambda)$ 
arises because one is trying to model a complicated function containing multiple branches, branch points, and
singularities using a simple polynomial of finite order.
A truncated Taylor series can only predict a single sheet and does not have enough 
flexibility to adequately describe functions such as the MP energy.
Alternatively, the description of complex energy functions can be significantly improved
by introducing Pad\'e approximants, \cite{Pade_1892} and related techniques. \cite{BakerBook,BenderBook}

A Pad\'e approximant can be considered as the best approximation of a function by a 
rational function of given order.
More specifically, a $[d_A/d_B]$ Pad\'e approximant is defined as 
\begin{equation}
	\label{eq:PadeApp}
	E_{[d_A/d_B]}(\lambda) = \frac{A(\lambda)}{B(\lambda)} 
    = \frac{\sum_{k=0}^{d_A} a_k\, \lambda^k}{1 + \sum_{k=1}^{d_B} b_k\, \lambda^k},
\end{equation}
where the coefficients of the polynomials $A(\lambda)$ and $B(\lambda)$ are determined by collecting 
\titou{and comparing terms for each power of $\lambda$ with the low-order terms in the Taylor series expansion}.
Pad\'e approximants are extremely useful in many areas of physics and 
chemistry\cite{Loos_2013,Pavlyukh_2017,Tarantino_2019,Gluzman_2020} as they can model poles, 
which appear at the roots of $B(\lambda)$. 
However, they are unable to model functions with square-root branch points 
(which are ubiquitous in the singularity structure of perturbative methods) 
and more complicated functional forms appearing at critical points 
(where the nature of the solution undergoes a sudden transition).
Despite this limitation, the successive diagonal Pad\'e approximants (\ie, $d_A = d_B $) 
often define a convergent perturbation series in cases where the Taylor series expansion diverges.

\begin{table}[b]
	\caption{RMP ground-state energy estimate at $\lambda = 1$ of the Hubbard dimer provided by various truncated Taylor 
    series and Pad\'e approximants at $U/t = 3.5$ and $4.5$.
    We also report the distance of the closest pole to the origin $\abs{\lc}$ provided by the diagonal Pad\'e approximants.
	\label{tab:PadeRMP}}
	\begin{ruledtabular}
		\begin{tabular}{lccccc}
            &			&	\mc{2}{c}{$\abs{\lc}$}	&	\mc{2}{c}{$E_{-}(\lambda = 1)$} \\
																		\cline{3-4} \cline{5-6}
			Method		&	Degree	&	$U/t = 3.5$	&	$U/t = 4.5$	&	$U/t = 3.5$	&	$U/t = 4.5$	\\
			\hline
			Taylor		&	2		&			&			&	$-1.01563$	&	$-1.01563$	\\
						&	3		&			&			&	$-1.01563$	&	$-1.01563$	\\
						&	4		&			&			&	$-0.86908$	&	$-0.61517$	\\
						&	5		&			&			&	$-0.86908$	&	$-0.61517$	\\
						&	6		&			&			&	$-0.92518$	&	$-0.86858$	\\
            \hline
			Pad\'e		&	[1/1]	&	$2.29$	&	$1.78$	&	$-1.61111$	&	$-2.64286$	\\
						&	[2/2]	&	$2.29$	&	$1.78$	&	$-0.82124$	&	$-0.48446$	\\
						&	[3/3]	&	$1.73$	&	$1.34$	&	$-0.91995$	&	$-0.81929$	\\
						&	[4/4]	&	$1.47$	&	$1.14$	&	$-0.90579$	&	$-0.74866$	\\
						&	[5/5]	&	$1.35$	&	$1.05$	&	$-0.90778$	&	$-0.76277$	\\
			\hline
			Exact		&			&	$1.14$	&	$0.89$	&	$-0.90754$	&	$-0.76040$	\\
		\end{tabular}
	\end{ruledtabular}
\end{table}

Figure~\ref{fig:PadeRMP} illustrates the improvement provided by diagonal Pad\'e 
approximants compared to the usual Taylor expansion in cases where the RMP series of 
the Hubbard dimer converges ($U/t = 3.5$) and diverges ($U/t = 4.5$).
More quantitatively, Table \ref{tab:PadeRMP} gathers estimates of the RMP ground-state 
energy at $\lambda = 1$ provided by various truncated Taylor series and Pad\'e 
approximants for these two values of the ratio $U/t$.
While the truncated Taylor series converges laboriously to the exact energy as the truncation 
degree increases at $U/t = 3.5$, the Pad\'e approximants yield much more accurate results.
Furthermore, the distance of the closest pole to the origin $\abs{\lc}$ in the Pad\'e approximants 
indicate that they provide a relatively good approximation to the position of the 
true branch point singularity in the RMP energy.
For $U/t = 4.5$, the Taylor series expansion performs worse and eventually diverges,
while the Pad\'e approximants still offer relatively accurate energies and recovers
a convergent series.

\begin{figure}[t]
    \includegraphics[width=\linewidth]{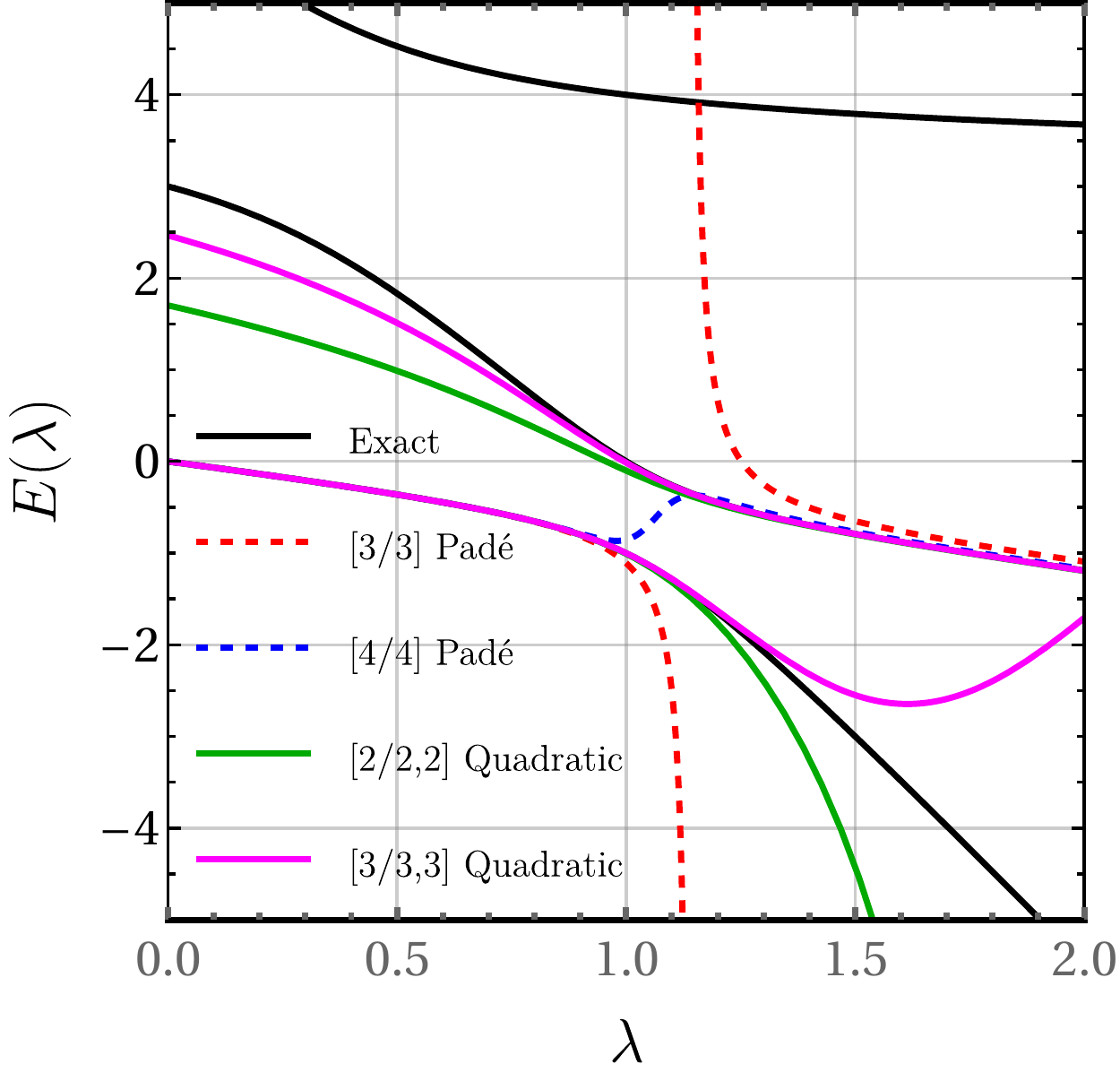}
    \caption{\label{fig:QuadUMP}
    UMP energies in the Hubbard dimer as a function of $\lambda$ obtained using various approximants at $U/t = 3$.}
\end{figure}

We can expect the UMP energy function to be much more challenging 
to model properly as it contains three connected branches 
(see Figs.~\ref{subfig:UMP_3} and \ref{subfig:UMP_7}).
Figure~\ref{fig:QuadUMP} and Table~\ref{tab:QuadUMP} indicate that this is indeed the case. 
In particular, Fig.~\ref{fig:QuadUMP} illustrates that the Pad\'e approximants are trying to model
the square root branch point that lies close to $\lambda = 1$ by placing a pole on the real axis
(\eg, [3/3]) or with a very small imaginary component (\eg, [4/4]).
The proximity of these poles to the physical point $\lambda = 1$ means that any error in the Pad\'e 
functional form becomes magnified in the estimate of the exact energy, as seen for the low-order
approximants in Table~\ref{tab:QuadUMP}.
However, with sufficiently high degree polynomials, one obtains
accurate estimates for the position of the closest singularity and the ground-state energy at $\lambda = 1$,
even in cases where the convergence of the UMP series is incredibly slow 
(see Fig.~\ref{subfig:UMP_cvg}).

\subsection{Quadratic Approximant}
Quadratic approximants are designed to model the singularity structure of the energy 
function $E(\lambda)$ via a generalised version of the square-root singularity 
expression \cite{Mayer_1985,Goodson_2011,Goodson_2019}
\begin{equation}
	\label{eq:QuadApp}
	E_{[d_P/d_Q,d_R]}(\lambda) = \frac{1}{2 Q(\lambda)} \qty[ P(\lambda) \pm \sqrt{P^2(\lambda) - 4 Q(\lambda) R(\lambda)} ],
\end{equation}
with the polynomials 
\begin{align}
	\label{eq:PQR}
	P(\lambda) & = \sum_{k=0}^{d_P} p_k \lambda^k,
	&
	Q(\lambda) & = \sum_{k=0}^{d_Q} q_k \lambda^k, 
	&
	R(\lambda) & = \sum_{k=0}^{d_R} r_k \lambda^k,
\end{align}
defined such that $d_P + d_Q + d_R = n - 1$, and $n$ is the truncation order of the Taylor series of $E(\lambda)$.
Recasting Eq.~\eqref{eq:QuadApp} as a second-order expression in $E(\lambda)$, \ie,
\begin{equation}
	Q(\lambda) E^2(\lambda) - P(\lambda) E(\lambda) + R(\lambda) \sim \order*{\lambda^{n+1}},
\end{equation}
and substituting $E(\lambda$) by its $n$th-order expansion and the polynomials by 
their respective expressions \eqref{eq:PQR} yields $n+1$ linear equations for the coefficients 
$p_k$, $q_k$, and $r_k$ (where we are free to assume that $q_0 = 1$).
A quadratic approximant, characterised by the label $[d_P/d_Q,d_R]$, generates, by construction, 
$n_\text{bp} = \max(2d_p,d_q+d_r)$ branch points at the roots of the polynomial 
$P^2(\lambda) - 4 Q(\lambda) R(\lambda)$ and $d_q$ poles at the roots of $Q(\lambda)$.

Generally, the diagonal sequence of quadratic approximant, 
\ie, $[0/0,0]$, $[1/0,0]$, $[1/0,1]$, $[1/1,1]$, $[2/1,1]$, 
is of particular interest as the order of the corresponding Taylor series increases on each step.
However, while a quadratic approximant can reproduce multiple branch points, it can only describe 
a total of two branches.
This constraint can hamper the faithful description of more complicated singularity structures such as the MP energy surface.
Despite this limitation, Ref.~\onlinecite{Goodson_2000a} demonstrates that quadratic approximants 
provide convergent results in the most divergent cases considered by Olsen and 
collaborators\cite{Christiansen_1996,Olsen_1996} 
and Leininger \etal \cite{Leininger_2000}

As a note of caution, Ref.~\onlinecite{Goodson_2019} suggests that low-order 
quadratic approximants can struggle to correctly model the singularity structure when 
the energy function has poles in both the positive and negative half-planes. 
In such a scenario, the quadratic approximant will tend to place its branch points in-between, potentially introducing singularities quite close to the origin.
The remedy for this problem involves applying a suitable transformation of the complex plane (such as a bilinear conformal mapping) which leaves the points at $\lambda = 0$ and $\lambda = 1$ unchanged. \cite{Feenberg_1956}

\begin{table}[b]
    \caption{Estimate for the distance of the closest singularity (pole or branch point) to the origin $\abs{\lc}$
    in the UMP energy function of the Hubbard dimer provided by various truncated Taylor series and approximants at $U/t = 3$ and $7$.
	The truncation degree of the Taylor expansion $n$ of $E(\lambda)$ and the number of branch 
    points $n_\text{bp} = \max(2d_p,d_q+d_r)$ generated by the quadratic approximants are also reported.
	\label{tab:QuadUMP}}
	\begin{ruledtabular}
		\begin{tabular}{lccccccc}
            &			&			&					&	\mc{2}{c}{$\abs{\lc}$}	&	\mc{2}{c}{$E_{-}(\lambda = 1)$}			\\
																		\cline{5-6}\cline{7-8}
			\mc{2}{c}{Method}		&	$n$		&	$n_\text{bp}$	&	$U/t = 3$	&	$U/t = 7$	&	$U/t = 3$	&	$U/t = 7$	\\
			\hline
			Taylor		&	     	&	2		&					&	    		&	    		&	$-0.74074$	&	$-0.29155$	\\
                        &	     	&	3		&					&	    		&	    		&	$-0.78189$	&	$-0.29690$	\\
			         	&	     	&	4		&					&	    		&	    		&	$-0.82213$	&	$-0.30225$	\\
						&	     	&	5		&					&	    		&	    		&	$-0.85769$	&	$-0.30758$	\\
                        &	     	&   6		&					&	    		&	    		&	$-0.88882$	&	$-0.31289$	\\
			\hline
			Pad\'e		&	[1/1]	&	2		&					&	$9.000$		&	$49.00$		&	$-0.75000$	&	$-0.29167$	\\
                        &	[2/2]	&	4		&					&	$0.974$		&	$1.003$		&	$\hphantom{-}0.75000$	&	$-17.9375$	\\
			         	&	[3/3]	&	6		&					&	$1.141$		&	$1.004$		&	$-1.10896$	&	$-1.49856$	\\
						&	[4/4]	&	8		&					&	$1.068$		&	$1.003$		&	$-0.85396$	&	$-0.33596$	\\
                        &	[5/5]	&	10		&					&	$1.122$		&	$1.004$		&	$-0.97254$	&	$-0.35513$	\\
            \hline
			Quadratic	&	[2/1,2]	&	6		&	4				&	$1.086$		&	$1.003$		&	$-1.01009$	&	$-0.53472$	\\
						&	[2/2,2]	&	7		&	4				&	$1.082$		&	$1.003$		&	$-1.00553$	&	$-0.53463$	\\
						&	[3/2,2]	&	8		&	6				&	$1.082$		&	$1.001$		&	$-1.00568$	&	$-0.52473$	\\
						&	[3/2,3]	&	9		&	6				&	$1.071$		&	$1.002$		&	$-0.99973$	&	$-0.53102$	\\
                        &	[3/3,3]	&	10		&	6				&	$1.071$		&	$1.002$		&	$-0.99966$	&	$-0.53103$	\\[0.5ex]
            (pole-free)	&	[3/0,2]	&	6		&	6				&	$1.059$	    &	$1.003$ 	&	$-1.13712$	&	$-0.57199$	\\
		                &	[3/0,3]	&	7		&	6				&	$1.073$		&	$1.002$		&	$-1.00335$	&	$-0.53113$	\\
						&	[3/0,4]	&	8		&	6				&	$1.071$		&	$1.002$		&	$-1.00074$	&	$-0.53116$	\\
						&	[3/0,5]	&	9		&	6				&	$1.070$		&	$1.002$		&	$-1.00042$	&	$-0.53114$	\\
						&	[3/0,6]	&	10		&	6				&	$1.070$		&	$1.002$		&	$-1.00039$	&	$-0.53113$	\\
			\hline
			Exact		&			&			&					&	$1.069$		&	$1.002$		&	$-1.00000$	&	$-0.53113$	\\
		\end{tabular}
	\end{ruledtabular}
\end{table}

\begin{figure*}
    \begin{subfigure}{0.32\textwidth}
	\includegraphics[height=0.85\textwidth]{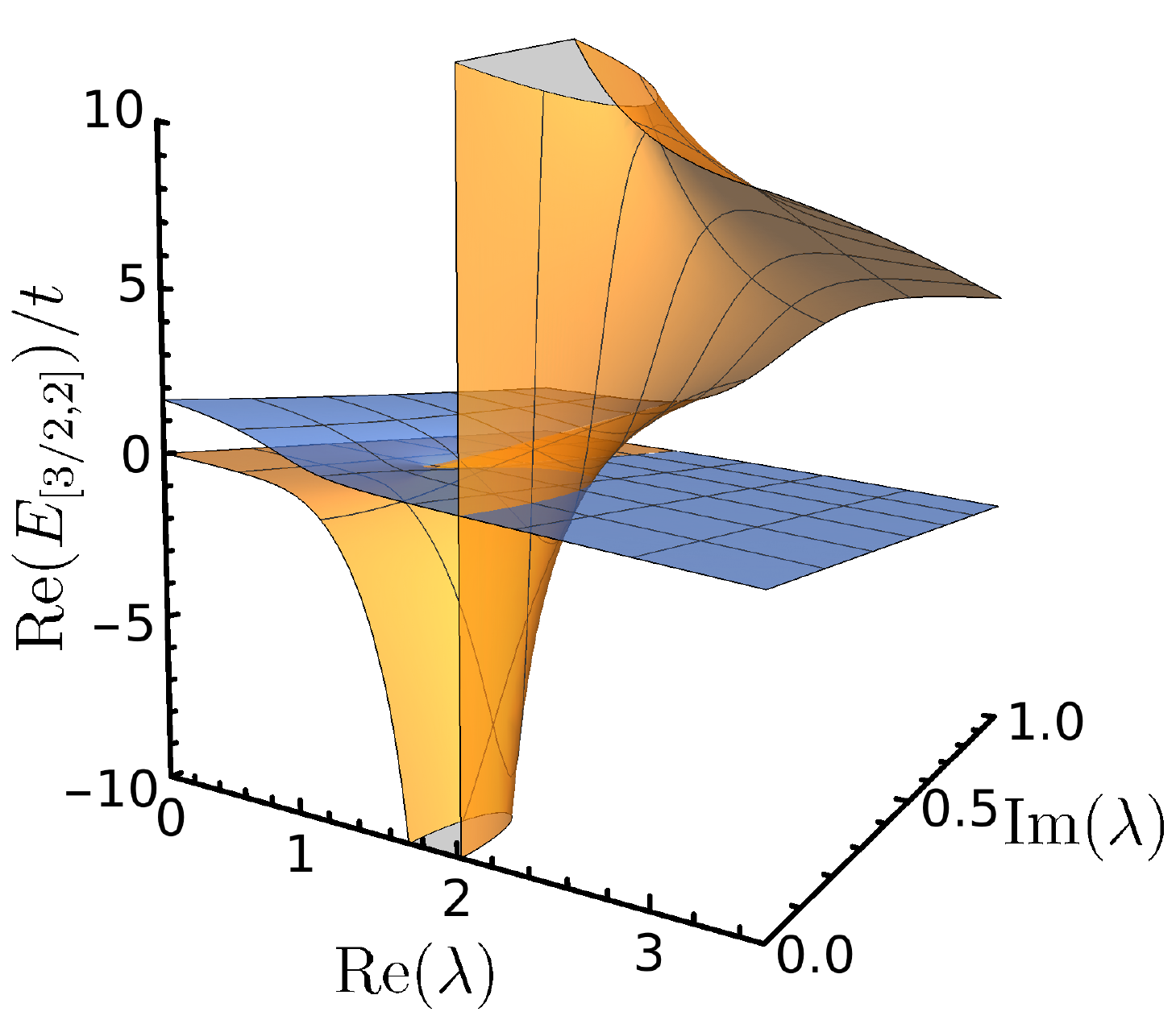}	
    \subcaption{\label{subfig:322quad} [3/2,2] Quadratic}
    \end{subfigure}
    \begin{subfigure}{0.32\textwidth}
	\includegraphics[height=0.85\textwidth]{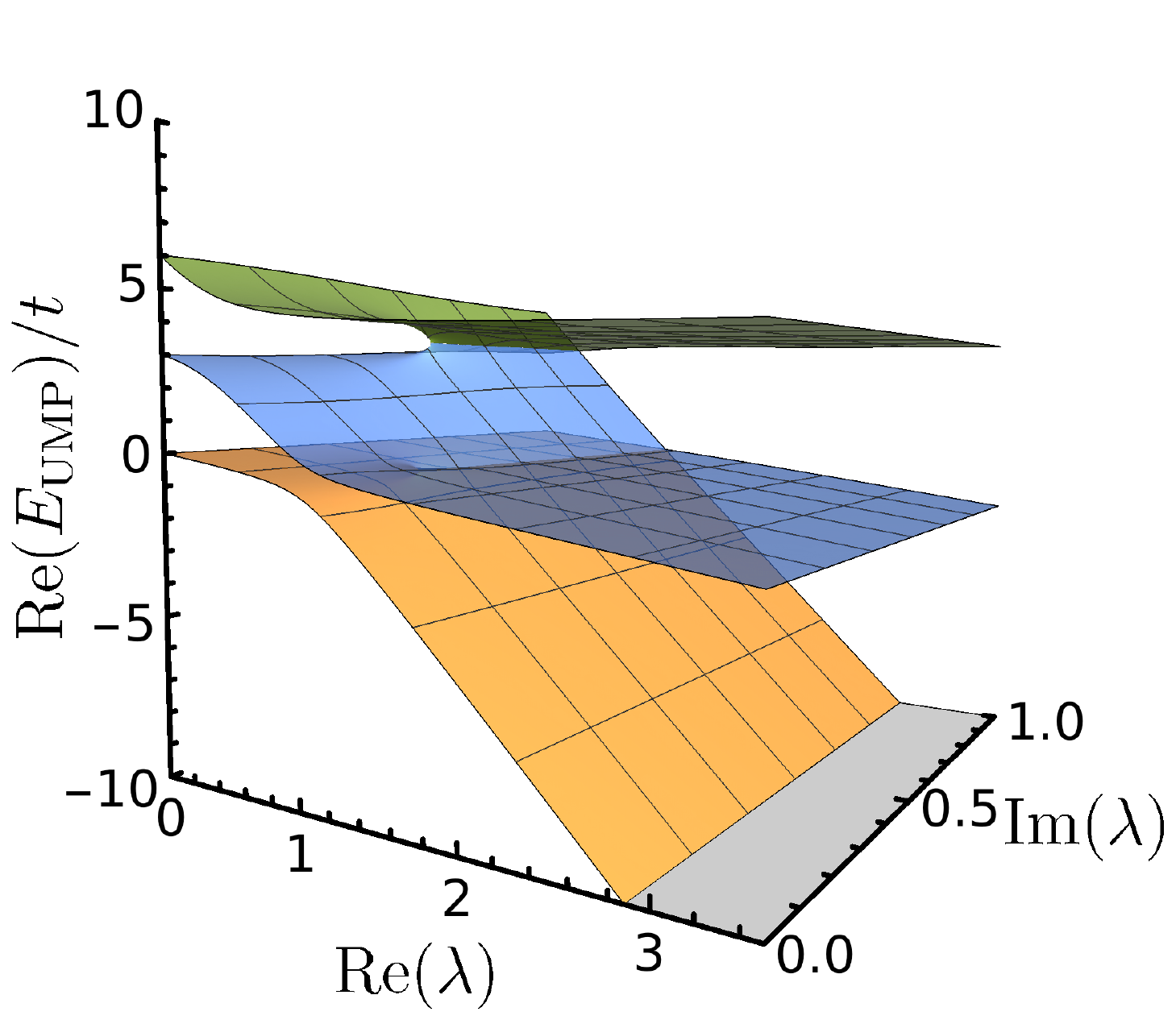}
		\subcaption{\label{subfig:exact} Exact}
    \end{subfigure}
	\begin{subfigure}{0.32\textwidth}
        \includegraphics[height=0.85\textwidth]{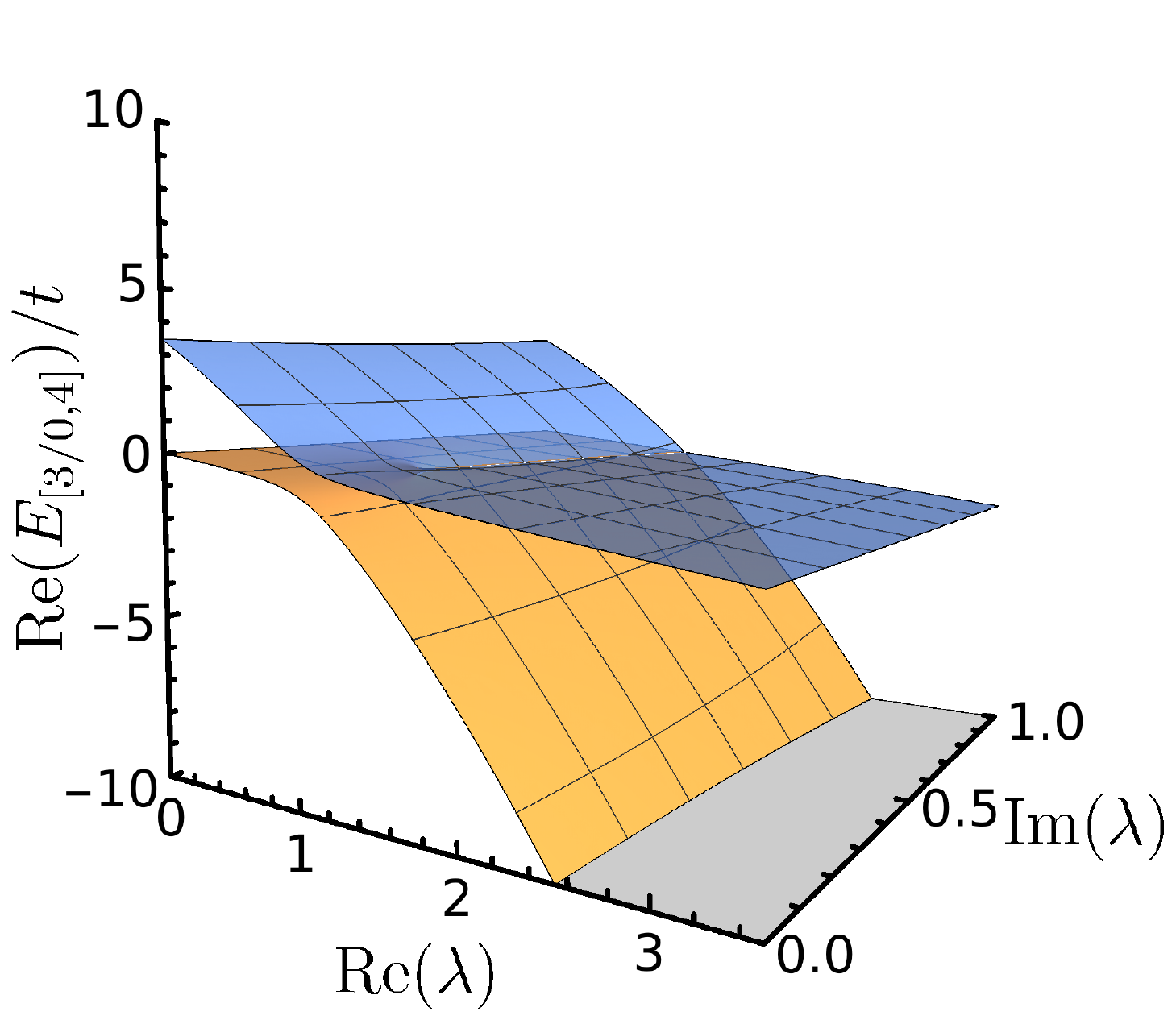}	
    \subcaption{\label{subfig:304quad} [3/0,4] Quadratic}
    \end{subfigure}
\caption{%
Comparison of the [3/2,2] and [3/0,4] quadratic approximants with the exact UMP energy surface in the complex $\lambda$ 
plane in the Hubbard dimer with $U/t = 3$. 
Both quadratic approximants correspond to the same truncation degree of the Taylor series and model the branch points 
using a radicand polynomial of the same order.
However, the [3/2,2] approximant introduces poles into the surface that limits it accuracy, while the [3/0,4] approximant
is free of poles.}
\label{fig:nopole_quad}
\end{figure*}

For the RMP series of the Hubbard dimer, the $[0/0,0]$ and $[1/0,0]$ quadratic approximants 
are quite poor approximations, but the $[1/0,1]$ version perfectly models the RMP energy
function by predicting a single pair of EPs at $\lambda_\text{EP} = \pm \i 4t/U$.
This is expected from the form of the RMP energy [see Eq.~\eqref{eq:E0MP}], which matches 
the ideal target for quadratic approximants.
Furthermore, the greater flexibility of the diagonal quadratic approximants provides a significantly
improved model of the UMP energy in comparison to the Pad\'e approximants or Taylor series.
In particular, these quadratic approximants provide an effective model for the avoided crossings 
(Fig.~\ref{fig:QuadUMP}) and an improved estimate for the distance of the 
closest branch point  to the origin.
Table~\ref{tab:QuadUMP} shows that they provide remarkably accurate 
estimates of the ground-state energy at $\lambda = 1$.

While the diagonal quadratic approximants provide significantly improved estimates of the 
ground-state energy, we can use our knowledge of the UMP singularity structure to develop
even more accurate results.
We have seen in previous sections that the UMP energy surface
contains only square-root branch cuts that approach the real axis in the limit $U/t \to \infty$. 
Since there are no true poles on this surface, we can obtain more accurate quadratic approximants by
taking $d_q = 0$ and increasing $d_r$ to retain equivalent accuracy in the square-root term [see Eq.\eqref{eq:QuadApp}].
Figure~\ref{fig:nopole_quad} illustrates this improvement for the pole-free [3/0,4] quadratic
approximant compared to the [3/2,2] approximant with the same truncation degree in the Taylor
expansion.
Clearly, modelling the square-root branch point using $d_q = 2$ has the negative effect of
introducing spurious poles in the energy, while focussing purely on the branch point with $d_q = 0$
leads to a significantly improved model.
Table~\ref{tab:QuadUMP} shows that these pole-free quadratic approximants
provide a rapidly convergent series with essentially exact energies at low order.

Finally, to emphasise the improvement that can be gained by using either Pad\'e, diagonal quadratic,
or pole-free quadratic approximants, we collect the energy and error obtained using only the first 10 terms of the UMP
Taylor series in Table~\ref{tab:UMP_order10}.
The accuracy of these approximants reinforces how our understanding of the MP
energy surface in the complex plane can be leveraged to significantly improve estimates of the exact
energy using low-order perturbation expansions.

\begin{table}[h]
	\caption{
    Estimate and associated error of the exact UMP energy of the Hubbard dimer at $U/t = 7$ for 
    various approximants using up to ten terms in the Taylor expansion.
	\label{tab:UMP_order10}}
	\begin{ruledtabular}
		\begin{tabular}{lccc}
            \mc{2}{c}{Method}	 &	$E_{-}(\lambda = 1)$ & \% Abs.\ Error \\ 
			\hline
            Taylor		         &	 10	         &	$-0.33338$      &  $37.150$ \\
            Pad\'e               &   [5/5]       &  $-0.35513$      &  $33.140$ \\
            Quadratic (diagonal) &   [3/3,3]     &  $-0.53103$      &  $\hphantom{0}0.019$ \\
            Quadratic (pole-free)&   [3/0,6]     &  $-0.53113$      &  $\hphantom{0}0.005$ \\
			\hline
            Exact		         &	        	 &	$-0.53113$	     &            \\
		\end{tabular}
	\end{ruledtabular}
\end{table}

\subsection{Shanks Transformation}
\label{sec:Shanks}

While the Pad\'e and quadratic approximants can yield a convergent series representation
in cases where the standard MP series diverges, there is no guarantee that the rate of convergence
will be fast enough for low-order approximations to be useful.
However, these low-order partial sums or approximants often contain a remarkable amount of information
that can be used to extract further information about the exact result.
The Shanks transformation presents one approach for extracting this information
and accelerating the rate of convergence of a sequence.\cite{Shanks_1955,BenderBook}

Consider the partial sums
$S_n = \sum_{k=0}^{n} s_k$
defined from the truncated summation of an infinite series 
$S = \sum_{k=0}^{\infty} s_k$.
If the series converges, then the partial sums will tend to the exact result 
\begin{equation}
	\lim_{n \to \infty} S_n = S. 
\end{equation}
The Shanks transformation attempts to generate increasingly accurate estimates of this
limit by defining a new series as
\begin{equation}
    T(S_n) = \frac{S_{n+1} S_{n-1} - S_{n}^2}{S_{n+1} - 2 S_{n} + S_{n-1}}.
\end{equation}
This series can converge faster than the original partial sums and can thus provide greater
accuracy using only the first few terms in the series.
However, it is only designed to accelerate converging partial sums with 
the approximate form $S_n \approx S + \alpha\,\beta^n$.
Furthermore, while this transformation can accelerate the convergence of a series, 
there is no guarantee that this acceleration will be fast enough to significantly
improve the accuracy of low-order approximations.

To the best of our knowledge, the Shanks transformation has never previously been applied
to accelerate the convergence of the MP series.
We have therefore applied it to the convergent Taylor series, Pad\'e approximants, and quadratic
approximants for RMP and UMP in the symmetric Hubbard dimer. 
The UMP approximants converge too slowly for the Shanks transformation
to provide any improvement, even in the case where the quadratic approximants are already 
very accurate.
In contrast, acceleration of the diagonal Pad\'e approximants for the RMP cases 
can significantly improve the estimate of the energy using low-order perturbation terms, 
as shown in Table~\ref{tab:RMP_shank}.
Even though the RMP series diverges at $U/t = 4.5$, the combination
of diagonal Pad\'e approximants with the Shanks transformation reduces the absolute error in
the best energy estimate to 0.002\,\% using only the first 10 terms in the Taylor series.
This remarkable result indicates just how much information is contained in the first few
terms of a perturbation series, even if it diverges.

\begin{table}[th]
	\caption{
    Acceleration of the diagonal Pad\'e approximant sequence for the RMP energy
    of the Hubbard dimer at $U/t = 3.5$ and $4.5$ using the Shanks transformation.
	\label{tab:RMP_shank}}
	\begin{ruledtabular}
		\begin{tabular}{lcccc}
            &			&	&	\mc{2}{c}{$E_{-}(\lambda = 1)$} \\
												\cline{4-5} 
            Method  &   Degree	&  Series Term &   $U/t = 3.5$ & $U/t = 4.5$ \\
			\hline
            Pad\'e	&	[1/1]	& $S_1$        &  $-1.61111$   &  $-2.64286$ \\
                    &   [2/2]   & $S_2$        &  $-0.82124$   &  $-0.48446$ \\
                    &   [3/3]   & $S_3$        &  $-0.91995$   &  $-0.81929$ \\
                    &   [4/4]   & $S_4$        &  $-0.90579$   &  $-0.74866$ \\
                    &   [5/5]   & $S_5$        &  $-0.90778$   &  $-0.76277$ \\
            \hline
            Shanks  &	        &  $T(S_2)$    &  $-0.90898$   &  $-0.77432$ \\
                    &           &  $T(S_3)$    &  $-0.90757$   &  $-0.76096$ \\
                    &           &  $T(S_4)$    &  $-0.90753$   &  $-0.76042$ \\
			\hline
            Exact	&	        & 	           &  $-0.90754$	&  $-0.76040$ \\
		\end{tabular}
	\end{ruledtabular}
\end{table}

\subsection{Analytic Continuation}

Recently, Mih\'alka \etal\ have studied the effect of different partitionings, such as MP or EN theory, on the position of 
branch points and the convergence properties of Rayleigh--Schr\"odinger perturbation theory\cite{Mihalka_2017b} (see also
Refs.~\onlinecite{Szabados_1999,Surjan_2000,Szabados_2003}).
Taking the equilibrium and stretched water structures as an example, they estimated the radius of convergence using quadratic
Pad\'e approximants.
The EN partitioning provided worse convergence properties than the MP partitioning, which is believed to be
because the EN denominators are generally smaller than the MP denominators.
To remedy the situation, they showed that introducing a suitably chosen level shift parameter can turn a 
divergent series into a convergent one by increasing the magnitude of these denominators.\cite{Mihalka_2017b}
However, like the UMP series in stretched \ce{H2},\cite{Lepetit_1988} 
the cost of larger denominators is an overall slower rate of convergence.

\begin{figure}
    \includegraphics[width=\linewidth]{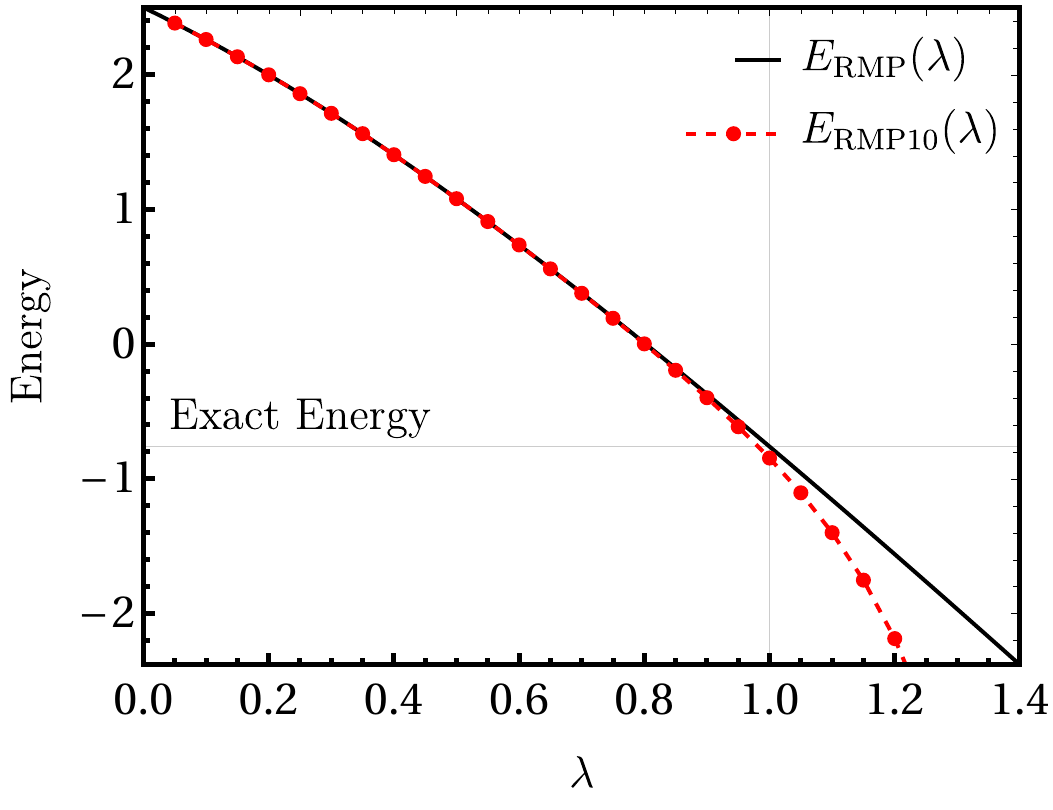}
    \caption{%
        Comparison of the scaled RMP10 Taylor expansion with the exact RMP energy as a function
        of $\lambda$ for the Hubbard dimer at $U/t = 4.5$. 
        The two functions correspond closely within the radius of convergence.
    }
    \label{fig:rmp_anal_cont}
\end{figure}

In a later study by the same group, they used analytic continuation techniques 
to resum a divergent MP series such as a stretched water molecule.\cite{Mihalka_2017a}
Any MP series truncated at a given order $n$ can be used to define the scaled function
\begin{equation}
   E_{\text{MP}n}(\lambda) = \sum_{k=0}^{n} \lambda^{k} E_\text{MP}^{(k)}.
\end{equation}
Reliable estimates of the energy can be obtained for values of $\lambda$ where the MP series is rapidly
convergent (\ie, for $\abs{\lambda} < \rc$), as shown in Fig.~\ref{fig:rmp_anal_cont} for the RMP10 series 
of the symmetric Hubbard dimer with $U/t = 4.5$.
These values can then be analytically continued using a polynomial- or Pad\'e-based fit to obtain an 
estimate of the exact energy at $\lambda = 1$.
However, choosing the functional form for the best fit remains a difficult and subtle challenge.

This technique was first generalised using complex scaling parameters to construct an analytic
continuation by solving the Laplace equations.\cite{Surjan_2018}
It was then further improved by introducing Cauchy's integral formula\cite{Mihalka_2019}
\begin{equation}
	\label{eq:Cauchy}
    E(\lambda) = \frac{1}{2\pi \i} \oint_{\mathcal{C}} \frac{E(\lambda')}{\lambda' - \lambda},
\end{equation}
which states that the value of the energy can be computed at $\lambda$ inside the complex 
contour $\mathcal{C}$ using only the values along the same contour.
Starting from a set of points in a ``trusted'' region where the MP series is convergent, their approach 
self-consistently refines estimates of the $E(\lambda')$ values on a contour that includes the physical point 
$\lambda = 1$.
The shape of this contour is arbitrary, but there must be no branch points or other singularities inside
the contour.
Once the contour values of $E(\lambda')$ are converged, Cauchy's integral formula Eq.~\eqref{eq:Cauchy} can 
be invoked to compute the value at $E(\lambda=1)$ and obtain a final estimate of the exact energy.
The authors illustrate this protocol for the dissociation curve of \ce{LiH} and the stretched water 
molecule and obtained encouragingly accurate results.\cite{Mihalka_2019} 

\section{Concluding Remarks}
\label{sec:ccl}

To accurately model chemical systems, one must choose a computational protocol from an ever growing 
collection of theoretical methods.
Until the Schr\"odinger equation is solved exactly, this choice must make a compromise on the accuracy
of certain properties depending on the system that is being studied.
It is therefore essential that we understand the strengths and weaknesses of different methods, 
and why one might fail in cases where others work beautifully.
In this review, we have seen that the success and failure of perturbation-based methods are 
directly connected to the position of exceptional point singularities in the complex plane.

We began by presenting the fundamental concepts behind non-Hermitian extensions of quantum chemistry into the complex plane, 
including the Hartree--Fock approximation and Rayleigh--Schr\"odinger perturbation theory.
We then provided a comprehensive review of the various research that has been performed 
around the physics of complex singularities in perturbation theory, with a particular focus on M{\o}ller--Plesset theory. 
Seminal contributions from various research groups have revealed highly oscillatory,
slowly convergent, or catastrophically divergent behaviour of the restricted and/or unrestricted MP perturbation series.%
\cite{Laidig_1985,Knowles_1985,Handy_1985,Gill_1986,Laidig_1987,Nobes_1987,Gill_1988,Gill_1988a,Lepetit_1988,Malrieu_2003} 
In particular, the spin-symmetry-broken unrestricted MP series is notorious
for giving incredibly slow convergence.\cite{Gill_1986,Nobes_1987,Gill_1988a,Gill_1988}
All these behaviours can be rationalised and explained by the position of exceptional points
and other singularities that arise when perturbation theory is extended across the complex plane.

The classifications of different convergence types developed by Cremer and He,\cite{Cremer_1996} 
Olsen \etal,\cite{Christiansen_1996,Olsen_1996,Olsen_2000,Olsen_2019} 
or Sergeev and Goodson\cite{Goodson_2000a,Goodson_2000b,Goodson_2004,Sergeev_2005,Sergeev_2006} are particularly
worth highlighting.
In Cremer and He's original classification, ``class A'' systems exhibit monotonic convergence and generally
correspond to weakly correlated electron pairs, while ``class B'' systems show erratic convergence after initial 
oscillations and generally contain spatially dense electron clusters.\cite{Cremer_1996}
Further insights were provided by Olsen and coworkers
who employed a two-state model to understand the various convergence behaviours of Hermitian and non-Hermitian 
perturbation series.\cite{Christiansen_1996,Olsen_1996,Olsen_2000,Olsen_2019}
The careful analysis from Sergeev and Goodson later refined these classes depending on the position of the
singularity closest to the origin, giving $\alpha$ singularities which have large imaginary component, 
and $\beta$ singularities which have a very small imaginary component.%
\cite{Goodson_2000a,Goodson_2000b,Goodson_2004,Sergeev_2005,Sergeev_2006}
Remarkably, the position of $\beta$ singularities close to the real axis can be justified as a critical 
point where one (or more) electron is ionised from the molecule, creating a quantum phase transition.\cite{Stillinger_2000}
We have shown that the slow convergence of symmetry-broken MP approximations can also be driven by a $\beta$ 
singularity and is closely related to these quantum phase transitions.

We have also discussed several resummation techniques that can be used to improve energy estimates
for both convergent and divergent series, including Pad\'e and quadratic approximants.
Furthermore, we have provided the first illustration of how the Shanks transformation can accelerate
convergence of MP approximants to improve the accuracy of low-order approximations.
Using these resummation and acceleration methods to turn low-order truncated MP series into convergent and
systematically improvable series can dramatically improve the accuracy and applicability of these perturbative methods.
However, the application of these approaches requires the evaluation of higher-order MP coefficients 
(\eg, MP3, MP4, MP5, etc) that are generally expensive to compute in practice.
There is therefore a strong demand for computationally efficient approaches to evaluate general terms in the MP 
series, and the development of stochastic,\cite{Thom_2007,Neuhauser_2012,Willow_2012,Takeshita_2017,Li_2019}
or linear-scaling approximations\cite{Rauhut_1998,Schutz_1999}
may prove fruitful avenues in this direction.

The present review has only considered the convergence of the MP series using the RHF or UHF 
reference orbitals.
However, numerous recent studies have shown that the use of orbitals optimised in the presence of the MP2 
correction\cite{Bozkaya_2011,Neese_2009,Lee_2018} or Kohn--Sham density-functional theory (DFT) orbitals 
can significantly improve the accuracy of the MP3 correction,\cite{Bertels_2019,Rettig_2020}
particularly in the presence of symmetry-breaking.
Beyond intuitive heuristics, it is not clear why these alternative orbitals provide such accurate results, 
and a detailed investigation of their MP energy function in the complex plane is therefore bound to provide
fascinating insights.
Furthermore, the convergence properties of the excited-state MP series using orbital-optimised higher energy 
HF solutions\cite{Gilbert_2008,Barca_2014,Barca_2018a,Barca_2018b} remains entirely unexplored.\cite{Lee_2019,CarterFenk_2020}

Finally, the physical concepts and mathematical tools presented in this manuscript have been illustrated 
on the symmetric (or asymmetric in one occasion) Hubbard dimer at half-filling.
Although extremely simple, these illustrations highlight the incredible versatility of the Hubbard model
for understanding the subtle features of perturbation theory in the complex plane, alongisde other examples 
such as Kohn-Sham DFT, \cite{Carrascal_2015,Cohen_2016} linear-response theory,\cite{Carrascal_2018} 
many-body perturbation theory,\cite{Romaniello_2009,Romaniello_2012,DiSabatino_2015,Hirata_2015,Tarantino_2017,Olevano_2019} 
ensemble DFT, \cite{Deur_2017,Deur_2018,Senjean_2018,Sagredo_2018,Fromager_2020} thermal DFT,\cite{Smith_2016,Smith_2018} 
wave function methods,\cite{Stein_2014,Henderson_2015,Shepherd_2016} and many more.
In particular, we have shown that the Hubbard dimer contains sufficient flexibility to describe 
the effects of symmetry breaking, the MP critical point, and resummation techniques, in contrast to the more 
minimalistic models considered previously.
We therefore propose that the Hubbard dimer provides the ideal arena for further developing our fundamental understanding
and applications of perturbation theory.

Perturbation theory isn't usually considered in the complex plane. 
But when it is, a lot can be learnt about the performance of perturbation theory on the real axis.
These insights can allow incredibly accurate results to be obtained using only the lowest-order terms in a perturbation series.
Yet perturbation theory represents only one method for approximating the exact energy, and few other methods
have been considered through similar complex non-Hermitian extensions.
There is therefore much still to be discovered about the existence and consequences of exceptional points
throughout electronic structure theory.

\begin{acknowledgements}
This project has received funding from the European Research Council (ERC) under the European Union's Horizon 2020 research and innovation programme (Grant agreement No.~863481).
HGAB gratefully acknowledges New College, Oxford for funding through the Astor Junior Research Fellowship.
\end{acknowledgements}

\bibliography{EPAWTFT}

\end{document}